\documentclass[letterpaper,11pt]{article}
\pdfoutput=1

\usepackage{xspace}
\usepackage{slashed}
\usepackage{array,multirow}
\usepackage{graphicx}
\usepackage{graphics}
\usepackage{epsfig}
\usepackage{amsfonts}
\usepackage{amsmath}
\usepackage{amssymb}
\usepackage{dsfont}
\usepackage{color}
\usepackage{xcolor}
\usepackage{cite}
\usepackage{pdflscape}
\usepackage{pifont}
\usepackage{pgfplots}
\usepackage{tikz} 
\usepackage{pgfplotstable}

\usepackage{bm}  %                                          

\newcommand{\beq}{\begin{equation}}
\newcommand{\eeq}{\end{equation}}
\newcommand{\bea}{\begin{eqnarray}}
\newcommand{\eea}{\end{eqnarray}}
\newcommand{\ben}{\begin{eqnarray*}}
\newcommand{\een}{\end{eqnarray*}}

\newcommand{\boldtau}{\mbox{\boldmath $\tau$}}
\newcommand{\boldsigma}{\mbox{\boldmath $\sigma$}}
\newcommand{\boldpi}{\mbox{\boldmath $\pi$}}

\renewcommand{\vec}[1]{{\mathbf #1}} 

\newcommand{\bma}{\begin{pmatrix}}
\newcommand{\ema}{\end{pmatrix}}

\def\lixo#1{}

\def\slashchar#1{\setbox0=\hbox{$#1$}           % set a box for#1
  \dimen0=\wd0                                    % and get its size
  \setbox1=\hbox{/} \dimen1=\wd1                  % get size of/
  \ifdim\dimen0>\dimen1                           % #1 is bigger
    \rlap{\hbox to \dimen0{\hfil/\hfil}}            % so center / in box
    #1                                             % and print #1
  \else                                          % / is bigger
    \rlap{\hbox to \dimen1{\hfil$#1$\hfil}}        % so center #1
    /                                           % and print/
 \fi}                                           %

\newcommand{\ep}{\epsilon}

\newcommand{\al}{\alpha}
\newcommand{\bt}{\beta}
\newcommand{\g}{\gamma}

\newcommand{\dt}{\delta}

\newcommand{\simu}{\sigma^{\mu\nu}}

\newcommand{\mpi}{m_{\pi}}

\newcommand{\Or}{\mathcal O}

\newcommand{\vL}{\ensuremath{\mathcal{L}}}

\newcommand{\sq}{^{2}}

\newcommand{\dslash}[1]{#1 \llap{/\kern-0.5pt}}
\newcommand{\Dslash}[1]{#1 \llap{/\kern+1.5pt}}
\newcommand{\DDslash}[1]{#1 \llap{/\kern+2.3pt}}
\newcommand{\dslashh}[1]{#1 \llap{/\kern+1pt}}

\newcommand{\nn}{\nonumber}

\newcommand{\NLDBD}{$0 \nu \beta \beta$}

\begin{document}

\begin{titlepage}

\begin{flushright}
LA-UR-17-27799\\
Nikhef 2017-039
\end{flushright}

\vspace{2.0cm}

\begin{center}
{\LARGE  \bf 
Neutrinoless double beta decay in\\ \vspace{3mm} chiral effective field theory:
\\

\vspace{3mm}

lepton number violation at dimension seven

}
\vspace{2cm}

{\large \bf  V. Cirigliano$^a$, W. Dekens$^{a,b}$, J. de Vries$^{c}$, \\ \vspace{3mm}M. L. Graesser$^a$, and E. Mereghetti$^a$ } 
\vspace{0.5cm}

\vspace{0.25cm}

{\large 
$^a$ 
{\it Theoretical Division, Los Alamos National Laboratory,
Los Alamos, NM 87545, USA}}

\vspace{0.25cm}
{\large 
$^b$ 
{\it 
New Mexico Consortium, Los Alamos Research Park, Los Alamos, NM 87544, USA
}}

\vspace{0.25cm}
{\large 
$^c$ 
{\it 
Nikhef, Theory Group, Science Park 105, 1098 XG, Amsterdam, The Netherlands
}}

\end{center}

\vspace{1cm}

\begin{abstract}
\vspace{0.1cm}
We analyze neutrinoless double beta decay ($0\nu\beta\beta$) within the framework of the Standard Model Effective Field Theory. Apart from the dimension-five Weinberg operator, the first contributions appear at dimension seven. We classify the operators and evolve them to 
the electroweak scale, where we match them to effective dimension-six, -seven, and -nine operators. In the next step, 
after renormalization group evolution to the QCD scale, we construct the chiral Lagrangian arising from these operators. We develop a power-counting scheme and derive the two-nucleon $0\nu\beta\beta$ currents up to leading order in the power counting for each lepton-number-violating operator. We argue that the leading-order contribution to the decay rate depends on a relatively small number of nuclear matrix elements. We test our power counting by comparing nuclear matrix elements obtained by various methods and by different groups. 
We find that the power counting works well for nuclear matrix elements calculated from a specific method, while, as in the case of light Majorana neutrino exchange, the overall magnitude of the matrix elements can differ by factors of two to three between methods. We calculate the constraints that can be set on dimension-seven lepton-number-violating operators from \NLDBD\ experiments and study the interplay between dimension-five and -seven operators, discussing how dimension-seven contributions affect 
 the interpretation of  \NLDBD\ in terms of the effective Majorana mass $m_{\beta \beta}$.

\end{abstract}

\vfill
\end{titlepage}

\tableofcontents

\section{Introduction}

The neutrino oscillation experiments of the last two decades have shown that neutrinos are massive particles, requiring an extension of the minimal version of the Standard Model (SM) of particle physics.
Neutrinos could have a  Dirac mass term, as all other fermions in the SM. This would require including sterile, right-handed neutrinos in the SM Lagrangian,  whose only purpose is to generate a neutrino mass. 
Yet neutrinos are the only observed  fundamental and charge-neutral fermions, so they could instead have a Majorana mass.  In the SM, a Majorana mass term is forbidden by the neutrino  $SU(2)_L \times U(1)_Y$ quantum numbers, making it impossible to construct a gauge-invariant, renormalizable  mass operator in terms of left-handed $\nu_L$ fields. Thus, in the SM one can distinguish neutrinos from antineutrinos, and  define a quantum number, lepton number  ($L$), which is conserved at the classical level.
$L$ is, however, an accidental symmetry of the SM. As soon as one introduces non-renormalizable operators, which parameterize physics at energy scales much larger than the electroweak scale,
$L$ is broken \cite{Weinberg:1979sa}, and neutrinos acquire a Majorana mass, inversely proportional to the scale of new physics $\Lambda$. 
The smallness of the neutrino mass might therefore offer a unique window on high-energy physics. 

Neutrinoless double beta decay ($0\nu\beta\beta$) experiments are the most sensitive probe of lepton number violation (LNV).
In this process two neutrons in a nucleus turn into two protons, with the emission of two electrons and no neutrinos, violating $L$ by two units.
The observation of $0\nu\beta\beta$ would have far reaching implications: it would demonstrate that neutrinos are Majorana fermions~\cite{Schechter:1981bd},  shed light on the mechanism of neutrino mass generation, 
and give insight on leptogenesis scenarios for the generation of the matter-antimatter asymmetry in the universe\cite{Davidson:2008bu}. 
The current experimental  limits on the half-lives are already 
impressive~\cite{Gando:2012zm,Agostini:2013mzu,Albert:2014awa,Alfonso:2015wka,Andringa:2015tza,Arnold:2015wpy,Elliott:2016ble,Arnold:2016ezh,Agostini:2017iyd,KamLAND-Zen:2016pfg},
at the level  of  $T^{0\nu}_{1/2} >  5.3\times10^{25}$~y  for $^{76}$Ge~\cite{Agostini:2017iyd}
and  $T^{0\nu}_{1/2} >  1.07\times10^{26}$~y  for $^{136}$Xe~\cite{KamLAND-Zen:2016pfg}, 
with  next generation ton-scale experiments aiming at a sensitivity of  $T^{0\nu}_{1/2} \sim 10^{27-28}$~y.

By itself, the observation of \NLDBD \   would not  immediately point to the underlying physical origin of  LNV. 
While \NLDBD  \ searches are  commonly interpreted in terms of the exchange of a light Majorana neutrino,  
in generic beyond-the-SM (BSM) models, \NLDBD\ receives contributions from several competing mechanisms 
(for a review see Ref.~\cite{Rodejohann:2011mu}). 
Well-studied examples are  left-right symmetric models \cite{Mohapatra:1974hk,Senjanovic:1975rk,Mohapatra:1983aa},  
which  contain an extended gauge and Higgs sector, as well as  heavy right-handed Majorana neutrinos.  
In these models light Majorana neutrinos  acquire mass via the type-I see-saw (via right-handed neutrinos) 
and / or the type-II see-saw (Higgs triplet) and can mediate \NLDBD.
In addition, however, \NLDBD\ receives contributions from the exchange of heavy right-handed neutrinos, mediated by the gauge boson of 
the additional $SU(2)_R$ gauge group,  from the mixing of light- and -heavy neutrinos  
or from the exchange of Higgs triplets~\cite{Doi:1985dx,Rodejohann:2011mu,Tello:2010am,Ge:2015yqa}.
Depending on the masses of the right-handed neutrinos and gauge boson, and on the Yukawa couplings  
of the left- and right-handed neutrinos to the Higgs, \NLDBD\ can be dominated by 
light-neutrino exchange, heavy-neutrino exchange, or receive several contributions of similar size.

Keeping  explicit model realizations in mind, in this paper we investigate \NLDBD\ in the framework of the SM Effective Field Theory (SM-EFT) \cite{Weinberg:1979sa,Buchmuller:1985jz}.
In this framework, the SM is complemented by higher-dimensional operators,  expressed in terms of SM fields and invariant under the SM gauge group. 
The coefficients of these operators are suppressed by powers of the scale $\Lambda$ at which new physics arises.  
There is a single gauge-invariant dimension-five operator~\cite{Weinberg:1979sa}. This operator violates $L$ by two units, and, as already mentioned, provides the first contribution to the neutrino Majorana mass.
Going further, there are no $\Delta L=2$ dimension-six operators \cite{Buchmuller:1985jz,Grzadkowski:2010es}, but there are several at dimension-seven \cite{Lehman:2014jma}, and -nine \cite{Prezeau:2003xn,Graesser:2016bpz}, and higher~\cite{deGouvea:2007qla}. \footnote{All $L=2$, $B=0$ operators have odd dimension \cite{Kobach:2016ami}.}
Notice that here we are not extending the SM field content with a light right-handed neutrino, but the construction of the effective operators can be generalized to include it   \cite{Liao:2016qyd}.

We systematically study the constraints on $SU(2)_L \times U(1)_Y$-invariant dimension-seven operators from \NLDBD. 
After defining the operator basis in Sec.~\ref{sec:operators}, in Sec. \ref{sec:lowE}  we integrate out heavy SM degrees of freedom, such as the Higgs and the $W$ boson, and match onto a low-energy $\Delta L=2$ Lagrangian
that only contains leptons and light quarks, suitable for the descriptions of low-energy processes such as double-beta decay.
The resulting Lagrangian contains the neutrino Majorana mass and transition magnetic moments,  dimension-six and -seven semileptonic four-fermion operators, as well as dimension-nine six-fermion operators. Of these operators, those of dimension-six and -seven  give rise to non-standard $\Delta L=2$ single beta decay
and to long-range neutrino-exchange contributions to \NLDBD\  not proportional to the neutrino mass. Instead,  the dimension-nine operators, which involve four quarks and two electrons, induce new \NLDBD\ contributions without the exchange of a neutrino.

In Sec.~\ref{ChiPT} we match the quark-level $\Delta L=2$ Lagrangian onto Chiral Perturbation Theory ($\chi$PT), the low-energy EFT of QCD,
and we discuss the hadronic input needed to constrain dimension-seven operators.  In Sec.~\ref{sec:2btOperators} we introduce a power counting and derive the neutrino potentials in $\chi$PT up to the first non-vanishing orders. The power counting reduces the number of matrix elements that are relevant at leading order in the chiral counting. 
The contribution of dimension-six $\Delta L=2$ operators to \NLDBD\ was considered in Refs. \cite{Doi:1985dx,Pas:1999fc,Deppisch:2012nb,Helo:2016vsi,Horoi:2017gmj},
while six-fermion dimension-nine were studied in Refs.  \cite{Vergados:1986td,Faessler:1996ph,Pas:2000vn,Prezeau:2003xn,Deppisch:2012nb, Bonnet:2012kh, Helo:2016vsi,Horoi:2017gmj}.
In Sec.~\ref{sec:2btOperators} we discuss similarities and differences between the neutrino potentials we obtain and the existing literature.

In Sec. \ref{Master} we obtain our main result which is the derivation of the master formula for $0 \nu \beta \beta$ half-life up to dimension-seven in the SM-EFT expansion and the first non-vanishing order in $\chi$PT. For earlier versions of such formula see, for example, Refs. \cite{Pas:1999fc,Pas:2000vn}.
The master formula  includes the following important effects: 
\begin{itemize} 
\item QCD renormalization group evolution of the dimension-seven operators from the high-energy scale to the weak scale, followed by the QCD evolution of the induced dimension-six, -seven, and -nine operators from the weak scale to the QCD scale. 
\item Up-to-date hadronic input for the low-energy constants, which are becoming increasingly under control. We find that nine low-energy constants are needed. Six of these are well-known from either experimental or lattice QCD (LQCD) input, while we estimate the remaining three with naive dimensional  analysis. The reader is referred to Table \ref{Tab:LECs} as well as Fig.\ \ref{fig:lecPlot} which illustrates the impact of the uncertainty on the unknown low-energy constants on the constraints on a particular $\Delta L=2$ Wilson coefficient.

\item Consistent power-counting in the chiral effective theory for the neutrino potentials induced by the dimension-seven operators, see Table~\ref{TabPC}. For some operators we find the first non-zero contributions in $0^+\rightarrow 0^+$ transitions to arise at next-to- or next-to-next-to-leading order in the chiral expansion. 
\item Long-distance contributions arising from either neutrino or pion exchange. When the latter is chirally suppressed, subleading short-range pion-nucleon and contact 4-nucleon contributions are considered. The full interference of all effects is included. 
\end{itemize} 

We find the master formula to depend on only a handful of nuclear matrix elements, a smaller set than typically considered, and we perform comparisons of calculations of the nuclear matrix elements elements already existing in the literature (see Table \ref{tab:comparison} and Figs.~\ref{FigNME} and \ref{FigNME2}). We test our power counting explicitly by comparing the sizes of different matrix elements and by comparing matrix elements related by symmetry. Bounds on the induced dimension-six, -seven, and -nine operators, as well as the original dimension-seven operators, are obtained in Sect.~\ref{Bounds} and presented in Tables \ref{tab:limits} and \ref{tab:limits2} and range from tens to hundreds of TeV, assuming a single dimension-seven operator (Tables \ref{tab:limits2} and \ref{fig:limitsChart}) or single induced operator (Table \ref{tab:limits}) turned on at a time. In Sect.~\ref{two-coupling-analysis} we discuss scenarios in which both a light Majorana neutrino mass and a dimension-seven operator contribute to the \NLDBD\ rate. We study what additional experimental input can be used to disentangle the various $\Delta L=2$ contributions to \NLDBD\ \!. We summarize, conclude, and give an outlook in Sect.~\ref{conclusions}.

\section{Dimension-seven SM-EFT operators}\label{sec:operators}

{\renewcommand{\arraystretch}{1.3}\begin{table}\small
\center
\begin{tabular}{||c|c||c|c||}
\hline Class $1$& $\psi^2 H^4$  & Class $5$ &  $\psi^4 D$\\
\hline
$\mathcal O^{}_{LH}$  & $\ep_{ij}\ep_{mn}(L_i^TCL_m )H_j H_n (H^\dagger H)$ & $\mathcal O^{(1)}_{LL\bar d u D}$&  $\ep_{ij} (\bar d \g_\mu u)(L_i^T C (D^\mu L)_j)$  \\\hline
\hline Class $2$&  $\psi^2 H^2 D^2$ & Class $6$ & $\psi^4 H$\\\hline
$\mathcal O^{(1)}_{LHD}$  & $\ep_{ij}\ep_{mn}(L_i^TC (D_\mu L)_j )H_m  (D^\mu H)_n$  & $\mathcal O_{LL \bar e H}$  & $\ep_{ij}\ep_{mn}(\bar e L_i)(L_j^TC L_m )H_n$\\
$\mathcal O^{(2)}_{LHD}$  & $\ep_{im}\ep_{jn}(L_i^TC (D_\mu L)_j )H_m  (D^\mu H)_n$  & $\mathcal O^{(1)}_{LL Q\bar d  H}$  &  $\ep_{ij}\ep_{mn}(\bar d L_i)(Q_j^TC L_m )H_n$ \\
\cline{1-2} Class $3$& $\psi^2 H^3 D$  & $\mathcal O^{(2)}_{LL Q\bar d  H}$  &  $\ep_{im}\ep_{jn}(\bar d L_i)(Q_j^TC L_m )H_n$\\\cline{1-2} 
$\mathcal O^{}_{LHDe}$ & $\ep_{ij}\ep_{mn}(L_i^TC \g_\mu e )H_j H_m  (D^\mu H)_n$ & $\mathcal O_{LL \bar Q u  H}$    &$\ep_{ij}(\bar Q_m u)(L_m^TC L_i )H_j$\\
\cline{1-2} \cline{1-2} Class $4$& $\psi^2 H^2 X $ & $\mathcal O^{}_{L e u \bar d  H}$   & $\ep_{ij}( L_i^T C\g_\mu e)(\bar d\g^\mu  u )H_j$\\\cline{1-2} 
$\mathcal O^{}_{LHB}$  & $\ep_{ij}\ep_{mn}g'(L_i^TC \simu L_m ) H_j H_n B_{\mu\nu}$ & & \\
$\mathcal O^{}_{LHW}$  & $\ep_{ij} (\ep\tau^I)_{mn} g(L_i^TC \simu L_m ) H_j H_n W^I_{\mu\nu}$ & & \\
\hline
\end{tabular}
\caption{Basis of $\Delta L=2$ baryon-number-conserving   dimension-seven operators derived in Ref.\ \cite{Lehman:2014jma}.  
} \label{dim7}
\end{table}}

The complete list of  dimension-seven $\Delta L=2$ operators, invariant under the gauge group of the Standard Model, was built in Ref.~\cite{Lehman:2014jma}, and it is  
summarized  in Table \ref{dim7}. A subset of the operators was published in Refs.~\cite{Babu:2001ex,Bell:2006wi}, and a few redundancies  were eliminated in Ref.~\cite{Liao:2016hru}.  
At the scale of new physics, $\Lambda$, we have the following $\Delta L=2$ Lagrangian
\begin{equation}\label{eq:1stLag}
\vL^{(\Delta L=2)} = \ep_{kl}\ep_{mn}(L_k^T\,\mathcal C^{(5)}\,CL_m )H_l H_n +\sum_i \mathcal C_i \mathcal O_i\,,\qquad v^3 \mathcal C_i = \mathcal O\left(\frac{v^3}{\Lambda^3}\right),
\end{equation}
where the first term is the dimension-five Weinberg operator, with $\mathcal C^{(5)}$ a $3\times 3$ matrix in flavor space. Furthermore, $i$ runs over the labels of the operators defined in Table \ref{dim7}. 
In Table \ref{dim7},  $L$ and $Q$ denote the left-handed quark and lepton doublets, $L = (\nu_L, e_L)^T$, $Q = (u_L, d_L)^T$, while $u_R$ and $d_R$ are right-handed quarks, singlet under $SU(2)_L$.
$H$ denote the scalar doublet
\begin{equation}
H = \frac{v}{\sqrt{2}} U(x) \left(\begin{array}{c}
0 \\
1 + \frac{h(x)}{v}
\end{array} \right)\,,
\end{equation}
where $v=246$ GeV is the scalar field vacuum expectation value (vev),  $h(x)$ is the Higgs field, and $U(x)$ is a $SU(2)$ matrix that encodes the three Goldstone bosons.
The covariant derivative $D_\mu$ is defined as $D_\mu  = \partial_\mu -ig_s t^a G^a_\mu-g \frac{\tau^I}{2}W^I_\mu-g' Y B_\mu$,
where $t^a$ and $\tau^{I}/2$ are $SU(3)$ and $SU(2)$ generators, in the representation of the field on which the derivative acts. $Y$ is the hypercharge quantum number, $Y = -1/2$ for $L$ 
and $Y=1/2$ for $H$. $\epsilon$ is a completely antisymmetric tensor, with $\epsilon_{12} = +1$.
$C$ is the charge conjugation  matrix, $C=i\g_2\g_0$, which, in this basis, satisfies $C=-C^T=-C^\dagger=-C^{-1}$.

All the couplings $\mathcal C_i$  have lepton flavor indices, which we omit unless explicitly needed, while the couplings of the four-fermion 
operators  in Classes 5 and 6 also carry indices for the quark flavors. Here we are only concerned with couplings to the first generation of quarks.

There are a few special cases in the above operator basis.
Firstly, the dimension-five  operator and $\mathcal O_{LH}$ trivially contribute to  $0\nu \beta\bt$ as they simply gives rise to a Majorana mass term below the electroweak scale, 
$\mathcal C^{(5)}\mathcal O^{(5)}+\mathcal C_{LH}\mathcal O_{LH}\to \frac{v^2}{2}(\mathcal C^{(5)}+\frac{v^2}{2} \mathcal C_{LH}) \nu^T C \nu$. 
The operator $\mathcal O_{LHB}$, and the component of $\mathcal O_{LHW}$ that is antisymmetric with respect to the lepton flavor indices, do not give rise to $0\nu \beta\bt$ at tree level,
but are well constrained by the transition magnetic moments of the neutrinos, as we discuss further in Section \ref{ref:magnetic-moments}. Also, both 
$\mathcal{O}^{(2)}_{LHD}$ and
$\mathcal O_{LL\bar e H}$  do not induce \NLDBD\ at tree level. 
For these two operators, in Section \ref{ref:neutrino-masses} we consider radiative corrections, such as the
one-loop mixing  onto the neutrino mass ($\mathcal O_{LH}$) and magnetic moment ($\mathcal O_{LHB}$ and $\mathcal O_{LHW}$)
operators. The effects of $\mathcal O_{LL\bar e H}$ are however suppressed by three and one power of the electron Yukawa coupling, respectively. 
Alternatively, one can study $\Delta L =2 $  decays such as $\mu^+ \rightarrow e^+  \bar{\nu}_e \bar{\nu}_\mu$ \cite{Armbruster:2003pq}.
We briefly discuss bounds on $\mathcal C_{LL\bar e H}$ arising from muon decay in Sec. \ref{ref:muon-decay}.

The remaining operators in Table \ref{dim7} --namely, the following 8 operators ${\mathcal O}^{(1)}_{LHD}$, ${\mathcal O}_{LHDe}$, ${\mathcal O}_{LHW}$, 
${\mathcal O}^{(1)}_{LL\bar d u D}$, ${\mathcal O}^{(1),(2)}_{LLQ\bar d H}$, ${\mathcal O}_{LLQ\bar u H}$ and ${\mathcal O}_{Leu\bar d H}$ --
induce tree-level corrections to \NLDBD. 
Before discussing the effects generated by these operators at the electroweak scale, we briefly comment on the QCD running between the scale $\Lambda$ and $\mu\sim m_W$.
As the majority of the dimension-seven operators do not involve quarks, or only involve a quark vector or axial current, most of these operators do not run under QCD at one loop. The only exceptions are $\mathcal O^{(1,2)}_{LL Q\bar d  H}$ and $\mathcal O_{LL \bar Q u  H}$. The latter  runs like a scalar current, while the former two operators can be written as combinations of tensor and scalar currents,
\bea
\sum_{i=1}^2\mathcal C_{LLQ\bar d H}^{(i)}\mathcal O_{LLQ\bar d H}^{(i)}=\sum_{i=1}^2\left[C_S^{(i)} O_S^{(i)}+C_T^{(i)} O_T^{(i)}\right]\,,
\eea
with $O_S^{(1)} =\frac{1}{2}\epsilon_{ij}\epsilon_{mn} (\bar d Q_j)(L_i^{ T}C L_m )H_n$ and $O_T^{(1)} =\frac{1}{8}\epsilon_{ij}\epsilon_{mn} (\bar d \simu Q_j)(L_i^{ T}C \sigma_{\mu\nu} L_m )H_n$ and $O_{S,T}^{(2)}$ can be obtained by replacing $\epsilon_{ij}\epsilon_{mn}\to \epsilon_{im}\epsilon_{jn}$. The couplings of these operators are given by,
\bea
C^{(1),ij}_{S,T} &=& -\frac{\mathcal C_{LLQ\bar d H}^{(1),ij} \pm\mathcal C_{LLQ\bar d H}^{(1),ji} }{2}\,,\quad C^{(2),ij}_{S,T} = -\frac{\mathcal C_{LLQ\bar d H}^{(1),ij} \mp\mathcal C_{LLQ\bar d H}^{(1),ji} }{4}-\frac{\mathcal C_{LLQ\bar d H}^{(2),ij} \mp\mathcal C_{LLQ\bar d H}^{(2),ji} }{2}\,.
\eea
Here the $i$ and $j$ indicate the generation of the left- and right-most lepton fields, respectively. The running is then given by
\bea \label{eq:dim7RG}
\frac{d}{d\ln \mu} \mathcal C_{LL \bar Q u  H} &=& -6 C_F\, \frac{\al_s}{4\pi}  \mathcal C_{LL \bar Q u  H}\,,\quad \nn
\frac{d}{d\ln \mu} C_{S}^{(1,2),ij} = -6 C_F\, \frac{\al_s}{4\pi} C_{S}^{(1,2),ij}\,,\\
\frac{d}{d\ln \mu} C_{T}^{(1,2),ij} &=& 2 C_F\, \frac{\al_s}{4\pi} C_{T}^{(1,2),ij}\,,
\eea
where $C_F=(N_c\sq-1)/2N_c$, and $N_c = 3$ is the number of colors. The analytic solutions to these equations are discussed in Appendix \ref{AppRG}, where we also give numerical relations between $\mathcal C_i(\Lambda)$ and $\mathcal C_i(m_W)$.

Note that Eq.\ \eqref{eq:dim7RG} only takes into account the QCD running, which should be the dominant contribution to the RG up to scales, $\mu\sim 10$ TeV. For larger renormalization scales, which one is sensitive to if $\Lambda$ is significantly above the electroweak scale, electroweak contributions could become relevant as well (since $\al_2(\mu)\simeq \frac{1}{2}\al_{s}(\mu)$ for $\mu\simeq 10$ TeV). However, as the largest RG effects result from relatively low scales, $\mu<$ TeV, and the electroweak RGEs are currently not known in the literature, we neglect their effects here.
\section{Low-energy Lagrangian}\label{sec:lowE}

After the breaking of electroweak symmetry, the low-energy $\Delta L = 2$ Lagrangian contains neutrino Majorana masses and transition magnetic moments. In addition, there appear  
several dimension-six and -seven four-fermion operators as well as dimension-nine six-fermion operators, which give long- and short-distance contributions to $0\nu \beta\beta$ decay, respectively. We write
\begin{eqnarray}\label{DL2lag}
\mathcal L_{\Delta L=2} =  -  \frac{1}{2}  (m_\nu)_{ij} \nu^T_{L,\, i} \, C \nu_{L, \, j}  +  \mu_{i j}\, \nu^{T}_{L,\, i} \, C \sigma^{\mu \nu} \nu_{L,\,j }\, e F_{\mu \nu} +  \mathcal L^{(6)}_{\Delta L = 2} +   \mathcal L^{(7)}_{\Delta L = 2} + \mathcal L^{(9)}_{\Delta L = 2}\,.
\end{eqnarray}
We choose to work in the mass basis of the charged leptons, but the flavor basis of the neutrinos. This implies that the charged-current interaction and the charged-lepton Yukawa matrix are flavor diagonal, while the neutrino Majorana mass matrix in Eq.~\eqref{DL2lag} is not.
Thus the flavor indices $i,j$ in Eq.~\eqref{DL2lag}, and in what follows, run over the charged leptons, $ i,j \in \{e,\mu,\tau\}$.

The neutrino mass and magnetic moment terms are discussed in Sec. \ref{Bounds}, and here we focus on the operators that mediate 
$0\nu\bt\bt$. 
Below the electroweak scale the gauge-invariant dimension-seven operators of Table \ref{dim7} induce the following dimension-six, -seven, and -nine operators
\bea
\mathcal L^{(6)}_{\Delta L = 2}& =& \frac{2 G_F}{\sqrt{2}} \Bigg\{ 
C^{(6)}_{\textrm{VL},ij} \,  \bar u_L \gamma^\mu d_L \, \bar e_{R,i} \,  \gamma_\mu \, C\bar \nu^T_{L,j} + 
C^{(6)}_{\textrm{VR},ij} \,  \bar u_R \gamma^\mu d_R \, \bar e_{R,i}\,  \gamma_\mu  \,C\bar\nu_{L,j}^T \label{lowenergy6}   \\
& & +
C^{(6)}_{ \textrm{SR},ij} \,  \bar u_L  d_R \, \bar e_{L,i}\, C  \bar \nu^T_{L,j} + 
C^{(6)}_{ \textrm{SL},ij} \,  \bar u_R  d_L \, \bar e_{L,i} \, C  \bar\nu_{L,j}^T + 
C^{(6)}_{ \textrm{T},ij} \,  \bar u_L \sigma^{\mu\nu} d_R \, \bar e_{L,i}  \sigma_{\mu\nu}  \, C\bar\nu_{L,j}^T
\Bigg\}  +{\rm h.c.}\nn
\\
\mathcal L^{(7)}_{\Delta L = 2} &=& \frac{2 G_F}{\sqrt{2} v} \Bigg\{ 
C^{(7)}_{\textrm{VL},ij} \,  \bar u_L \gamma^\mu d_L \, \bar e_{L,i} \, C \,  i \overleftrightarrow{\partial}_\mu \bar \nu_{L,j}^T  +
C^{(7)}_{\textrm{VR},ij} \,  \bar u_R \gamma^\mu d_R \, \bar e_{L,i} \, C i \overleftrightarrow{\partial}_\mu \bar \nu^T_{L,j}  \Bigg\}  +{\rm h.c.}\label{lowenergy7}
\\
\mathcal L^{(9)}_{\Delta L = 2}  &=& \frac{\bar e_{L,i} C \bar e_{L,j}^T}{v^5}\bigg\{C^{(9)}_{1,ij}\,\bar u_L \gamma^\mu d_L\,  \bar u_L \gamma_\mu d_L +C^{(9)}_{4,ij}\, \bar u_L \gamma^\mu d_L\,  \bar u_R \gamma_\mu d_R+C^{(9)}_{5,ij}\, \bar u^\al_L \gamma^\mu d^\bt_L\,  \bar u^\bt_R \gamma_\mu d^\al_R\bigg\}\nn\\&& + {\rm h.c.}\label{lowenergy9}
\eea
The coefficients $C^{(6,7,9)}_{ij}$ are all defined to be dimensionless.

Keeping the  lepton flavor structure, the matching coefficients for the dimension-six operators at the electroweak scale are given by\footnote{Note that the operators in Eqs. \eqref{lowenergy6}, 
\eqref{lowenergy7}, and \eqref{lowenergy9} are defined to give rise to $d \rightarrow u$ transitions, whereas the opposite convention is used for the dimension-seven operators in Table \ref{dim7}.}
\begin{eqnarray}\label{match6}
\frac{1}{v^{3}}\, C^{(6)}_{\textrm{VL},ij} & = & - \frac{i}{\sqrt{2}} V_{ud}  \mathcal C_{LHDe,ji}^* + 4  V_{ud} \frac{m_e}{v} \mathcal C_{LHW,ji}^*\,, \nonumber \\ 
\frac{1}{v^{3}}\,C^{(6)}_{\textrm{VR},ij} &=& \frac{1}{\sqrt{2}}  \mathcal C_{Leu\bar dH,ji}^*\,,  \nonumber \\
\frac{1}{v^{3}}\,C^{(6)}_{\textrm{SR},ij} & =& \frac{1}{2\sqrt{2}} \left(\mathcal C^{(2)}_{LL Q \bar d H,ij}-\mathcal C^{(2)}_{LL Q \bar d H,ji}+ \mathcal C^{(1)}_{LL Q \bar{d} H ,ij} \right)^* \nn\\
&&+ \frac{V_{ud}}{2}\frac{m_d}{v} \left(\mathcal C^{(1)}_{LHD,ij}-\mathcal C^{(1)}_{LHD,ji}-\mathcal C^{(2)}_{LHD,ji} \right)^*
-\frac{i}{2}\frac{m_u}{v}\left(\mathcal C^{(1)}_{LL \bar d u D,ij}-\mathcal C^{(1)}_{LL \bar d u D,ji}\right)^*\,, \nonumber \\  
\frac{1}{v^{3}}\,C^{(6)}_{\textrm{SL},ij}  &=&  
 \frac{1}{\sqrt{2}}  \mathcal C_{LL \bar Q u H,ij}^* \nn\\
 &&- \frac{V_{ud}}{2}\frac{m_u}{v} \left(\mathcal C^{(1)}_{LHD,ij}-\mathcal C^{(1)}_{LHD,ji}-   \mathcal C^{(2)}_{LHD,ji} \, \right)^*
+\frac{i}{2}\frac{m_d}{v}\left( \mathcal C^{(1)}_{LL \bar d u D,ij}-\mathcal C^{(1)}_{LL \bar d u D,ji}\right)^*\,,  \nonumber \\
\frac{1}{v^{3}}\,C^{(6)}_{\textrm{T},ij } &= &  \frac{1}{8 \sqrt{2}} \left(\mathcal C^{(2)}_{LL Q \bar d H,ij}+\mathcal C^{(2)}_{LL Q \bar d H,ji}+ \mathcal C^{(1)}_{LL Q \bar d H,ij}\right)^* \,\,. 
\end{eqnarray}
For the dimension-seven operators we have
\begin{eqnarray}\label{match7}
\frac{1}{v^{3}}\,C^{(7)}_{\textrm{VL},ij} &=& -  \frac{V_{ud}}{2}\left( \mathcal \,C^{(1)}_{LHD,ij}+\mathcal C^{(1)}_{LHD,ji} +   \mathcal C^{(2)}_{LHD,ji} + 8 \mathcal C_{LHW,ji} \right)^* \,, \nonumber \\ 
\frac{1}{v^{3}}\,C^{(7)}_{ \textrm{VR},ij} &=& - \frac{i}{2}  \, \left(\mathcal C^{(1)}_{LL \bar d u D,ij}+\mathcal C^{(1)}_{LL \bar d u D,ji}\right)^*\, ,  
\end{eqnarray}
while the matching conditions for the dimension-nine operators are
\bea\label{match9}
\frac{1}{v^{3}}\,C^{(9)}_{1,ij} &=& - 2 V_{ud}^2 \left(  \mathcal C^{(1)}_{LHD,ij}+ 4 \mathcal C_{LHW,ij}\right)^*\,, \nn\\
\frac{1}{v^{3}}\,C^{(9)}_{4,ij} &=& - 2 i V_{ud} \,   \mathcal C^{(1)*}_{LL \bar d u D,ij}\,,\qquad \frac{1}{v^{3}}\,C^{(9)}_{5,ij}=0\, .
\eea
Although we explicitly kept the lepton flavors in the matching coefficients, only one of the elements will actually contribute to $0\nu\bt\bt$. This is due to the fact that we require two electrons in the final state, which for the dimension-nine operators implies only the $C^{(9)}_{i,\, ee}$ element can contribute. In addition, this means that the long-range contributions of the dimension-six and -seven operators have to be mediated by $\nu_e$ (since the SM weak current has to produce an electron), implying that only the $C_{i,\,ee}^{(6),(7)}$ component can contribute as well. In the following we therefore drop the flavor indices and use the shorthand, $C_{i,\, ee}\to C_{i}$.

The coefficients in Eqs.\ \eqref{match6}, \eqref{match7}, and \eqref{match9} need to be evolved from the matching scale $\mu \sim m_W$ to scales $\mu \sim 2$ GeV, where the matching to chiral perturbation theory and LQCD calculations is performed. The vector operators, $C^{(6)}_{\textrm{ VL,\,VR}}$ and $C^{(7)}_{\textrm{VL,\,VR}}$, 
consisting of quark non-singlet axial and vector currents, do not run in QCD\footnote{In the $\overline{\textrm{MS}}$ scheme,
the renormalization factor of the non-singlet axial current $Z^{A}_{\overline{\textrm{MS}}}$ receives non-vanishing contributions starting at two loops 
\cite{Larin:1991tj}. It is however always possible to introduce a finite renormalization that restores the non-renormalization of the flavor non-singlet current \cite{Collins:1984xc}. }.
The renormalization group equations (RGEs) of the scalar and tensor operators below $\mu=m_W$ are given by
\bea
\frac{d}{d\ln \mu} C^{(6)}_{\textrm{SL\,(SR)}} &=& -6 C_F\, \frac{\al_s}{4\pi}  C^{(6)}_{\, \textrm{SL\,(SR)}}\,,\quad \nn
\frac{d}{d\ln \mu}  C^{(6)}_{\textrm T} = 2 C_F\, \frac{\al_s}{4\pi} C^{(6)}_{\textrm T}\,.
\eea
Here we have suppressed the flavor indices as the QCD running is independent of them. The above RGEs correct the anomalous dimensions derived in Ref.\ \cite{Arbelaez:2016zlt}.
The RGEs of the dimension-nine operators are given by \cite{Buras:2000if,Buras:2001ra}
\bea
\frac{d}{d\ln \mu} C^{(9)}_{1} &=& 6\left(1-\frac{1}{N_c}\right)\, \frac{\al_s}{4\pi}  C^{(9)}_{1}\,,\nn\\
\frac{d}{d\ln \mu}  \bma C^{(9)}_{4}\\C^{(9)}_{5}\ema  &=&  \frac{\al_s}{4\pi}\,\bma 6/N_c&0\\-6&-12 C_F\ema \bma C^{(9)}_{4}\\C^{(9)}_{5}\ema\,.
\eea
The analytic (and numerical) relations between $C_i(m_W)$ and $C_i(2\, {\rm GeV})$ that result from the above RGEs are discussed in Appendix \ref{AppRG}. 

\section{Chiral Perturbation Theory}\label{ChiPT}
Having obtained the relevant $\Delta L=2$ interactions around $2$ GeV, we want to study their manifestation at even lower energies. We do so by applying  the framework of chiral perturbation theory ($\chi$PT) \cite{Weinberg:1978kz,Gasser:1983yg,Bernard:1995dp}, and its generalization to multi-nucleon systems, chiral EFT ($\chi$EFT) \cite{Weinberg:1990rz,Weinberg:1991um,Ordonez:1992xp,Epelbaum:2008ga}.
$\chi$PT is the low-energy EFT of QCD and consists of the interactions among the relevant low-energy degrees of freedom (mesons, baryons, photons, and leptons) that incorporate the symmetries of the underlying microscopic theory: QCD supplemented by electroweak four-fermion interactions and, in our case, $\Delta L=2$ operators. 

A particularly important role at low energy is played by the approximate symmetry of QCD under the
chiral group $SU (2)_L \times SU(2)_R$. Since it is not manifest in the spectrum, which instead exhibits
an approximate isospin symmetry, chiral symmetry must be spontaneously broken down to the isospin
subgroup $SU (2)_{I}$. The corresponding Goldstone bosons can be identified with the pions.
Chiral symmetry and its spontaneous breaking strongly constrain the form of the interactions among
nucleons and pions. In particular, in the limit of vanishing quark masses and charges, when chiral
symmetry is exact, pion interactions are derivative, allowing for an expansion in 
$p/\Lambda_\chi$, where $p$ is the typical momentum scale in a process and $\Lambda_\chi \sim m_N \sim 1$ GeV 
is the intrinsic mass scale of QCD. These constraints are captured by  $\chi$PT. 

The $\chi$PT Lagrangian is  obtained by constructing all chiral-invariant interactions between nucleons and pions.
In principle, an infinite number of interactions exist, but they can be ordered by a power-counting scheme. We use the chiral index $\Delta = d + n/2-2$, where $d$ counts the number of derivatives and $n$ counts the number of nucleon fields \cite{Weinberg:1978kz}. The higher the chiral index, the more suppressed the effects of a coupling are by factors of $p/\Lambda_\chi \sim m_\pi/\Lambda_\chi \sim \epsilon_\chi$, where we introduced $\epsilon_\chi = m_\pi/\Lambda_\chi$.
Chiral symmetry is explicitly broken by the quark masses and charges, and, in our case, by electroweak and $\Delta L=2$ operators, 
but the explicit breaking is small, and can be systematically included in the power counting by considering  $ m_q\sim\mpi^2\sim p^2$.
Because the $\Delta L=2$ interactions are associated with very small parameters, we only consider operators linear in the $\Delta L=2$ couplings. 

The coupling constants of the effective interactions in $\chi$PT, usually called low-energy constants (LECs), are not fixed by symmetry, and they
capture the nonperturbative nature of low-energy QCD. 
In principle these LECs are unknown but their sizes can be estimated from naive dimensional analysis (NDA) \cite{Manohar:1983md}, or, preferably, fitted to data or calculated from QCD directly for instance by using lattice simulations. As we discuss below, for $0\nu\beta\beta$ processes most LECs are relatively well known although there are some exceptions.

In the mesonic and single-nucleon sector, all momenta and energies are typically $\sim p$. The perturbative expansion of the $\chi$PT Lagrangian then implies that the scattering amplitudes can also be expanded in $p/\Lambda_\chi$, with every loop (using $4\pi F_\pi \sim \Lambda_\chi$, where $F_\pi$ is the pion decay constant) or insertions of subleading terms in the $\chi$PT Lagrangian causing further suppression.

\begin{figure}
\center
\includegraphics[width=0.75\textwidth]{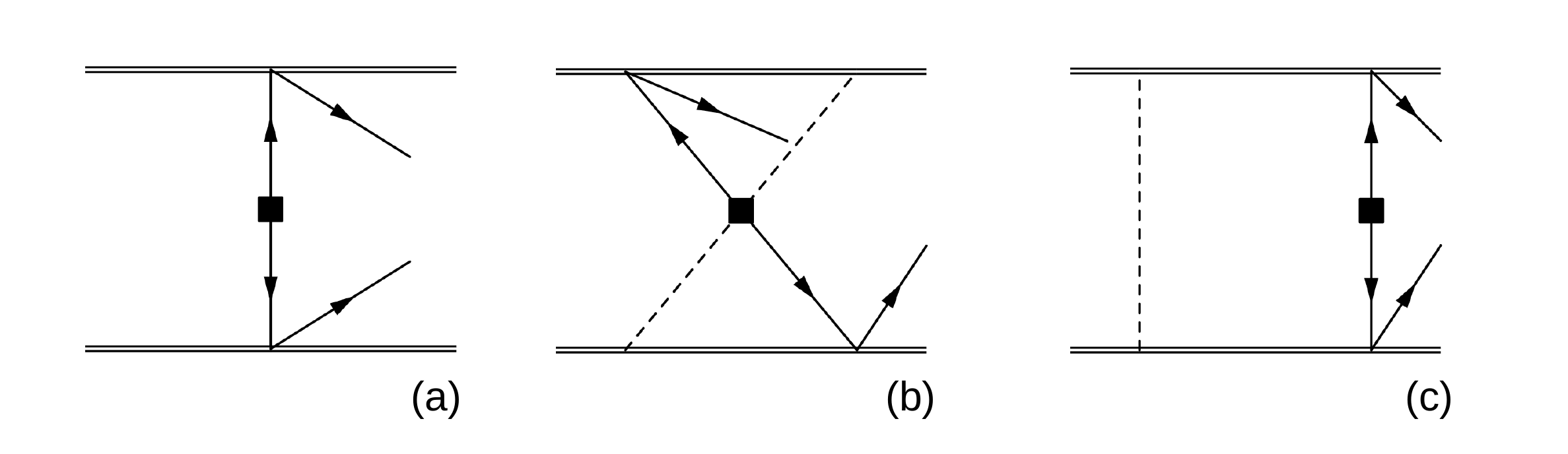}
\caption{
Examples of irreducible (diagrams $(a)$ and $(b)$) and reducible (diagram $(c)$) two-nucleon LNV diagrams. Double and single lines denote, respectively, nucleon and lepton fields. The black square denotes an insertion of the neutrino Majorana mass. Notice that diagram $(c)$ is non planar, i.e. the pions ``go around'' the neutrino line.   
The first two diagrams respect the $\chi$PT power counting, and their scaling is determined by the chiral index $\Delta$ of the vertices  and by the number of loops. 
The sum of two-nucleon irreducible diagrams defines the $0\nu\beta\beta$ two-nucleon transition operator, or ``neutrino potential''.  
In the third diagram the nucleons can be close to their mass shell, and the diagram is enhanced by $m_N/p$ with respect to the $\chi$PT
power counting. This diagram is included by taking the matrix element of the neutrino potential between the nuclear bound-state wavefunctions. 
}\label{fig:red}
\end{figure}

For system with two or more nucleons, in addition to the momentum  $p$, the energy scale  $p^2/2m_N$ becomes relevant.
Nucleon-nucleon amplitudes therefore do not have an homogeneous scaling in $p $, and the perturbative expansion of the $\chi$PT interactions   
does not guarantee a perturbative expansion of the amplitudes \cite{Weinberg:1990rz,Weinberg:1991um}.
In Fig.\ \ref{fig:red} we show two types of contributions to the amplitude. Diagram $(c)$ represents the so-called ``reducible" diagrams,
in which the intermediate state consists purely of propagating nucleons. In these diagrams the contour of integration for integrals over the 0th components of loop momenta
cannot be deformed in way to avoid the poles of the nucleon propagators, thus picking up energies
$\sim p^2/m_N$ from nucleon recoil, no longer a subleading effect, rather than $\sim p$. These diagrams are therefore enhanced by factors of $m_N/p$ with respect to the $\chi$PT
power counting and need to be resummed, typically by solving a Schr\"odinger equation. The resummation leads to the appearance of shallow bound states in systems with two or more nucleons.

Diagrams $(a)$ and $(b)$ exemplify ``irreducible'' diagrams, whose intermediate states contain interacting nucleons and pions. These diagrams do not suffer from this infrared enhancement,
and here nucleon recoil remains a small effect. 
Irreducible diagrams involving pions and nucleons  
follow the $\chi$PT power counting~\cite{Weinberg:1990rz,Weinberg:1991um}  (commonly called  ``Weinberg power counting''), while
the situation is more complicated
for contact interactions, where different schemes exist such as ``KSW" \cite{Kaplan:1998tg} or pionless EFT \cite{Bedaque:2002mn}, where the $N\!N$ interactions become relatively enhanced.

Reducible diagrams are then obtained by patching together irreducible diagrams with intermediate states consisting of $A$ free-nucleon propagators.
This is equivalent to solving the Schr\"{o}dinger equation with a potential $V$ defined by the sum of irreducible diagrams. Notice, in particular, that the potential is only sensitive to the scale
$p$, and does not depend on properties of the bound states such as the binding energy.   
For external currents, such as the electromagnetic and weak currents, one can similarly  identify irreducible contributions, that can be organized in an expansion in $p/\Lambda_\chi$,
and separate them from the effects that arise from the iteration of the strong-interaction potential. 
For example, diagrams such as Fig.\ \ref{fig:red}$(c)$ are taken into account by taking the matrix element of the neutrino-exchange potential, induced by the irreducible diagrams,  between the wavefunctions of the nuclear bound states.

In the following subsections we construct the chiral Lagrangian relevant for $0\nu\beta\beta$ processes, and discuss 
the hadronic input needed to determine its couplings.
The Lagrangian contains charged-current operators with an electron and an explicit neutrino, which is later exchanged between two nucleons (see Fig.~\ref{twobody}(b)) to give rise to long-range neutrino-exchange contributions to \NLDBD.
For these operators the hadronic input consists of the vector, axial, scalar, pseudoscalar, and tensor nucleon form factors, which, with the exception of a subleading LEC in the tensor form factor, 
are well determined either experimentally or via LQCD calculations.

In addition, the Lagrangian has operators with pions, nucleons, and two electrons, but no neutrinos (see Fig.~\ref{twobody}(c)), which give pion-exchange and short-range contact contributions to
\NLDBD.  
In this case new LECs arise from the hadronization of four-quark operators. In the case of purely mesonic operators, these LECs are well determined \cite{Cirigliano:2017ymo,Nicholson:2016byl}.
For pion-nucleon and nucleon-nucleon operators at the moment they can only be estimated with NDA. 

In Sec. \ref{sec:2btOperators} we then use the Lagrangian constructed in Sec. \ref{ChiPT} to derive the two-nucleon operators (the so-called ``neutrino potentials'') that mediate \NLDBD.

\subsection{The $\Delta L =2$ chiral Lagrangian}\label{sec:ChiLag}
After evolving the $\Delta L=2$ operators to low energies, $\mu\sim 2$ GeV, we match them to $\chi$PT. The construction of the chiral Lagrangian closely follows that of the standard $\chi$PT Lagrangians \cite{Gasser:1983yg}. We describe the pions by 
\begin{equation}\label{eq:2.3}
U=u^2   = \exp\left(\frac{ i \boldpi\cdot \boldtau}{F_0}\right)\,, 
\end{equation}
where $\tau_i$ are the Pauli matrices, $F_0$ is the pion decay constant in the chiral limit, and we use $F_\pi=92.2$ MeV for the physical decay constant. We also introduce the nucleon doublet $N= (p\, n)^T$ in terms of the proton ($p$) and neutron ($n$) fields. The pions transform as $U\to LUR^\dagger$ and $u\to LuK^\dagger=KuR^\dagger$ under $SU(2)_L\times SU(2)_R$ transformations, while the nucleon doublet transforms as $N\to KN$. Additional ingredients are external scalar, vector, and tensor sources in the quark-level Lagrangian, which, for our purposes, take the following form
\bea \label{eq:sources}
s + i p &=& -\frac{2 G_F}{\sqrt{2}} \left[C^{(6)}_{ \textrm{SL}} \, ( \tau^+)  \,  \bar e_L C  \bar \nu_L^T +C^{(6)*}_{\textrm{SR}} \, ( \tau^-)  \,  \nu^T_L C   e_L\right], \nn\\
s - i p &=& -\frac{2 G_F}{\sqrt{2}}\left[ C^{(6)}_{ \textrm{SR}} \, ( \tau^+)  \,  \bar e_L C  \bar \nu_L^T+ C^{(6)*}_{ \textrm{SL}} \, ( \tau^-)  \,  \nu_L^T Ce_L\right],\nn \\
l_\mu  &=&  \frac{2 G_F}{\sqrt{2}v }( \tau^+ )\bigg[ -2v V_{ud}\bar e_L\g_\mu \nu_L+v\,C_{\textrm{VL}}^{(6)} \,  \,  \bar e_R \gamma_\mu C\bar \nu_L^T +C^{(7)}_{\textrm{VL}}  \, \bar e_L \, C i\overleftrightarrow{\partial}_\mu \bar \nu_L^T  \bigg] +{\rm h.c.}\,\,, \nn\\
r_\mu  &=&  \frac{2 G_F}{\sqrt{2}v}( \tau^+)\bigg[ v\, C_{ \textrm{VR}}^{(6)} \,   \,  \bar e_R \gamma_\mu C \bar\nu_L^T +
C^{(7)}_{\textrm{VR}} \,   \bar e_L \, C {i}\overleftrightarrow{\partial}_\mu \bar \nu^T_L\bigg] +{\rm h.c.}\,\,,\nonumber \\
t^{\mu\nu}_R &=& \frac{2G_F}{\sqrt{2}}(\tau^+)C_T^{(6)}\bar e_L\simu C \bar\nu^T_L\, ,
\eea
where $\tau^\pm = (\tau_1\pm i\tau_2)/2$.
The chiral Lagrangian is then given by chiral invariants constructed from the meson and baryon fields and the above spurions, which transform as follows, $r_\mu \to Rr_\mu R^\dagger$, $l_\mu \to Ll_\mu L^\dagger$, $s+i p \to R(s+i p) L^\dagger$, $s-i p \to L(s-i p) R^\dagger$, and $t^{\mu\nu}_R\to Lt^{\mu\nu}_R R^\dagger$. The dimension-$9$ operators, $C_{1}^{(9)}$ and $C_{4,5}^{(9)}$, can not be written in terms of the above sources and additional chiral constructions are required. The former transforms as $5_L\times 1_R$ while $C_{4,5}^{(9)}$ transform as $3_L\times 3_R$. We will discuss their chiral representations separately below. 

\subsection{Mesonic sector}
In the meson sector the interactions that are responsible for long-range neutrino-exchange contributions  arise from the standard leading-order (LO) chiral Lagrangian 
\bea\label{Lpi}
\vL_\pi &=& \frac{F_0\sq}{4}{\rm Tr}\left[ (D_\mu U)^\dagger D^\mu U\right] +\frac{F_0\sq}{4}{\rm Tr}\left[ U^\dagger \chi+ U\chi^\dagger\right]\, ,
\eea
where
\bea
D_\mu U = \partial_\mu U-il_\mu U +i U r_\mu,\qquad \chi = 2B(M+s-ip),\qquad M={\rm diag}(m_u,m_d)\,\,.
\eea
$B$ is the quark condensate, related to the pion mass by $m_\pi^2 = B(m_u + m_d)$. We use $(m_u+m_d)/2 = (3.5_{-0.3}^{+0.7})$ MeV \cite{Olive:2016xmw}, such that $B\simeq 2.8$ GeV. 
The dimension-six and -seven operators enter through the external sources, $l_\mu,\,r_\mu,\, s$, and $p$. Contributions from the dimension-six tensor operator require two additional derivatives which increase the chiral index by two. As such, the dominant contribution from $C_{\rm T}^{(6)}$ comes from the pion-nucleon sector which is discussed below. 

One of the advantages of the chiral notation is its compactness, which, however, has the downside of making it more difficult to see to which processes the operators contribute. Here we expand the $\Delta L=2$ interactions in Eq.~\eqref{Lpi} up to terms linear in the pion field which provide the main contribution to $0\nu\beta\beta$ processes
\begin{eqnarray}
\vL_\pi &=& -i F_0 G_F B\left(\pi^-\right)\left[ \left(C_{ \textrm{SL}}^{(6)} - C_{ \textrm{SR}}^{(6)}\right)\left(\bar e_L C  \bar \nu_L^T\right)\right]\\
&&
-F_0 G_F \left( \partial^\mu \pi^- \right)\left[ \left(C_{ \textrm{VL}}^{(6)} - C_{ \textrm{VR}}^{(6)}\right)\left(\bar e_R \gamma_\mu C \bar\nu_L^T\right)+ \frac{1}{v} \left(C_{ \textrm{VL}}^{(7)} - C_{ \textrm{VR}}^{(7)}\right)\left(\bar e_L \, C i\overleftrightarrow{\partial}_\mu \bar \nu_L^T\right)\right]+\mathrm{h.c.}\nonumber
\end{eqnarray}

In addition, the dimension-nine operators give rise to contributions that do not involve the exchange of a neutrino. In this case, the higher-dimensional operators induce interactions that convert two pions ($\pi^-$) into two electrons. 
Following  Refs.\ \cite{Savage:1998yh,Prezeau:2003xn,Cirigliano:2017ymo} we write the chiral representations of these interactions as
\bea\label{eq:pipi}
\vL^{(9)}_\pi &=&\frac{F_0^4}{4}\left[\left(g_{8\times 8}C_{4}^{(9)}+g_{8\times 8}^{\rm mix}C_{5}^{(9)}\right) {\rm Tr}\left[ U\tau^+U^\dagger\tau^+\right]+\frac{5}{3}g_{27\times 1}C_{1}^{(9)} L_{21}^\mu L_{21\,\mu}\right]\frac{\bar e_L C\bar e_L^T}{v^5}\nn\\
&= &\frac{F_0^2}{2}\left[\left( C^{(9)}_{4} g_{8 \times  8} + C^{(9)}_{5} g^{\textrm{mix}}_{8 \times  8}\right)  \pi^- \pi^- + \frac{5}{3}C^{(9)}_{1} g_{27 \times 1} \,  \partial_\mu \pi^- \partial^\mu \pi^-   \right]\frac{\bar e_L C\bar e_L^T}{v^5}+\dots\,\,,\eea
where $L_{ij}^\mu = i\left(U\partial^\mu U^\dagger\right)_{ij}$ and the dots stand for terms involving more than two pions.
By dimensional analysis the low-energy constants $g_{8\times 8}^{\rm (mix)}$ scale as $\mathcal O(\Lambda_\chi^2)$, while $g_{27\times 1} = \mathcal O(1)$.
We follow the notation of Ref.~\cite{Cirigliano:2017ymo}, in which 
these three low-energy constants  were estimated using $SU(3)$-$\chi$PT relations and LQCD calculations. The values of the LECs we use are given in Table~\ref{Tab:LECs},
and are in reasonable agreement with naive dimensional analysis.

\begin{table}
\center
\begin{tabular}{|c|cc||c|cc|}
\hline
$g_{27 \times 1}$   		& $ 0.38 \pm 0.08 $         & \cite{Cirigliano:2017ymo}  & $g_A$  & $1.272 \pm 0.002$ &\cite{Olive:2016xmw}  \\
    $g_{8 \times 8}$   		& $-(3.1 \pm 1.3)$ GeV$^2$  & \cite{Cirigliano:2017ymo}     & $g_S$  & $0.97 \pm 0.13$   &\cite{Bhattacharya:2016zcn}\\
  $g^{\textrm{mix}}_{8 \times 8}$ & $-(11 \pm 4)$   GeV$^2$     & \cite{Cirigliano:2017ymo}            & $g_T$  & $0.99 \pm 0.06 $  &\cite{Bhattacharya:2016zcn} \\
  $|g^{\pi N}_{27 \times 1}|$ & $\mathcal{O}(1)$ & - &  $|g^\prime _T|$ & $\mathcal{O}(1)$ & - \\
 $ |g^{N N}_{27 \times 1}|$ & $\mathcal{O}(1)$ & - & &  &   \\
\hline
\end{tabular}
\caption{Hadronic input for the LECs $g_S$, $g_T$, $g_{27 \times 1}$, $g_{8 \times 8}$,
and $g^{\textrm{mix}}_{8 \times 8}$, at the scale $\mu = 2$ GeV. 
Currently we lack experimental or LQCD input for the LECs $ g^{N \pi}_{27 \times 1}$, $ g^{N N}_{27 \times 1}$, and $g_T^\prime$, and we follow naive dimensional analysis. 
}\label{Tab:LECs}
\end{table}

\subsection{Nucleon sector}\label{sec:NuclLag}

The LO nucleon Lagrangian responsible for long-range neutrino exchange is given by
\begin{eqnarray}\label{eq:3.1}
\mathcal L^{(1)}_{\pi N}& =&  i\bar N  v \cdot \mathcal D N + g_A \bar N S\cdot u N 
+ c_5\,\bar N \hat \chi_+ N  -\left(2g_T \ep_{\mu\nu\al\bt}v^\al\,\bar N  S^\bt (u^\dagger t_R^{\mu\nu} u^\dagger) N+{\rm h.c.}\right).
\end{eqnarray} 
Here $v^\mu$ and $S^\mu$ are the nucleon velocity and spin, $v^{\mu} = (1,\vec{0})$ and $S^{\mu} = (0,\boldsigma/2)$ in the nucleon rest frame, and $\hat \chi_+ = \chi_+ - \mathrm{Tr}(\chi_+)/2$ where $\chi_\pm$ is defined below.
We have applied the heavy-baryon framework to remove the nucleon mass from the LO Lagrangian \cite{Jenkins:1990jv}.  The values of the couplings $g_A$ and $g_T$ are given in Table~\ref{Tab:LECs}. The LEC $c_5$ 
is related to the strong proton-neutron mass splitting and we give its value below.  The chiral covariant derivative is defined as $
\mathcal D_{\mu} N  =( \partial_{\mu} +  \Gamma_{\mu}) N$, where
\begin{eqnarray}
\Gamma_\mu &=& \frac{1}{2} \left[ u^\dagger \left(\partial_\mu - i l_\mu \right) u + \, u \left( \partial_\mu - i r_\mu\right) u^\dagger  \right] ,\nonumber \\
u_\mu & = &- i \left[ u^\dagger \left( \partial_\mu - i l_\mu \right) u - u \left( \partial_\mu - i r_\mu  \right) u^\dagger  \right] , \nonumber\\
\chi_{\pm} &=& u^{\dagger} \chi u^\dagger \pm u^{} \chi^\dagger u \,\,.
\end{eqnarray}
The first two terms in Eq.\ \eqref{eq:3.1} involve contributions from the vector operators $C_{\rm VL(VR)}^{(6,7)}$, while the last two terms involve contributions from the scalar couplings $C_{\rm SL(SR)}^{(6)}$. The last
term is generated by the tensor interaction  $C_{\rm T}^{(6)}$. Eq.\ \eqref{eq:3.1} turns out to capture the dominant contributions from $C_{\rm SL(SR)}^{(6)}$ and $C_{\rm VL(VR)}^{(7)}$. However, for both the dimension-six vector and tensor operators, the LO terms do not contribute to the $0\nu\bt\bt$ $0^+\rightarrow 0^+$ transitions and non-vanishing interactions only appear at next-to-leading order (NLO). 

The relevant NLO corrections can be written as follows
\bea
\vL^{(2)}_{\pi N} &=& \frac{1}{2m_N}\left(v^\mu v^\nu-g^{\mu\nu}\right)\left(\bar N\mathcal D_\mu \mathcal D_\nu N\right)- \frac{i g_A}{2 m_N}\bar N \{S\cdot \mathcal D, v\cdot u\}N
-\frac{g_M}{4m_N}\ep^{\mu\nu\al\bt}v_\al\,\bar N S_\bt f^+_{\mu\nu}N
\nn\\
&&-\left( \frac{g_T}{m_N} \ep_{\mu\nu\al\bt}\, \bar N S_\bt \lbrace u^\dagger t_R^{\mu\nu}u^\dagger,\, i \mathcal D_\al \rbrace N -\frac{g_T'}{m_N} v_\mu\,\bar N \left[  u^\dagger t_R^{\mu\nu}u^\dagger,\,  \mathcal D_\nu\right] N+{\rm h.c.}\right)\,,
\eea
where the coefficients of the first two and fourth operators are fixed by reparametrization invariance \cite{Luke:1992cs} in terms of the LO nucleon Lagrangian, 
$g_M=1+\kappa_1$ with $\kappa_1\simeq 3.7$ the anomalous isovector nucleon  magnetic moment, and $g_T'$ is the only unknown LEC at this chiral order\footnote{That is, to NLO the tensor matrix element depends on only two form factors. This counting agrees with the general relativistic expression for the matrix element $\langle p | \overline{u} \sigma^{\mu \nu} d | n \rangle$, which depends on four form factors. However, one of these form factors vanishes in the isospin limit and the other involves two derivatives and appears at N$^2$LO in the chiral expansion. In the notation of Ref.\ \cite{Adler:1975he}, which is commonly used in the literature \cite{Hirsch:1995cg,Pas:1999fc,Deppisch:2012nb}, 
we can identify 
$g_T'=2 \hat T_2^{(3)} - T_1^{(3)}$.
Using the estimates of Ref. \cite{Adler:1975he}, $\hat T_2^{(3)} = -0.62$ and $T_1^{(3)} = 1.38$, we would find $g_T^\prime = -2.62$,
compatible with the NDA estimate of Table \ref{Tab:LECs}. 
Some literature uses $\hat T_2^{(3)} = -4.54$, which, however, does not appear in Ref. \cite{Adler:1975he}.}  , which by NDA scales as $g_T^\prime = \mathcal O(1)$. 
Furthermore, $f_{\mu\nu}^\pm = u^\dagger L_{\mu\nu} u\pm u R_{\mu\nu} u^\dagger$, with 
\bea
L_{\mu\nu}=\partial_\mu l_\nu-\partial _\nu l_\mu -i[l_\mu,\, l_\nu ] \, ,\quad R_{\mu\nu}=\partial_\mu r_\nu-\partial _\nu r_\mu -i[r_\mu,\, r_\nu ] \,.
\eea
This is the most general chiral-invariant Lagrangian at this order,  that is also hermitian, as well as reparametrization, parity, and time-reversal invariant.

Apart from long-range neutrino-exchange contributions, the nucleon sector mediates short-range contributions induced by the dimension-nine operators. These can involve a single pion exchange, through vertices of the form $\bar p n \,\pi^- ee$, or through nucleon-nucleon interactions of the form $\bar p n \,\bar p n\, ee$. For the $C_{4,5}^{(9)}$ couplings, the short-range contributions to $0\nu \bt\bt$  are suppressed in the chiral power counting with respect to the long-range pion-exchange terms from Eq.\ \eqref{eq:pipi}. However, for the $C_{1}^{(9)}$ coupling, the $\pi N$ and $NN$ interactions  contribute at the same level as the $\pi\pi $ terms of Eq.\ \eqref{eq:pipi}~\cite{Prezeau:2003xn,Graesser:2016bpz}. Thus, for $C_{1}^{(9)}$  all three mechanisms have to be considered. 

Starting with the chiral realization of the pion-nucleon couplings there is one relevant operator, 
\bea\label{eq:dim9PiN}
\vL^{\pi N}_{27\times 1}& =& g_Ag^{\pi N}_{27\times 1}C_1^{(9)}F_0^2\left[\,\bar N S^\mu u^\dagger\tau^+ u N\,{\rm Tr}\left(u_\mu u^\dagger \tau^+ u \right)\right]\frac{\bar e_L C\bar e_L^T}{v^5}\nn\\
&=&\sqrt{2}g_A g^{\pi N}_{27\times 1}C_1^{(9)}F_0 \left[\bar p\, S\cdot (\partial \pi^-)n\right] \, \frac{\bar e_L C\bar e_L^T}{v^5}+\dots\,\,,
\eea
where the dots stand for terms involving additional pions and $g^{\pi N}_{27\times 1}$ is a LEC of $\Or(1)$. For later convenience we have pulled out a factor of $g_A$ in our 
definition of $g^{\pi N}_{27\times 1}$.
For the nucleon-nucleon interactions we also find a single relevant operator 
\bea\label{eq:dim9NN}
\vL_{27\times 1}^{NN} &=&C_1^{(9)} g_V\sq g_{27\times 1}^{NN} \, (\bar N u^\dagger \tau^+ u N)(\bar N u^\dagger \tau^+ u N)\, \frac{\bar e_L C\bar e_L^T}{v^5}\nn\\
& =& C_1^{(9)}g_V\sq g_{27\times 1}^{NN}\left(\bar pn\right )\,\left(\bar pn\right )\, \frac{\bar e_L C\bar e_L^T}{v^5}+\dots\,\,,
\eea
where the dots again stand for terms involving additional pions, and $g_{27\times 1}^{NN} \simeq \mathcal O(1)$ is another unknown LEC. 
As for the previous LEC, we have pulled out a factor of $g^2_V$ in our 
definition of $g^{N N}_{27\times 1}$.
Additional structures, such as  $ \bar p S^\mu n \, \bar p S_\mu n$, can be eliminated using Fierz identities and are not independent.

We note here that the distinction between long- and short-distance contributions loses its meaning as one goes to sufficient high order in the construction of the $\chi$PT Lagrangian. 
For example, the operators in Eqs.\ \eqref{eq:dim9PiN} and \eqref{eq:dim9NN}
receive a contribution from the neutrino Majorana mass, proportional to $m_{\beta\beta}/\Lambda_\chi^2$, induced by the exchange of hard neutrinos, with momentum 
$|\vec q|  > \Lambda_\chi$, which are integrated out in $\chi$PT
\cite{AWL}.
Similarly, the operators $C^{(6)}_{i}$ and $C^{(7)}_{i}$ in Eqs. \eqref{lowenergy6} and \eqref{lowenergy7} will induce $\Delta L =2$ operators without 	neutrinos in the $\chi$PT Lagrangian. These contributions  appear at N$^{2}$LO,  and  we neglect them here.

\subsection{One-body currents for $\beta$ decays}\label{OneBodyCurrent}

We now summarize the single $\bt$ decay amplitude, which provides the building blocks necessary to construct the full $0\nu\bt\bt$ amplitude. The single $\bt$ decay amplitude involves  two types of diagrams, 
which either involve a single vertex or a single pion exchange between the lepton and nucleon line. Using the Lagrangians constructed in the previous sections, we write the amplitude as
\bea
\mathcal A^{n\to pe^- \nu} = \bar N \tau^+ \left[\frac{l_\mu+r_\mu}{2}J_V^\mu+ \frac{l_\mu-r_\mu}{2}J_A^\mu -s\,J_S +i p\, J_P +t_{R\, \mu\nu}\,J_T^{\mu\nu}\right] N\,\,,
\eea
with the sources given in Eq.\ \eqref{eq:sources}. As discussed in Section \ref{sec:NuclLag}, for some operators we will need expressions through NLO in the chiral expansion. 
Up to NLO, the currents become
\bea \label{eq:currents}
J^\mu_V  &=& g_V(\vec q^2) \left( v^\mu + \frac{p^\mu + p^{\prime \mu}}{2m_N} \right) + \frac{i  g_M(\vec q^2)}{m_N} \varepsilon^{\mu \nu \alpha \beta} v_\alpha S_\beta q_\nu  \,, \nn\\
J^\mu_A  &=& - g_A(\vec q^2)  \left( 2 S^\mu  - \frac{v^\mu}{2 m_N}\, 2 S \cdot (p + p^\prime) \right) + \frac{g_P(\vec q^2)}{2 m_N} 2 q^\mu \, S \cdot q\,, \nn\\
J_S &=&g_S(\vec q^2)\,, \nn\\
 J_P &=& B \frac{g_P(\vec q^2)}{m_N} S\cdot q\,,\nn\\
J^{\mu \nu}_T &=&  - 2 g_T(\vec q^2)  \varepsilon^{\mu \nu \alpha \beta} \left( v_{\alpha} + \frac{p_\alpha + p^\prime_\alpha}{2 m_N} \right) S_{\beta}  -i
  \frac{g_{T}^\prime(\vec q^2)}{2 m_N} (v^\mu q^\nu - v^\nu q^\mu) \,.
\eea
Here $p$ and $p'$ stand for the momentum of the incoming neutron and outgoing proton, respectively, and $q^\mu=(q^0,\, \vec q) = p^\mu-p^{\prime\,\mu}$. Furthermore, $ \varepsilon^{\mu \nu \alpha \beta}$ is the totally antisymmetric tensor, with $\varepsilon^{0123}=+1$. At LO in  $\chi$PT the form factors are given by
\bea \label{eq:FF}
g_V(\vec q^2) &=& g_V = 1\, ,\qquad g_A(\vec q^2) = g_A = 1.27\,,\qquad g_M(\vec q^2) = 1+\kappa_1\, ,\nn\\
g_S(\vec q^2) &=&-4 Bc_5 = \frac{(m_n-m_p)_{\rm str}}{m_d-m_u},\qquad g_P(\vec q^2) =-g_A\frac{2m_N}{\vec q\sq+m_\pi\sq}\,,\nn\\
g_T(\vec q^2) &=& g_T^d-g_T^u \simeq  1\,,\qquad g_T'(\vec q^2) \simeq 1\, ,
\eea
where we followed the normalization of Ref.\ \cite{Engel:2016xgb}. 

Vector current conservation enforces $g_V(0) = 1$, up to small isospin-breaking corrections. For $g_A$ and $\kappa_1$ we used the experimental values \cite{Olive:2016xmw}.  
There is some disagreement in the literature on the value of $g_M(0)$, with some authors  using $g_M(0) = \kappa_1 =3.7$,
rather than the correct $g_M(0) = 1 + \kappa_1= 4.7$. The error appears to stem from
one of the first papers that studied  the contribution of weak magnetism \cite{Simkovic:1999re},
which did not account for the non-anomalous contribution to the isovector nucleon magnetic moment in the non-relativistic limit. 
We notice that earlier papers, such as \cite{Doi:1985dx,Tomoda:1990rs}, correctly use $g_M(0) = 4.7$.
The isovector scalar charge $g_S(0)$ is related to the quark mass contribution to the neutron-proton mass splitting \cite{Gonzalez-Alonso:2013ura}. Using 
$(m_n - m_p)|_{\textrm{str}} = 2.32$ MeV \cite{Brantley:2016our} and $m_d - m_u = 2.5$ MeV \cite{Olive:2016xmw} gives $g_S(0) = 0.93$, at the renormalization scale $\mu = 2$ GeV, in very good agreement with the direct LQCD calculation of Ref.~\cite{Bhattacharya:2016zcn}. For the isovector tensor charge $g_T(0)$ we use the results of Ref.\ \cite{Bhattacharya:2015wna,Bhattacharya:2016zcn}. 
The numerical input we use is listed in Table \ref{Tab:LECs}. 

The expression of the currents in Eq.~\eqref{eq:currents} in terms of the form factors $g_{V,A,M,S,P,T}(\vec q^2)$, while traditional,
somewhat blurs the $\chi$PT expansion of the various contributions. For instance, at LO in $\chi$PT
only the pseudoscalar form factor $g_P(\vec q^2)$ has non-trivial momentum dependence, due to the pion propagator, while 
all other form factors are purely static.
Furthermore, the standard notation in Eq.~\eqref{eq:currents} makes the power counting less apparent by artificially hiding a factor of $m_N$ in $g_P$. This means $\vec q^2\, g_P(\vec q^2)/m_N=\Or(1)$ is actually a LO contribution, while the magnetic contribution, $g_M/m_N$,  is  suppressed by $1/\Lambda_\chi$, such that pieces proportional to $g_M$ are higher order in the chiral counting. Thus, at LO in  $\chi$PT, we could drop the magnetic contributions in Eq.~\eqref{eq:currents} and use Eq.\ \eqref{eq:FF} for $g_{V,A,P}(\vec q\sq)$.  

The form factors $g_{V,A}(\vec q^2)$ and $g_A(\vec q^2)$ acquire momentum dependence at N${}^2$LO in $\chi$PT. At this order this momentum dependence is encoded in the nucleon isovector charge and axial radii, respectively, $r_V = 0.76$ fm \cite{Olive:2016xmw} and $r_A = 0.49$ fm \cite{Rajan:2017lxk}, corresponding to vector and axial masses $\Lambda_V = 0.9$ GeV and $\Lambda_A = 1.4$ GeV in a dipole parameterization of the form factors.
This subset of N$^{2}$LO corrections is usually taken into account in the calculation of $0\nu\beta\beta$ matrix elements by including a dipole form factor for $g_V$ and $g_A$, with different vector and axial masses \cite{Engel:2016xgb}. 
 While including such corrections does not formally improve the accuracy of the calculation, as other N${}^2$LO contributions, 
such as pion-neutrino loops or short-range  nucleon-nucleon contributions, are not considered, the numerical impact of the axial and vector form factors  
is not negligible,  giving an $\mathcal O(10-20\%)$ correction \cite{Simkovic:1999re,Menendez:2008jp,Barea:2009zza}. 
This suggests that it might be important to consistently include all N$^{2}$LO corrections to \NLDBD\ .

While the momentum dependence of the $g_{V,A,S,T}$ form factors only enters at N${}^2$LO in the chiral expansion, the magnetic form factor has a correction at NLO with respect to Eq.~\eqref{eq:currents}, 
due to pion loops\footnote{Since the magnetic moment itself appears at NLO, the momentum dependence of the magnetic FF enters at the same order as that of the vector and axial FF.} \cite{Bernard:1995dp}. The treatment of the magnetic form factor $g_M(\vec q^2)$ in the $0 \nu \beta \beta$ decay literature is at odds with this result, as it is often assumed $g_M(\vec q^2) = g_M(0) g_V(\vec q^2)$, which is not justified in $\chi$PT \cite{Bernard:1995dp}.  

To conclude this section, we stress that while most of the currents in Eq.~\eqref{eq:currents} have been studied up to N$^2$LO,
here we do not include these corrections in the construction of the two-nucleon operators that mediate $0\nu\beta\beta$,
as consistency requires the inclusion of other, unknown, contributions, such as the pion-neutrino loops mentioned above. Thus, 
even when we use calculations that include partial N$^{2}$LO corrections, our results are formally valid at LO in $\chi$PT.

\section{$0\nu \beta \beta$ operators}\label{sec:2btOperators}

\begin{figure}
\begin{center}
\includegraphics[width=14cm]{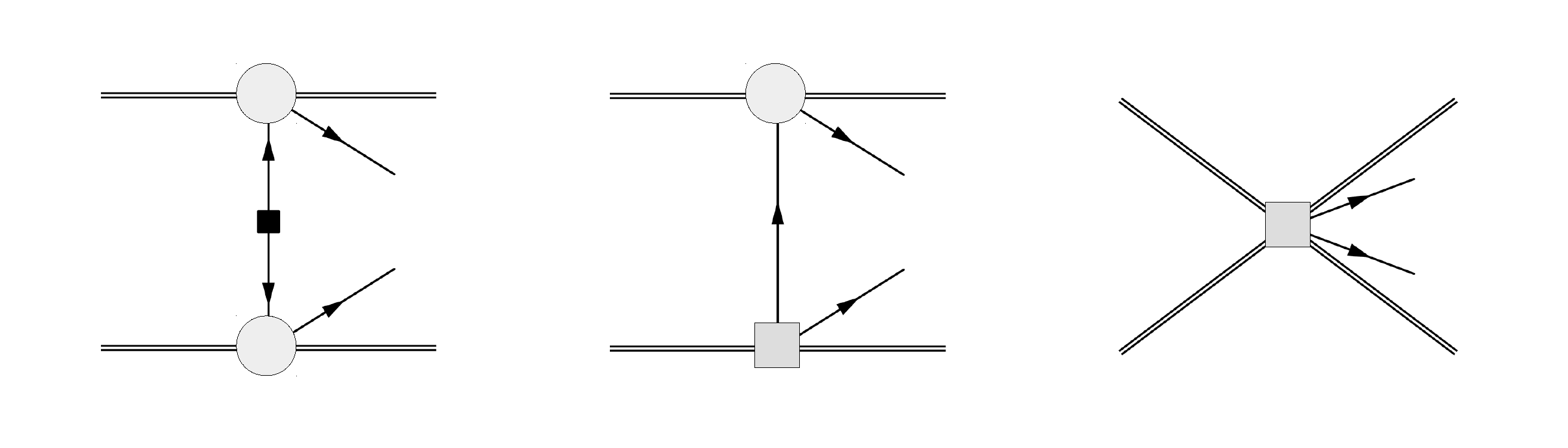}
\end{center}
\caption{Schematic representation of the diagrams contributing to the neutrino potentials. Double and single lines denote, respectively, nucleon and lepton fields. The black square denotes an insertion of the neutrino 
Majorana mass, while the gray squares denote the $\Delta L =2$ interactions between nucleons, pion, and leptons induced by the dimension-seven operators discussed in Sec. \ref{sec:ChiLag}.
The gray circle denotes SM interactions between nucleons, pion, and leptons.
}\label{twobody}
\end{figure}

The ingredients derived in the previous section allow us to construct the two-nucleon operators that mediate $0\nu \beta \beta$ decays. 
Fig.~\ref{twobody} shows three possible contributions. The first diagram depicts the standard contribution proportional to the neutrino Majorana mass. 
The second diagram depicts long-range neutrino-exchange contributions that arise from the $\Delta L =2$ charged current interactions in Eqs.\ \eqref{lowenergy6} and \eqref{lowenergy7}. 
These contributions are obtained by combining the one-body currents of the previous section.
Finally, operators such as $\mathcal O_{LHD}^{(1)}$ and $\mathcal O_{LL\bar d u D}$ induce six-fermion dimension-nine operators at the GeV scale, whose contribution to $0\nu \beta\beta$ decays is represented by the third diagram in Fig.~\ref{twobody}. These diagrams do not involve the exchange of a neutrino.

For each operator, we will construct the dominant contribution to $0^+ \rightarrow 0^+$ transitions, within the framework of chiral EFT.
The application of chiral EFT is justified by the separation of the scales involved in \NLDBD\, where the typical momentum exchange between the nucleons  is of similar size as the Fermi momentum within nuclei $q  \sim k_F \sim m_\pi = \mathcal O(100 \, \textrm{MeV})$, which is much larger than the reaction $Q$ value, typically around a few MeV. 

For the  diagrams in Fig.~\ref{twobody}(a) and (b), the LO neutrino potential is obtained by tree-level neutrino exchange. This involves the single-nucleon currents, represented by the gray circle and square in Fig.~\ref{twobody}, at the lowest order that yields non-vanishing results. Analogously to the strong-interaction potential, the two-body transition operators in chiral EFT are 
only sensitive to the momentum scale $q \sim k_F$, and are therefore independent of the properties 
of the bound states. 
In particular this implies that the transition operators do not depend on the often used ``closure energy'' $\bar E$, which encodes the average energy difference between intermediate and initial states.
This can be understood from Fig.~\ref{twobody}. An insertion of the strong-interaction potential between the emission and absorption of the neutrino in Fig. \ref{twobody}(a) or (b) would generate a diagram which, in the language of Sec. \ref{ChiPT}, is irreducible. That is, it is always possible to choose the contour of integration such that 
the energy and momentum of the nucleons in the loop have to be $\sim k_F$, and the nucleon is far from on-shell. 
Insertions of the strong interaction potential between the emission and absorption of the neutrino,
which would give rise to intermediate nuclear states, are therefore suppressed and can be ignored at LO.
Instead, in chiral EFT the dependence on the intermediate states arises from the region where the neutrino momentum is very soft $q_0\sim |\vec q| \ll k_F$. The exchange of soft neutrinos gives rise to effects that are suppressed
by $\bar E/k_F$ \cite{AWL}.
Notice that the situation is different from $2\nu\beta\beta$ decay, where insertions of the strong interaction potential between the two points where the neutrinos are emitted are not suppressed (in between the first and second neutrino emission, there are only propagating nucleons and the diagrams are ``reducible"), and  the intermediate states do need to be considered.

For neutrino-exchange contributions, the LO chiral EFT potential is very similar to standard results. In fact, as we will see, the chiral EFT potential reduces to results in the literature 
in the limit where the closure energy vanishes, $\bar E \rightarrow 0$. The advantage of chiral EFT is that it is possible to systematically consider subleading corrections. These consist of corrections to single-nucleon currents, which are often included in the literature via momentum-dependent form factors, but also genuine two-body effects, such as loop corrections to Fig.~\ref{twobody}(a) and (b), which induce short-range neutrino potentials  even for the standard mechanism  \cite{AWL}, and three-body effects \cite{Menendez:2011qq}. 

Diagram \ref{twobody}(c) does not involve the exchange of a neutrino. In this case the resulting LO potential is of pion range, $\sim 1/m_\pi$, or shorter range, $\sim 1/\Lambda_\chi$. We work at LO in this case as well, but  it is straightforward to include subleading corrections in chiral EFT.

In deriving the neutrino potential we take advantage of the fact that the $Q$ value and the electron energies $E_{1,2}$ have typical size $\mathcal O(5\, {\rm MeV})$  and are thus much smaller than $k_F$. We assign  the scaling $Q \sim  E_{1,2} \sim \mpi \epsilon_\chi^2$ such that these scales can be incorporated in the standard $\chi$EFT power counting. The assigned counting generally allows us to neglect the lepton momenta, nuclear recoil,  and soft-neutrino exchange, except in a few cases where
the matrix element of the LO operator vanishes for $0^+ \rightarrow 0^+$ transitions. In these cases we consider subleading contributions in the $\chi$PT power counting.

Before discussing the contributions in Fig.~\ref{twobody}(b) and (c) from the dimension-six, -seven, and -nine operators, we first recall the potential generated by light Majorana-neutrino exchange to establish our notation. For definiteness, we define the neutrino potentials as $- \mathcal A$, where $\mathcal A$ is the amplitude for the process $n n \rightarrow p p e^- e^-$.

\subsection{Light Majorana-neutrino  exchange}
In momentum space, the neutrino potential induced by light Majorana-neutrino exchange is 
\begin{eqnarray}
V_{\nu}(\vec q) &=& - (\tau^{(1) +} \tau^{(2) +})  (4 \,g_A\sq G_F^2V_{ud}\sq) \frac{m_{\beta\beta}}{\vec q^2} \, \Bigg\{ - \frac{g_V\sq}{g_A\sq}h_F(\vec q^2)  + \boldsigma^{(1)}\cdot \boldsigma^{(2)}  \, h_{GT}(\vec q^2)   
+  S^{(12)}\,  h_T(\vec q^2) \Bigg\} \nn \\
&&\times \bar u(k_1) P_RC \bar u^T(k_2)\,,
\end{eqnarray}
where  $k_{1,2}\sim Q$ are the electron momenta, $\hat{\vec q} = \vec q/|\vec q|$, and the tensor operator  is given by $S^{(12)}= -\left( 3\,\boldsigma^{(1)} \cdot  \hat{ \vec q} \, \boldsigma^{(2)} \cdot  \hat{ \vec q} - \boldsigma^{(1)}\cdot \boldsigma^{(2)} \right)$. In addition, $m_{\bt\bt}=(m_\nu)_{ee}=\sum m_{\nu_i}U_{ei}\sq$ where $m_{\nu_i}$ are the neutrino mass eigenvalues and $U$ is the PMNS matrix.
The Fermi (F) function only receives contributions from the vector currents at leading order.
In contrast, the Gamow-Teller (GT) and tensor (T) functions receive contributions from the nucleon axial current, including the induced pseudoscalar contribution dominated by the pion pole,
and, at higher order, from the nucleon magnetic moment.  Here we follow Refs.~\cite{Simkovic:1999re,Menendez:2008jp,Barea:2009zza,Hyvarinen:2015bda} and separate the direct axial, induced pseudoscalar, and magnetic contributions. We then have the following expressions for $h_F$, $h_{GT}$, and $h_T$
\begin{eqnarray}\label{eq:hK(q)}
h_F(\vec q^2) &=&   \frac{g_V^2(\vec q^2)}{g_V\sq}\,,\nn\\ 
h_{GT}(\vec q^2) &=&  h^{AA}_{GT}(\vec q^2) + h^{AP}_{GT}(\vec q^2) + h^{PP}_{GT}(\vec q^2) + h^{MM}_{GT}(\vec q^2)\,,\nn\\
h_{T}(\vec q^2)  &=&  h^{AP}_{T}(\vec q^2) + h^{PP}_{T}(\vec q^2) + h^{MM}_{T}(\vec q^2)\,.
\end{eqnarray}
For the GT and T functions, we have
\begin{eqnarray}\label{smff}
h^{AA}_{GT,T}(\vec q^2) &=&  \frac{g_A^2(\vec q^2)}{g_A\sq}\,, \quad \, h_{GT}^{AP}(\vec q^2) = \frac{g_P(\vec q^2)}{g_A\sq} \, g_A(\vec q^2) \frac{\vec q^2}{3 m_N}\,,\quad  \, h_{GT}^{PP}(\vec q^2) = \frac{ g^2_P(\vec q^2)}{g_A\sq} \frac{\vec q^4}{12 m_N^2}\,, \nonumber \\
h^{MM}_{GT}(\vec q^2) &=&  g_M^2(\vec q^2) \frac{\vec q^2}{6g_A\sq m_N^2}\,,
\end{eqnarray}
and $h^{AP}_T(\vec q^2) = -h^{AP}_{GT}(\vec q^2)$, $h^{PP}_T(\vec q^2) = - h^{PP}_{GT}(\vec q^2)$, and  $h^{MM}_{T}(\vec q^2) = h^{MM}_{GT}(\vec q^2)/2$. 
In order to compare with the $0\nu\beta\beta$ literature, we express the long-range neutrino-exchange potentials in terms of $g_{V,A,P,M}(\vec q\sq)$ where it is implied that they follow the $\chi$PT relations in Eq.\ \eqref{eq:FF}.

\subsection{Neutrino exchange without mass insertion}

\subsubsection{$\mathcal O_{SR,SL}^{(6)}$ and $\mathcal O_{VR,VL}^{(7)}$}\label{MEscalar}
The dimension-six scalar operators $C^{(6)}_{\textrm{SL}}$ and $C^{(6)}_{\textrm{SR}}$, 
and dimension-seven  vector operators, $ C^{(7)}_{\textrm{VL}}$ and $C^{(7)}_{\textrm{VR}}$, give a potential that is very similar to the one that is induced by light Majorana-neutrino exchange. At LO in $\chi$PT
\begin{eqnarray}\label{ps.1}
V(\vec q^2) =  & &  \tau^{(1) +} \tau^{(2) +} \,  \, 4g_A\sq G_F^2V_{ud} \, \left(  B  \left( C^{(6)}_{\textrm{SL}} - C^{(6)}_{\textrm{SR}} \right)  + \frac{m^2_\pi}{ v}  \left( C^{(7)}_{\textrm{VL}} - C^{(7)}_{\textrm{VR}} \right)  \right)  \frac{1}{\vec q^2} \,   \bar u(k_1) P_RC \bar u^T(k_2) \nonumber \\ 
& & \Bigg\{   \boldsigma^{(1)}\cdot \boldsigma^{(2)}  \, \left(\frac{1}{2} h^{AP}_{GT}(\vec q^2) + h^{PP}_{GT}(\vec q^2) \right)    + S^{(12)} \, \left(\frac{1}{2} h^{AP}_{T}(\vec q^2) + h^{PP}_{T}(\vec q^2) \right)  \Bigg\}\,\,.
\end{eqnarray}
Here we used Eq.\ \eqref{eq:FF} to rewrite the potential that is induced by the dimension-seven operators, $h_{\rm GT,7}(\vec q \sq)$, as follows
\begin{eqnarray}\label{ps.2}
h^{}_{GT,\, 7}(\vec q^2)  \equiv - \frac{\vec q^2}{3 m_\pi^2}  \, \left( g_A(\vec q^2) + \frac{\vec q^2}{2 m_N} g_P(\vec q^2)  \right)^2 = -g_A^2  \frac{\vec q^2}{3} \frac{m_\pi^2}{(\vec q^2 + m_\pi)^2}\, ,
\end{eqnarray}
which is equal to $\frac{1}{2} h^{AP}_{GT}(\vec q^2) + h^{PP}_{GT}(\vec q^2)  $ at LO in $\chi$PT.

The vector component $C^{(7)}_{\textrm{VL}} + C^{(7)}_{\textrm{VR}}$ does not contribute at LO because of vector current conservation.
The scalar current $C^{(6)}_{\textrm{SL}} + C^{(6)}_{\textrm{SR}}$, combined with the standard model axial current, gives a contribution that is suppressed by $\vec q/\Lambda_\chi$, and, in addition, 
is parity odd and does not contribute to $0^+ \rightarrow 0^+$ transitions. The first non-vanishing contributions from the scalar current appear at $\mathcal O(\epsilon_\chi^2)$.

The pseudoscalar contribution in Eq.~\eqref{ps.1} has been considered in the literature \cite{Hirsch:1995cg,Pas:1999fc,Deppisch:2012nb,Horoi:2017gmj}, 
while the $C^{(7)}_{VL,VR}$ terms have not, even though they appear at the same chiral order. In these works, the neutrino potential is derived  
by considering the pseudoscalar form factor  at $\vec q =0$, and by neglecting the induced pseudoscalar component of the axial current. For the pseudoscalar density at zero momentum the value $F^{(3)}_P = 4.4$ is used, which is obtained from a quark-model calculation \cite{Adler:1975he}.
These approximations have two consequences. First of all, as pointed out already in Ref.~\cite{Adler:1975he}, the value $F^{(3)}_P = 4.4$ fails to reproduce the pion pole dominance of the pseudoscalar density, which in $\chi$PT gives the much larger $F^{(3)}_P = 2g_A B m_N/m_\pi^2 \simeq 300$. The value of Ref.~\cite{Adler:1975he} thus corresponds to using a pion mass of $1100$ MeV such that $m_\pi  \sim \Lambda_\chi$. Secondly, neglecting the momentum dependence of the pion propagator in Eqs.\ \eqref{ps.1} and \eqref{ps.2} implies that the neutrino potential is of much shorter range than the typical pion range, affecting the value of the nuclear matrix elements.

\subsubsection{$\mathcal O_{T}^{(6)}$}\label{MEtensor}
At lowest order in $\chi$PT, the tensor operator $\mathcal O^{(6)}_{\textrm{T}}$ induces two operators whose matrix elements vanish in $0^+ \rightarrow 0^+$ transitions.
Including the NLO corrections to the tensor, axial, and vector currents outlined in Section \ref{sec:NuclLag}, we obtain 
\begin{eqnarray}\label{tensor}
V(\vec q^2) &=&  4 g_A\sq \tau^{(1) +} \tau^{(2) + } \, 2  G_F^2V_{ud} \,   m_N C^{(6)}_{\textrm{T}} \frac{1}{\vec q^2}\, \bar u(k_1) \,P_R C \bar u^T(k_2)\, 
\Bigg\{  \frac{ g^\prime_{T}(\vec q^2) g_V(\vec q^2)}{g_A\sq} \frac{\vec q^2}{m_N^2} \nonumber \\
& & - 4\frac{ g_T(\vec q^2) }{g_M(\vec q\sq)}   \left( h_{GT}^{MM}(\vec q\sq)\boldsigma^{(1)} \cdot \boldsigma^{(2)} 
+ h_{T}^{MM}(\vec q\sq)S^{(12)} \right) 
\Bigg\}\,.
\end{eqnarray}
In addition we find a recoil piece (see Appendix \ref{AppRecoil}), which we neglect in our results below. These contributions involve \NLDBD\ operators that depend on the nucleon momenta and whose nuclear matrix elements are unknown. We expect these unknown contributions to be small, however, with respect to Eq.~\eqref{tensor} because they are not enhanced by the large isovector nucleon magnetic moment.

Our expressions for the neutrino potentials induced by  tensor currents disagree with the literature in two respects. 
First of all, together with $\mathcal O_{T}^{(6)}$, another tensor structure is commonly considered,  
$ \mathcal O_{T}^{(6) \prime} = \bar u_R \sigma^{\mu\nu} d_L \, \bar e_{L,i}  \sigma_{\mu\nu}  \, C\bar\nu_{L,j}^T$ \cite{Hirsch:1995cg,Pas:1999fc,Deppisch:2012nb, Bonnet:2012kh}.
This operator however is identically zero (see Appendix \ref{AppEps}). This is in disagreement with Refs. \cite{Hirsch:1995cg,Pas:1999fc,Deppisch:2012nb} that
find a non-zero neutrino potential for this tensor structure. Secondly, the first term in Eq.~\eqref{tensor} is sometimes erroneously associated with $\mathcal O_{T}^{(6)\, \prime}$ \cite{Hirsch:1995cg,Pas:1999fc,Deppisch:2012nb}.

\subsubsection{$\mathcal O_{VL,VR}^{(6)}$}\label{MEvector}

The LO operators induced by $C^{(6)}_{ \textrm{VR}}$ and  $C^{(6)}_{\textrm{VL}}$ also turn out to give vanishing contributions to $0^+ \rightarrow 0^+$ transitions.
By employing the NLO vector and axial currents in Eq.\ \eqref{eq:currents} and taking into account the electron momenta and the  equations of motion for the electrons, we obtain
\bea\label{eq:vlvr}
V(\vec q^2) &=&  
\tau^{(1) +} \tau^{(2) + } \,  \,  g_A\sq G_F^2V_{ud} \,   \frac{1}{\vec q\sq} \Bigg\{  \bar u(k_1)  \gamma_0  C\bar u^T(k_2) \, (k_1^0 - k_2^0) \left[C^{(6)}_{\textrm{VL}}\, M^{(1)}_L+C^{(6)}_{\textrm{VR}}\, M^{(1)}_R\right]\\
&& +2 m_e \, \bar u(k_1) C   \bar u^T(k_2) \, \left[C^{(6)}_{\textrm{VL}}\, M^{(2)}_L+C^{(6)}_{\textrm{VR}}\, M^{(2)}_R\right]\nn\\
&&+ \bar u(k_1)  \gamma_0 \g_5 C \bar u^T(k_2)\,C^{(6)}_{\textrm{VL}}\,8 m_N \frac{g_A(\vec q\sq)}{g_M(\vec q\sq)}\left[h_{GT}^{MM}(\vec q\sq)\,\boldsigma^{(1)} \cdot \boldsigma^{(2)}+h_T^{MM}(\vec q\sq)S^{12} \right]
\Bigg\}\,\,,\nn
\eea
where
\bea
M_{L,R}^{(1)} &=& - \frac{4}{3} \frac{g_V\sq }{g_A\sq}h_F(\vec q\sq)\mp \frac{8}{9} h_{GT}^{AA}(\vec q^2)\,  \boldsigma^{(1)} \cdot \boldsigma^{(2)}  \mp \frac{4}{9}\, h_{T}^{AA}(\vec q^2) \, S^{(12)}\,\,,\nn\\
M_{L,R}^{(2)} &=&   \frac{1}{3} \frac{g_V\sq }{g_A\sq}h_F(\vec q\sq)\mp \left(\frac{1}{9}  h_{GT}^{AA}+h_{GT}^{AP}(\vec q^2)+h_{GT}^{PP}(\vec q^2)\right)\,  \boldsigma^{(1)} \cdot \boldsigma^{(2)}\nn\\  &\pm& \left(\frac{4}{9}\,  h_{T}^{AA}-h_{T}^{AP}(\vec q^2)-h_{T}^{PP}(\vec q^2)\right) \, S^{(12)}\,\,.
\eea
These expressions agree with  Ref.~\cite{Doi:1985dx,Muto:1989cd},  in the limit $|\vec q| \gg \bar E$, where $\bar E$ is the closure energy, $\bar E = \mathcal O(10\,\,\mathrm{MeV})$.
In principle there is an additional recoil contribution for the left-handed current $C^{(6)}_{\textrm{VL}}$,  see Appendix \ref{AppRecoil}. We neglected this term in the above as it turns out to be suppressed  with respect to the magnetic-moment contributions contained in the $h_{GT,T}^{MM}$ terms \cite{Muto:1989cd}.

For $C^{(6)}_{\rm VR}$, the first tree-level two-body contribution is proportional to the electron mass or energy and thus of order  $\mathcal O(\epsilon_\chi^2)$ in the power counting.
At the same order one should consider pion-neutrino loops, i.e.\ the contributions of $C^{(6)}_{\rm VR}$ to short-range $\Delta L =2$ operators without neutrinos, and three-body operators. While we leave a more detailed study for future work, we stress that the limits we obtain on $C^{(6)}_{\rm VR}$, and, consequently, on $\mathcal C_{Leu\bar{d}H}$, should be taken as order-of-magnitude estimates, rather than rigorous bounds.

\subsection{Dimension-nine operators}
Finally, we discuss the contributions from the dimension-nine operators. 
In the case of $C^{(9)}_{4,5}$, the most important operators are the pionic ones, while the pion-nucleon and nucleon-nucleon interactions are suppressed by two powers of $\epsilon_\chi$. In contrast, the pionic, pion-nucleon, and nucleon-nucleon couplings all enter at the same order for the operator  $C_1^{(9)}$.
The relevant terms are included in the Lagrangians of Eq.\ \eqref{eq:pipi}, \eqref{eq:dim9PiN}, and \eqref{eq:dim9NN}, which give rise to the following potential 
\bea\label{short}
V(\vec q^2) &=& - \tau^{(1) +} \tau^{(2) + } \, g_A\sq \,  \frac{4 G_F^2}{v} \,\bar u(k_1)P_RC\bar u^T(k_2) \bigg\{-\frac{C^{(9)}_{4} g_{8 \times  8} + C^{(9)}_{5} g^{\textrm{mix}}_{8 \times  8}}{2m_\pi\sq}\nn\\
&&\times \bigg[\left(h_{GT}^{PP}(\vec q\sq)+\frac{h_{GT}^{AP}(\vec q\sq)}{2}\right) \, \boldsigma^{(1)} \cdot \boldsigma^{(2)}+\left(h_{T}^{PP}(\vec q\sq)+\frac{h_{T}^{AP}(\vec q\sq)}{2}\right) \, S^{(12)}  \bigg]\nn\\
& & +  C^{(9)}_1 \bigg[ 2g_{27\times 1}^{NN}\frac{g_V\sq}{g_A\sq}h_F(\vec q\sq)-\frac{1}{2}g_{27\times 1}^{\pi N}\left( h_{GT}^{AP}(\vec q\sq) \, \boldsigma^{(1)} \cdot \boldsigma^{(2)}+h_T^{AP}(\vec q\sq)\, S^{(12)}\right)\nn\\
&&-\frac{5}{6}g_{27\times 1}\left( h_{GT}^{PP}(\vec q\sq) \, \boldsigma^{(1)} \cdot \boldsigma^{(2)}+h_T^{PP}(\vec q\sq)\, S^{(12)}\right)\bigg]\bigg\}\,\,.\eea

The above potential disagrees with parts of the existing literature in several aspects.
In Refs.\ \cite{Pas:2000vn,Deppisch:2012nb} the dimension-nine operators defined in Eq.\ \eqref{lowenergy9} appear as a subset of the most general set of dimension-nine 
four-quark two-electron operators. The conversion between $C^{(9)}_{1,4,5}$ and the coefficients $\varepsilon$ defined in Refs.~\cite{Pas:2000vn,Deppisch:2012nb} is given in App.\ \ref{AppEps}.
When considering the low-energy manifestations of these quark-level operators,  the authors of Refs.\ \cite{Pas:2000vn,Deppisch:2012nb} only take into account four-nucleon operators, which are of the same form as the one in Eq.\ \eqref{eq:dim9NN},
and estimate their coefficients by assuming factorization. This approach should provide a reasonable estimate for the bounds on $\varepsilon^{LLR}_3$ as this coupling is related to $C_1^{(9)}$, whose neutrino potential receives contributions of similar size from $\pi \pi$, $\pi N$, and $\bar N N$ operators. 
On the other hand, the contributions of the operators $O^{(9)}_{4,5}$, and thus the bounds on $\varepsilon^{LRR}_{3}$ and $\varepsilon^{RLR}_1$,  are severely underestimated. 
In these cases, the neutrino potential is dominated by the $\pi \pi$ contribution, given in Eq.\ \eqref{short}, and the $\bar N N$ pieces  are
suppressed by $\epsilon_\chi^2$. Thus, for $O^{(9)}_{4,5}$, the neutrino potentials of Refs.\ \cite{Pas:2000vn,Deppisch:2012nb} miss the dominant contributions to \NLDBD. 

The importance of the pion-exchange contributions for certain BSM mechanisms has long been recognized ~\cite{Faessler:1996ph,Vergados:2012xy}.	
Usually, however, pion exchange is included for the scalar-pseudoscalar operators  $\varepsilon^{LLR}_1$ and $\varepsilon^{RRR}_1$  \cite{Faessler:1996ph,Vergados:2012xy},
while its contribution for vector and axial operators, $\varepsilon^{LRR}_{3}$ and $\varepsilon^{RLR}_1$, has been largely ignored   \cite{Pas:2000vn,Deppisch:2012nb,Faessler:1996ph,Vergados:2012xy,Horoi:2017gmj}.
These issues were already addressed in Refs.\ \cite{Prezeau:2003xn,Graesser:2016bpz}, which performed a systematic power counting in $\chi$PT. The above expression is in agreement with the results of Refs.\ \cite{Prezeau:2003xn,Graesser:2016bpz}.

Finally we comment that in the literature the low-energy constants that describe the hadronization of the four-quark  operators
have often been estimated using the vacuum insertion approximation~\cite{Deppisch:2012nb,Vergados:2012xy,Faessler:1996ph}.
While, in those cases in which all the relevant hadronic channels are included, this leads to acceptable results,  
we remark that for the $\pi \pi$ channel more rigorous estimates exist, based on direct LQCD calculations \cite{Nicholson:2016byl}
and on $SU(3)$ $\chi$PT and LQCD \cite{Cirigliano:2017ymo}.

\section{Master formula for decay rate and nuclear matrix elements}\label{Master}
Using the potentials in the previous sections we can write down an expression for the inverse half-life for $0^+\rightarrow 0^+$ transitions \cite{Doi:1985dx,Bilenky:2014uka}
\bea \label{eq:InvHalfLife}
\left(T^{0\nu}_{1/2}\right)^{-1} = \frac{1}{8 \ln 2}\frac{1}{(2\pi)^5 }\int \frac{d^3k_1}{2E_1}\frac{d^3k_2}{2E_2} |\mathcal A |\sq F(Z,E_1)F(Z,E_2)\dt(E_1+E_2+E_f-M_i)\,\,,
\eea
where $E_{1,2}$ are the energies of the electrons, and $E_{f}$ and $M_i$ are the energy and mass of the final and initial nuclei in the rest frame of the decaying nucleus. The functions $F(Z,E_i)$ take into account the fact that the emitted electrons feel the Coulomb potential of the daughter nucleus and are therefore not plane waves. They take the following form
\bea \label{fermi}
F(Z,E) &=& \left[\frac{2}{\Gamma(2\g+1)}\right]\sq (2|\vec k|R_A)^{2(\g-1)}|\Gamma(\g+i y)|\sq e^{\pi y}\,,\nn\\
\g &=&\sqrt{1-(\al Z)\sq}\,,\qquad y = \al Z E/|\vec k|\,,\quad |\vec k | = \sqrt{E^2-m_e\sq}\,,
\eea
where $R_A = 1.2 \,A^{1/3}$ fm and $Z$ are, respectively, the radius and atomic number of the daughter nucleus. This procedure of calculating the Coulomb corrections assumes a uniform charge distribution in the nucleus and only the lowest-order terms in the expansion in $\vec r$, the electron position, 
factor, 
is taken into account. More precise calculations of the phase space factors apply exact Dirac wave functions \cite{Stefanik:2015twa} and the effect of electron screening \cite{Kotila:2012zza}. The use of exact wave functions leads to somewhat smaller phase space factors (up to $30\%$ for the heaviest nuclei) while the effects of electron screening are at the percent level \cite{Stefanik:2015twa}. In what follows we do not use Eq.~\eqref{fermi} to calculate the phase space factors but instead use the more accurate results of Ref.~\cite{Horoi:2017gmj} (see Table~\ref{Tab:phasespace}) which were found to be close to those of Ref.~\cite{Stefanik:2015twa}. We only use Eq.~\eqref{fermi} when calculating differential decay rates in Sect.~\ref{sec:differential}.

The Fourier-transformed amplitude is given by\footnote{
$V(\vec q^2)$ takes into account diagrams where the two nucleons are interchanged, which implies that the unrestricted sum in Eq.~\eqref{eq:FullAmp} counts each of these graphs twice. We correct for this double counting by inserting a factor of $1/4$ in the prefactor of Eq.~\eqref{eq:InvHalfLife}. An additional factor $1/2$ appears because of the two identical electrons in the final state, leading to an overall factor of $1/8$.}
\bea \label{eq:FullAmp}
\mathcal A = \langle 0^+| \sum_{m,n} \int \frac{d^3\vec q}{(2\pi)^3} e^{i\vec q \cdot \vec r}V(\vec q\sq) |0^+\rangle\,\,,
\eea
where $V(\vec q\sq)$ is the sum of the potentials discussed in Section \ref{sec:2btOperators}, and $\vec r = \vec r_n - \vec r_m$ is the distance between the $m^{\rm th}$ and $n^{\rm th}$ nucleon.  

Organizing the amplitude in Eq.\ \eqref{eq:FullAmp} according to the different leptonic structures, the contributions of a light Majorana neutrino mass and dimension-seven operators are given by
\bea\label{eq:TotAmp}
\mathcal A&=& \frac{g_A\sq G_F\sq m_e}{\pi R_A}\bigg[\mathcal{ A}_{\nu}\, \bar u (k_1)P_R C\bar u^T(k_2)+ \mathcal{ A}_{ E}\, \bar u (k_1)\g_0 C\bar u^T(k_2)\,\frac{E_1-E_2}{m_e}\nn\\
&&+ \mathcal{ A}_{ m_e}\, \bar u (k_1) C\bar u^T(k_2)+ \mathcal{ A}_{ M} \,\bar u (k_1)\g_0 \g_5 C\bar u^T(k_2)\bigg]\,.
\eea
Here we factored out the leptonic structures such that the $\mathcal A_{i}$ only depend on nuclear (and hadronic) matrix elements and the Wilson coefficients of the $\Delta L=2$ operators. These are discussed in much more detail below. 

With the definitions in Eq.~\eqref{eq:TotAmp}, the final form of the inverse half-life can be written as
\bea \label{eq:T1/2}
\left(T^{0\nu}_{1/2}\right)^{-1} &=& g_A^4 \Big\{ G_{01} \,|\mathcal A_{\nu}|\sq + 4G_{02} \,|\mathcal A_{E}|\sq+ 2 G_{04} \left[|\mathcal A_{m_e}|\sq+{\rm Re} \left(\mathcal A_{m_e}^* \mathcal A_{\nu}\right)\right]+ G_{09}\, |\mathcal A_{M}|\sq\nn\\
&&-2 G_{03}\,{\rm Re}\left( \mathcal A_{\nu} \mathcal A_{E}^*+2\mathcal A_{m_e} \mathcal A_{E}^*\right)+ G_{06}\, {\rm Re}\left( \mathcal A_{\nu} \mathcal A_{M}^*\right) \Big\}\,\,,
\eea
where the $G_{0i}$ are phase space factors given by
\bea\label{eq:PhaseSpace}
G_{0k}=\frac{1}{\ln 2}\frac{G_F^4m_e\sq}{64\pi^5 R_A\sq}\int  dE_1 dE_2 |\vec k_1| |\vec k_2| d\cos\theta \,b_{0k} \,F(Z,E_1)F(Z,E_2) \dt(E_1+E_2+E_f-M_i)\,.
\eea
Here $\theta$ is the angle between the electron momenta and we followed the standard normalization of Ref.~\cite{Doi:1985dx}. The $b_{0k}$ factors are obtained from the electron traces that result from taking the square of Eq.\ \eqref{eq:TotAmp}. They  are given by 
\bea\label{eq:b0k}
b_{01}&=&E_1 E_2-\vec k_1\cdot \vec k_2\,,\quad b_{02} = \left(\frac{E_1-E_2}{m_e}\right)\sq\frac{E_1 E_2 +\vec k_1\cdot \vec k_2 -m_e\sq}{2}\,,\quad b_{03} = (E_1-E_2)\sq\,,\nn\\
b_{04} &=& \left(E_1 E_2 -\vec k_1\cdot \vec k_2 -m_e\sq\right)\,,\quad  b_{06} = 2 m_e \left( E_1+E_2 \right)\,,\quad b_{09} = 2\left( E_1E_2+\vec k_1\cdot \vec k_2 +m_e\sq\right)\,. \nn\\
\eea
Here we kept terms proportional to $\vec k_1\cdot \vec k_2$, which are odd in $\cos\theta$ and therefore do not contribute to the total decay rate, but can potentially be observed in measurements of angular distributions.
The definitions in Eq.~\eqref{eq:b0k} follow for the most part the existing literature \cite{Doi:1985dx}. 
For $G_{06}$ and $G_{09}$, in order not to cloud the chiral scaling of the matrix element, we did not extract a factor of $(R_A m_e)^{-1}$ from $\mathcal A_M$, as commonly done in the literature \cite{Doi:1985dx}. The phase space factors $G_{06}$ and $G_{09}$ defined in Eqs.~\eqref{eq:PhaseSpace} and \eqref{eq:b0k} are obtained by multiplying the results in Ref.~\cite{Doi:1985dx,Horoi:2017gmj}
by $(m_e R_A)/2$ and $(m_e R_A/2)^2$, respectively. In addition, we removed a factor of  $2/9$ from the definition of $G_{04}$ in order to avoid small dimensionless factors.

The phase space factors are summarized in Table~\ref{Tab:phasespace}. These are extracted from the calculation of Ref.~\cite{Horoi:2017gmj}, with the trivial rescalings discussed above.
With the definitions of Eq.~\eqref{eq:b0k}, the different phase space factors for a given isotope are all of similar size, with no parametric enhancements or suppressions, such that the relative importance of different contributions is determined by the matching coefficients and by the nuclear matrix elements. With the modified phase space factors, we can now apply the $\chi$PT power counting purely on the level of nuclear matrix elements.

\begin{table}
\center
\begin{tabular}{|c|cccc|}
\hline
\hline
\cite{Horoi:2017gmj}	    & $^{76}$Ge & $^{82}$Se & $^{130}$Te & $^{136}$Xe \\ 

\hline
$G_{01}$    & 0.22 & 1. & 1.4 & 1.5 \\
$G_{02}$    & 0.35 & 3.2 & 3.2 & 3.2 \\
$G_{03}$    & 0.12 & 0.65 & 0.85 & 0.86 \\
$G_{04}$    & 0.19 & 0.86 & 1.1 & 1.2 \\
$G_{06}$    & 0.33 & 1.1 & 1.7 & 1.8 \\
$G_{09}$    & 0.48 & 2. & 2.8 & 2.8 \\\hline
\hline
$Q/{\rm MeV} $ \cite{Stoica:2013lka} & 2.04& 3.0&2.5 & 2.5 \\
\hline\hline
\end{tabular}
\caption{Phase space factors in units of $10^{-14}$ yr$^{-1}$ taken from Ref.~\cite{Horoi:2017gmj} apart from a rescaling of $G_{04}$, $G_{06}$, and $G_{09}$ discussed in the text. In addition the table shows the $Q$ values for the different isotopes, where $Q = M_i - M_f -2m_e$.}
\label{Tab:phasespace}
\end{table}

\subsection{Nuclear matrix elements}\label{NME}

To describe the nuclear parts of this amplitude, we follow standard conventions, e.g.\ those of Ref.~\cite{Hyvarinen:2015bda}, and define the following neutrino potentials\footnote{Note that we normalized $h^{ij}_{K,sd}(r)$ with a factor of $m_\pi^{-2}$ instead of $(m_N m_e)^{-1}$ as done in Ref.~\cite{Hyvarinen:2015bda}. Apart from this rescaling, these definitions agree with the literature once we drop the energy of the intermediate states, which is a subleading correction in $\chi$PT.}
\begin{equation}\label{eq:hK}
h^{ij}_K(r) = \frac{2}{\pi} R_A \int_0^{+\infty} d |\vec q|\,   h^{ij}_{K}(\vec q\sq) j_{\lambda} (|\vec q| r)\,,\qquad h^{ij}_{K,sd}(r) = \frac{2}{\pi} \frac{R_A}{m_\pi^2} \int_0^{+\infty} d |\vec q|\, \vec q\sq \,   h^{ij}_K(\vec q\sq) j_{\lambda} (|\vec q| r)\,,
\end{equation}
where $K \in \{F, GT, T\}$ and $h^{ij}_K(\vec q\sq)$ are defined in Eq.\ \eqref{eq:hK(q)}. 
The $h^{ij}_K(r)$ functions describe long-range contributions, while the $h^{ij}_{K,sd}(r)$ indicate short-range contributions. 
$j_{\lambda} (|\vec q| r)$ are  spherical Bessel functions, with $\lambda = 0$ for F and GT, and $\lambda =2$ for the tensor.
The factors of $R_A$ and $m_{\pi}$ have been inserted so that the neutrino potentials are dimensionless.
Having defined the neutrino potentials, we express the nuclear matrix elements (NMEs) as
\begin{eqnarray}\label{MEdef}
M_{F,(sd)}         &=&  \langle 0^+ | \sum_{m,n} h_{F,(sd)}(r) \tau^{+(m)} \tau^{+(n)}  | 0^+ \rangle\,, \nn\\
M^{ij}_{GT,(sd)} &=&  \langle 0^+ | \sum_{m,n} h^{ij}_{GT,(sd)}(r) \, \boldsigma^{(m)} \cdot \boldsigma^{(n)} \, \tau^{+(m)} \tau^{+(n)} | 0^+ \rangle\,, \nn\\
M^{ij}_{T,(sd)} &=&  \langle 0^+ | \sum_{m,n} h^{ij}_{ T,(sd)}(r) \, S^{(mn)}(\hat{\vec r}) \, \tau^{+(m)} \tau^{+(n)} | 0^+ \rangle\,, 
\end{eqnarray}
where the tensor in position space is defined by $S^{(mn)}(\hat{\vec r}) = \left( 3\,\boldsigma^{(m)} \cdot  \hat{ \vec r} \, \boldsigma^{(n)} \cdot  \hat{ \vec r} - \boldsigma^{(m)}\cdot \boldsigma^{(n)} \right)$. 
In the $\chi$PT power counting, the matrix elements defined in Eq.~\eqref{MEdef} are all expected to be $\mathcal O(1)$, with the exception of $M^{MM}_{GT}$ and  $M^{MM}_{T}$, which are suppressed by 
$\mathcal O(\epsilon_\chi^2)$. The latter suppression, however, is softened by the large isovector magnetic moment of the nucleon which numerically scales as $(1+\kappa_1)\epsilon_\chi \simeq \mathcal O(1)$.

The $\mathcal A_{i}$ that appear in Eq.~\eqref{eq:TotAmp} can be obtained from the potentials in Section \ref{sec:2btOperators}, and, for completeness, we give them explicitly in this section. 
$\mathcal A_{\nu}$  has the same leptonic structure as the amplitude induced by light Majorana-neutrino exchange. We can divide it in a component which is proportional to the Majorana mass $m_{\beta\beta}$,
a long-distance component $M_{\nu,\, \textrm{ld}}$ arising from the dimension-six and -seven operators in Eqs. \eqref{lowenergy6} and \eqref{lowenergy7}, and a short-distance component $M_{\nu,\, \textrm{sd}}$, proportional to the coefficients
of low-energy dimension-nine operators
\begin{eqnarray}\label{MSM1}
\mathcal{ A}_{\nu} = \frac{m_{\beta \beta} }{m_e}V_{ud}\sq M_{\nu}  +  \frac{m_N}{m_e}  V_{ud}M_{\nu,\, {ld}} +    \frac{m^2_N}{m_e v} M_{\nu,\, {sd}}\,.
\end{eqnarray}
The nuclear matrix element for light Majorana-neutrino exchange has the well-known form
\begin{eqnarray}\label{MSMstandard}
M_{\nu} = -  \left(  - \frac{g_V^2}{g_A^2} M_{F} +   M_{GT}   +  M_{T}   \right),
\end{eqnarray}
where the GT and T matrix element are, respectively, $M_{GT} =   M^{AA}_{GT} +  M^{AP}_{GT} + M^{PP}_{GT} + M^{MM}_{GT}$ and $M_T =  M^{AP}_{T} + M^{PP}_{T} + M^{MM}_{T}$.

The long-distance component  $M_{\nu,\, \textrm{ld}}$
receives contributions from the scalar operators $C^{(6)}_{\rm SL,\, SR}$, the tensor operator $C^{(6)}_{\rm T}$, and the dimension-seven vector operators $C^{(7)}_{\rm VL,\, VR}$.
The contributions of these operators are not proportional to the neutrino mass, which is replaced by a nuclear scale. We take this into account by factoring one power of the nucleon mass out of the nuclear matrix element in Eq.~\eqref{MSM1}. Combining the results of Secs. \ref{MEscalar} and \ref{MEtensor}, we obtain
\begin{eqnarray}\label{MSM}
M_{\nu,\, ld} &=&   
\left( \frac{B}{m_N} (C^{(6)}_{\rm SL}  - C^{(6)}_{\rm SR} ) + \frac{m^2_\pi}{m_N v} \left( C^{(7)}_{\rm VL} - C^{(7)}_{\rm VR}\right) \right) M_{PS} +  C^{(6)}_{\rm T} g_T M_{T6}\,, 
\end{eqnarray}
where 
\begin{eqnarray}
M_{PS} &=& \frac{1}{2} M^{AP}_{GT} + M^{PP}_{GT}  + \frac{1}{2} M^{AP}_{T} + M^{PP}_{T}, \\
M_{T6} &=& 2 \frac{g^\prime_T g_V}{g_T g_A^2} \frac{m^2_\pi}{m_N^2} M_{F,\, sd}  - \frac{8}{ g_M} \left( M_{GT}^{MM} + M_T^{MM} \right).
\end{eqnarray}
We see that  $C^{(6)}_{\rm SL,\, SR}$ give the largest contributions to  $M_{\nu,\, ld}$, 
followed by the  tensor operator $C^{(6)}_{\rm T}$ whose effects are formally suppressed by $m_\pi^2/\Lambda_\chi^2$, but again this suppression is somewhat mitigated by the large value of $g_M$. The  dimension-seven operators are severely suppressed by the Yukawa couplings of the light quarks (since the relative factor can be written as $m_\pi\sq/B v  = (m_u+m_d)/v$). 

The short-distance component arises from the dimension-nine operators in Eq.~\eqref{lowenergy9}, which always involve an additional power of $1/v$ with respect to the contribution from light Majorana-neutrino exchange. To compensate for this factor, and for the absence of the neutrino mass, we factored two powers of $m_N$ out of the short-distance nuclear matrix element in Eq.~\eqref{MSM1}. We then have
\begin{eqnarray}
& & M_{\nu,\, sd} = \left( \frac{g_{8 \times 8}}{2m_N^2}  C^{(9)}_{4} + \frac{g^{\rm mix}_{8 \times 8}}{2m_N^2}  C^{(9)}_{5} \right) M_{sd,\,1}  +   \frac{m_\pi^2}{m_N^2} g^{NN}_{27 \times 1}\, C^{(9)}_{1} M_{sd,\,2}  \, ,
\end{eqnarray}
where we defined
\begin{eqnarray}
M_{sd,\, 1} &=& \frac{1}{2} M^{AP}_{GT, sd} + M^{PP}_{GT, sd}
+ \frac{1}{2} M^{AP}_{T, sd} + M^{PP}_{T, sd}\,,\\
M_{sd,\, 2} &=&  - 2\frac{g_V^2}{g_A^2}  M_{F,\, sd} + \frac{1}{2} \frac{ g^{\pi N}_{27 \times 1}}{g^{NN}_{27 \times 1}} \left(M^{AP}_{GT, sd} + M^{AP}_{T, sd} \right) +\frac{5}{6} 
\frac{g_{27 \times 1}^{\pi\pi}}{g^{NN}_{27 \times 1}}
\left(M^{PP}_{GT, sd} + M^{PP}_{T, sd} \right)\,. \label{eq54}
\end{eqnarray}
In Eq.~\eqref{eq54} we factored the LEC $g^{NN}_{27 \times 1}$ out of  $M_{sd,\, 2}$  as to make the NME independent of the renormalization scale.
With the scaling of the LECs discussed in Sec. \ref{sec:ChiLag},  the left-right operators  $C^{(9)}_{4,5}$ give the largest contribution to 
$M_{\nu,\, sd}$, while contributions from the purely left-handed operator $C_{1}^{(9)}$ are suppressed by $\epsilon_\chi^2$.

The dimension-six vector and axial operators $C^{(6)}_{\rm VL,\, VR}$ induce the additional leptonic structures in Eq.~\eqref{eq:TotAmp}.
$\mathcal A_{ M}$  is generated through the nucleon magnetic moment and is proportional to $C^{(6)}_{\rm VL}$
\begin{eqnarray}\label{MM}
\mathcal A_{ M} =  \frac{m_N}{m_e} V_{ud} C^{(6)}_{\rm VL }   \, M_{M}\,, \qquad M_M = 2  \frac{g_A}{g_M}  \left(M_{GT}^{MM} + M_{T}^{MM} \right).
\end{eqnarray}

The terms proportional to the electron energies and to the electron mass receive contributions from both $C^{(6)}_{\rm VL}$
and $C^{(6)}_{\rm VR}$, and are given by
\begin{eqnarray}\label{MEme}
\mathcal A_{E} &=&  V_{ud} C^{(6)}_{\rm VL}\,M_{E,L} +  V_{ud} C^{(6)}_{\rm VR}\, M_{E,R} \, ,\nn \\
\mathcal A_{me} &=&  V_{ud} C^{(6)}_{\rm VL}\,M_{m_e,L} +  V_{ud} C^{(6)}_{\rm VR}\,M_{m_e,R}\,,
\end{eqnarray}
where
\begin{eqnarray}\label{MmeLR}
M_{E,L} &=&- \frac{1}{3}\left(  \frac{g^2_V}{g_A^2} M_F  + \frac{1}{3} \left( 2 M_{GT}^{AA} + M_T^{AA} \right)\right)\,, \nn \\
M_{E,R}& =&- \frac{1}{3}\left(  \frac{g^2_V}{g_A^2} M_F  - \frac{1}{3} \left( 2 M_{GT}^{AA} + M_T^{AA} \right)\right)\,, \nn \\
 M_{m_e,L} &=&  \frac{1}{6}\left(   \frac{g^2_V}{g_A^2} M_F  - \frac{1}{3} \left( M_{GT}^{AA} - 4 M^{AA}_{T}\right) - 3 \left( M^{AP}_{GT} + M^{PP}_{GT} + M^{AP}_{T} + M^{PP}_{T}\right)   \right)\, , \nn \\
 M_{m_e,R} &=&  \frac{1}{6}\left(   \frac{g^2_V}{g_A^2} M_F  + \frac{1}{3} \left( M_{GT}^{AA} - 4 M^{AA}_{T}\right) +  3 \left( M^{AP}_{GT} + M^{PP}_{GT} + M^{AP}_{T} + M^{PP}_{T}\right)   \right)\, .
\end{eqnarray}
One of the NME combinations is redundant as we can write $M_{m_e,R} = - ( M_{E,L}+M_{E,R} + 2 M_{m_e,L})/2$. We choose to eliminate $M_{m_e,R}$ in the sections below.

\subsection{Chiral power counting}

With these definitions we have introduced nine independent combinations of nuclear matrix elements that determine the $0\nu\beta\beta$ rate at LO in $\chi$PT arising from dimension-5 and -7 operators in the SM EFT. The combination of matrix elements $M_{\nu}$, $M_{PS}$, $M_{sd,\{1,2\}}$, $M_{E,\{L,R\}}$, $M_{m_e,L}$ are all expected to be $\mathcal O(1)$, while $M_M,\,M_{T6}$ scale as $\mathcal O(m_\pi^2/\Lambda_\chi^2)$ but are enhanced by a factor of $g_M$. Not all matrix elements contribute equally to the decay rate because of factors of $m_N/m_e$ and $m_\pi^2/m_N^2$ that appear in the definitions of the amplitudes $\mathcal A_i$ in Eqs.~\eqref{MSM}, \eqref{MM}, and \eqref{MEme}. 

The power-counting estimates of the amplitudes are summarized in Table \ref{TabPC}. As discussed in Sec. \ref{sec:2btOperators}, the smallness of the electron's mass  
and energy is accounted for in the power counting 
by assigning the scaling $E_1 \sim E_2 \sim m_e \sim m_\pi \epsilon_\chi^2 = \Lambda_\chi \, \epsilon_\chi^3$. 
The power counting suggests that $C^{(6)}_{\rm SL,\, SR}$ give the largest contribution to the inverse half-life, and thus are the most constrained from $0\nu\beta\beta$ experiments. This expectation is verified in Sect.~\ref{Bounds}. 
$C^{(6)}_{\rm T}$ and $C^{(6)}_{\rm VL}$ give contributions of similar size, suppressed by two powers of $\epsilon_\chi$. In both cases, the large nucleon isovector magnetic moment enhances 
the matrix elements leading to somewhat stronger bounds than expected. $C^{(6)}_{\rm VL}$ induces contributions to $\mathcal A_{E}$ and $\mathcal A_{me}$, which arise at $\mathcal O(\epsilon_\chi^3)$,
and thus can be neglected compared to $\mathcal A_{M}$. This expectation is very well confirmed when using realistic values of the nuclear matrix elements.  
In the case of $C^{(6)}_{\rm VR}$, there is no contribution to $\mathcal A_{M}$, and thus the first correction to the half-life is of $\mathcal O(\epsilon_\chi^3)$. As a consequence, the bound on this coefficient, which is particularly interesting for left-right symmetric models, is weaker than for the remaining dimension-six operators as is explicitly found in Sect.~\ref{Bounds}. 

Dimension-seven and -nine operators are further suppressed due to inverse powers of the electroweak scale. Contributions from $C^{(9)}_{4,5}$ are suppressed by $\Lambda_\chi/v$, while contributions from $C^{(9)}_{1}$ and the dimension-seven operators $C^{(7)}_{\rm VL,VR}$ by $ \Lambda_\chi\epsilon_\chi^2/v$.

\begin{table}
\center
\begin{tabular}{||c|cccccccc||}
 \hline \hline 
			    & $\nu$ &  $C^{(6)}_{\rm SL,\, SR}$       & $C^{(6)}_{\rm T}$ 		  &  $C^{(6)}_{\rm VL}$  & $C^{(6)}_{\rm VR}$  & $C^{(7)}_{\rm VL, \, VR}$  & $C^{(9)}_{1}$ & $C^{(9)}_{4, 5}$ \\
\hline
$m_e \mathcal A_{\nu}$   & $m_{\beta\beta}$ & $\Lambda_{\chi}$  & $\Lambda_{\chi} \epsilon^2_\chi$ &  $-$  &  $-$ & $ \frac{\Lambda_\chi^2}{v} \, \epsilon_\chi^2$  & $ \frac{\Lambda_\chi^2}{v} \, \epsilon_\chi^2$ & $ \frac{\Lambda_\chi^2}{v}$  \\
$m_e \mathcal A_{M}$        & $-$ 	       & $-$ 		   & $-$ &  $\Lambda_{\chi} \epsilon^2_\chi$ & $-$ & $-$ & $-$ & $-$    \\
$m_e \mathcal A_{E}$        & $-$ 	       & $-$  		   & $-$ &  $\Lambda_{\chi} \epsilon^3_\chi$ & $\Lambda_{\chi} \epsilon^3_\chi$ & $-$ & $-$ & $-$   \\
$m_e \mathcal A_{me}$       & $-$ 	       & $-$  		   & $-$ &  $\Lambda_{\chi} \epsilon^3_\chi$ & $\Lambda_{\chi} \epsilon^3_\chi$ & $-$ & $-$ & $-$    \\
\hline
\end{tabular}
\caption{Power-counting estimates of the contribution of low-energy dimension-six, -seven, and -nine operators, as well as $m_{\bt\bt}$ to the amplitudes in Eq.~\eqref{eq:TotAmp}.  
Here $\nu$ stands for the contribution of the light Majorana-neutrino exchange mechanism. Furthermore, $\epsilon_\chi \equiv m_\pi/\Lambda_\chi$, where $\Lambda_\chi \sim m_N \sim 1$ GeV is the symmetry-breaking scale. For the power counting, we consider
the electron mass and energies and to scale as $E_1 \sim E_2 \sim m_e \sim \Lambda_\chi \, \epsilon_\chi^3$.  }\label{TabPC}
\end{table}

Having discussed the  $\chi$PT power-counting expectations, in Table \ref{tab:comparison} we list the numerical values of the NMEs, 
which are obtained from the calculations of Refs.\ \cite{Hyvarinen:2015bda,Horoi:2017gmj,Javier,Barea:2015kwa,Barea}.
It is interesting that, with the exception of $M^{AA}_{T}$, all the NMEs that are needed to constrain the contributions of dimension-seven operators  
can be lifted from existing calculations of  \NLDBD\ mediated by light and heavy Majorana neutrino exchange, provided that these calculations 
include the contributions of weak magnetism and of the induced pseudoscalar form factor, and the results for the various components of $M_{GT}$
and $M_T$ in Eq.~\eqref{MSMstandard} (and in the analogous expression for heavy-neutrino exchange) are listed separately, as done for examples in Refs.\ \cite{Menendez:2008jp,Barea:2009zza,Hyvarinen:2015bda}\footnote{We thank J. Men\'endez and J. Barea for providing us with updated values of the NMEs for light- and heavy-neutrino exchange \cite{Javier,Barea}, 
with GT and T matrix elements separated in $AA$, $AP$, $PP$, and $MM$ components.}.
In Appendix \ref{MEconversion}  we discuss how to convert the nuclear matrix elements of the original references to the notation of Eqs.\ \eqref{eq:hK} and \eqref{MEdef} (see Table \ref{tab:NMEs}).
The NME $M^{AA}_{T}$ does not contribute to the light Majorana exchange mechanism, and thus requires a dedicated calculation.
This matrix element is important only for $C^{(6)}_{\rm VR}$ and, as we argue in Appendix \ref{MEconversion}, even in this case 
its contribution is numerically small. Therefore, in Sec. \ref{Bounds} we set $M^{AA}_{T}$ to zero.

\begin{table}
$\renewcommand{\arraystretch}{1.5}
\begin{array}{l||rrrr|rrr|rrr |rrr}
 \text{NMEs} & \multicolumn{4}{c|}{\text{}^{76} \text{Ge}} & \multicolumn{3}{c|}{\text{}^{82} \text{Se}} & \multicolumn{3}{c|}{ \text{}^{130} \text{Te}} & \multicolumn{3}{c}{  \text{}^{136} \text{Xe}}  \\
& \text{\cite{Hyvarinen:2015bda}} & \text{\cite{Horoi:2017gmj}} &  \text{\cite{Javier}}  
& \text{\cite{Barea:2015kwa,Barea}} & \text{\cite{Hyvarinen:2015bda}} & \text{\cite{Horoi:2017gmj}} &  \text{\cite{Javier}} & \text{\cite{Hyvarinen:2015bda}} & \text{\cite{Horoi:2017gmj}} &  \text{\cite{Javier}} & \text{\cite{Hyvarinen:2015bda}} & \text{\cite{Horoi:2017gmj}} &  \text{\cite{Javier}} \\
 \hline
 M_F 			   & $-$1.74    & $-$0.67 	&  $-$0.59 	& $-$0.68 
			   & $-$1.29 	& $-$0.63    	&  $-$0.55	
			   & $-$1.52    & $-$0.44    	&  $-$0.67	
			   & $-$0.89  	& $-$0.40  	&  $-$0.54 \\
 M_{GT}^{AA} & 5.48     	& 3.50	 	&  3.15	    	& 5.06 
			   & 3.87 	& 3.29     	&  2.97		 
			   & 4.28    	& 1.85    	&  2.97		        
			   & 3.16   	& 1.68  	&  2.45 \\
 M_{GT}^{AP} & $-$2.02  	& $-$0.25	 	& $-$0.94		& $-$0.92 
			   & $-$1.46    & $-$0.23     	& $-$0.89   			 
			   & $-$1.74 	&$-$ 0.19    	& $-$0.97			
			   & $-$1.19   	& $-$0.17  	& $-$0.79  \\
 M_{GT}^{PP} & 0.66  	& 0.33  	&  0.30	& 0.24 
			   & 0.48       & 0.31  	&  0.28 		  	 
			   & 0.59   	& 0.21 	&  0.31   				
			   & 0.39   	& 0.19 	&  0.25\\
 M_{GT}^{MM} & 0.51 	& 0.25 		& 0.22	& 0.17 
			   & 0.37       & 0.24    	& 0.20 
			   & 0.45 	& 0.17 		& 0.23 
			   & 0.31 	& 0.15 		& 0.19 \\
 M_T^{AA} 	   &  -    	& -	 	& - 		& - 
			   &  -         & - 		& -	
			   &  -     	& - 		& -      
			   &  -     	& - 		&  -	\\
 M_T^{AP} 	   & $-$0.35 	& 0.01 		& $-$0.01	& $-$0.31 
			   & $-$0.27    & 0.01  	& $-$0.01 				 
			   & $-$0.50	& $-$0.01 	&    0.01				 
			   & $-$0.28 	& 0.01 		&    0.01	\\
 M_T^{PP} 	   & 0.10     	& 0.00 		&    0.00	& 0.09
			   & 0.08       & 0.00 		&    0.00
			   & 0.16 	& 0.01  	& $-$0.01						 
			   & 0.09 	& $-$0.01 	& $-$0.01 	 \\
 M_T^{MM}           & $-$0.04	&  0.00		&   0.00	& $-$0.04
			   & $-$0.03    & $-$0.00  	&   0.00 				
			   & $-$0.06 	&  0.00  	&   0.00				         
			   & $-$0.03 	&  0.00		&   0.00  \\\hline
 M_{F,\, sd}  	   & $-$3.46 	& $-$1.55 	& $-$1.46		& $-$1.1  
			   & $-$2.53   	& $-$1.44    	& $-$1.37 					
			   & $-$2.97 	& $-$1.02	& $-$1.61 						 
			   & $-$1.53   	& $-$0.92    	& $-$1.28	\\
 M^{AA}_{GT,\, sd}  	   &    11.1 	& 4.03 		& 4.87		& 3.62
			   &  	7.98    & 3.72    	& 4.54 					
			   & 	10.1 	& 2.67 		& 5.31 						 
			   &    5.71    & 2.40    	& 4.25  \\
M^{AP}_{GT,\, sd}& $-$5.35 	& $-$2.37 		& $-$2.26 		& $-$1.37 
			   & $-$3.82    & $-$2.19		& $-$2.09 						
			   & $-$4.94 	& $-$1.61 		& $-$2.51 				 
			   & $-$2.80  	& $-$1.45  	& $-$1.99  \\
M^{PP}_{GT,\, sd}& 1.99 	& 0.85	& 0.82	& 0.42
			   & 1.42       & 0.79	& 0.77 			
			   & 1.86 	& 0.60	& 0.92 					 
			   & 1.06  	& 0.53       & 0.74\\
M^{AP}_{T,\, sd}& $-$0.85 	& 0.01		&  $-$0.05		&  $-$0.97
			   & $-$0.65    & 0.02		&  $-$0.05	
			   & $-$1.50 	& $-$0.07	&  0.07				 
			   & $-$0.92  	&  0.08 	&  0.05		\\  
M^{PP}_{T,\, sd} & 0.32 	& 0.00		&  0.02		&  0.38
			   & 0.24       & $-$0.01	&  0.02			
			   & 0.58 	& 0.03		&  $-$0.02					 
			   & 0.36  	& $-$0.03 	&  $-$0.02 \\
\end{array}$
\caption{Comparison of the different NMEs of Refs.\ \cite{Hyvarinen:2015bda,Javier,Barea:2015kwa,Barea,Horoi:2017gmj}, for the nuclei relevant for the GERDA \cite{Agostini:2017iyd}, NEMO \cite{SeNEMO}, CUORE \cite{Alfonso:2015wka}, and KamLAND-Zen \cite{KamLAND-Zen:2016pfg} experiments. To obtain $M_{F}$, $M_{GT}^{AA}$, $M_{GT}^{MM}$, $M_{F,\, sd}$, and $M^{AA}_{GT,\, sd}$ we used, respectively, $M_{F}$, $M_{GT\omega}$, $M_{GT'}$, $M_{FN}$, and $M_{GTN}$  of Ref.~\cite{Horoi:2017gmj}, see Appendix \ref{MEconversion} and Table \ref{tab:NMEs}.}
\label{tab:comparison}
\end{table}

A few comments are in order. First of all, the neutrino potentials derived in $\chi$PT are not sensitive to the closure energy  $\bar E$, 
where $\bar E \sim 1 -10$ MeV is much smaller than the typical Fermi momentum. The relations in Table \ref{tab:NMEs} are valid in the limit $\bar{E} \rightarrow 0$,
which should be a good approximation if the bulk of the nuclear matrix elements comes from the region $r \sim 1/k_F$. 
Secondly, the momentum dependence of the axial and vector form factors is an $\mathcal O(\epsilon_\chi^2)$ effect in $\chi$PT, and some of the relations in Table \ref{tab:NMEs} neglect the difference between the axial and vector dipole masses, which is justified at leading order.
Refs.~\cite{Hyvarinen:2015bda}, \cite{Horoi:2017gmj}, and \cite{Javier} computed NMEs that, with these assumptions, should be equal, up to higher-order corrections. By comparing these NMEs we can thus explicitly test the validity 
of the chiral power counting.

As a first example, if the momentum dependence of $g_V(\vec q^2)$ and $g_A(\vec q^2)$ is neglected, 
the short-distance matrix elements $M_{F, sd}$ and $M^{AA}_{GT, sd}$ are related by a Fierz identity
\begin{equation}\label{Fierz}
M^{AA}_{GT,\, sd} =  - 3  M_{F,\, sd}\,.
\end{equation}
Table~\ref{tab:comparison} shows that the results from Ref.~\cite{Hyvarinen:2015bda} obey
 Eq.~\eqref{Fierz} up to corrections that range from  $\sim10\%$ for $^{76}$Ge and $^{82}$Se to $\sim20\%$ for $^{136}$Xe, while  
in Refs.~\cite{Horoi:2017gmj} and ~\cite{Javier} the corrections are roughly $15\%$ and $10\%$ for all the nuclei that were considered.
The results of Ref. \cite{Barea} for $^{76}$Ge also respect Eq.~\eqref{Fierz} at the 10\% level. Once the momentum dependence of 
$g_V(\vec q^2)$ and $g_A(\vec q^2)$ is no longer neglected, the relation in Eq. (\ref{Fierz}) receives corrections at $\mathcal O(\epsilon_\chi^2)$ in $\chi$PT, 
a size consistent with these numerical results.

Furthermore, using the identity $\vec q\sq = (\vec q\sq +m_\pi\sq) -m_\pi\sq$, and again neglecting the momentum dependence of $g_V(\vec q^2)$ and $g_A(\vec q^2)$, we can derive the following relations between short- and long-distance matrix elements, 
\bea\label{relAP}
M_{GT,sd}^{PP}&=&-\frac{1}{2}M_{GT,sd}^{AP}-M_{GT}^{PP}\,,\qquad M_{T,sd}^{PP}=-\frac{1}{2}M_{T,sd}^{AP}- M_{T}^{PP}\,,\nn\\
M_{GT,sd}^{AP}&=&-\frac{2}{3}M_{GT,sd}^{AA}- M_{GT}^{AP}\,,\qquad M_{GT}^{MM}=\frac{g_M\sq m_\pi^2}{6 g_A^2\, m^2_N}M_{GT,sd}^{AA}\,,
\eea 
that are valid through NLO in the chiral counting.

The NMEs of Refs.~\cite{Hyvarinen:2015bda}, \cite{Javier} and \cite{Barea} respect the first three relations to $5\%$ accuracy, the fourth to $10\%$. 
For Ref. \cite{Horoi:2017gmj}, $M^{AP, PP}_{GT}$ and $M^{AP, PP}_{GT, sd}$ were constructed from pion-range NMEs using 
the relations of  Table~\ref{tab:NMEs}, which make the first two and the fourth equations in Eq.~\eqref{relAP} trivial identities.
The third relation in Eq.~\eqref{relAP} is non-trivial, and it is well respected by the NMEs in Ref. \cite{Horoi:2017gmj}.
These numerical results  confirm that the relations in Eq. (\ref{relAP}) are accurate up to $(5$-$10)\%$ corrections, which is of the same size as the expected $\mathcal O(\epsilon_\chi^2)$ $\chi$PT effects.

The large number of NMEs computed in  Ref.~\cite{Horoi:2017gmj} allows for additional consistency checks, which we discuss in Appendix \ref{MEconversion}.
In general, for the consistency checks performed in Appendix \ref{MEconversion}, we observe that various relations between NMEs are respected up to $20\%$-$30\%$ corrections, the level one would expect from LO $\chi$PT.  We conclude that the power counting is working satisfactory although stronger conclusions would require the explicit inclusion of NLO corrections.

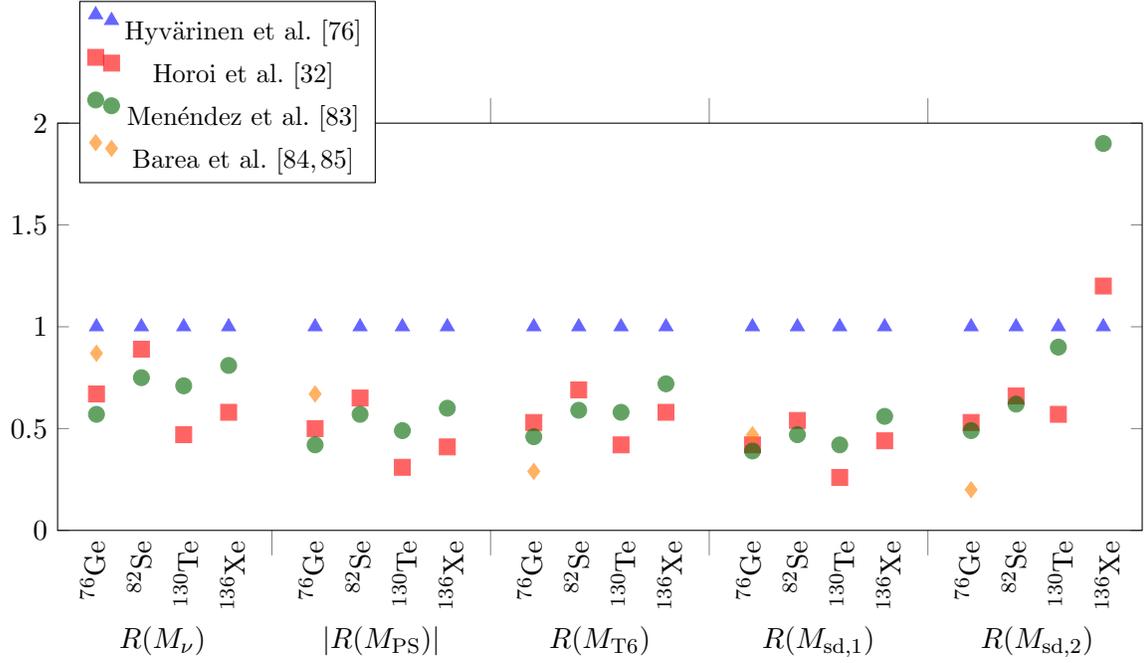
\begin{figure}
\center
\makeatletter
\pgfplotsset{
    calculate offset/.code={
        \pgfkeys{/pgf/fpu=true,/pgf/fpu/output format=fixed}
        \pgfmathsetmacro\testmacro{(\pgfplotspointmeta *10^\pgfplots@data@scale@trafo@EXPONENT@y)*\pgfplots@y@veclength)}
        \pgfkeys{/pgf/fpu=false}
    },
    every node near coord/.style={
        /pgfplots/calculate offset,
        yshift=-\testmacro
    },
}

\pgfplotstableread{
0 1 1 1 1
1 1 1 1 1 
2 1 1 1 1 
3 1 1 1 1 
4 1 1 1 1 
}\finnish

\pgfplotstableread{
0 .67 .89 .47 .58
1 -.5 -.65 -.31 -.41
2 .53 .69 .42 .58
3 -.42 -.54 -.26 -.44
4 1.6 1.9 1.6 3.9
}\horoi

\pgfplotstableread{
0 .67 .89 .47 .58
1 .49 .65 .31 .41
2 .53 .69 .42 .58
3 .42 .54 .26 .44
4 .53 .66 .57 1.2
}\horoiPrime

\pgfplotstableread{
0 .67 .89 .47 .58
1 .5 .65 .31 .41
2 .53 .69 .42 .58
3 .42 .54 .26 .44
4 .53 .66 .57 1.2
}\horoiFINAL

\pgfplotstableread{
0 .57 .75 .71 .81
1 .42 .57 .49 .6
2 .46 .59 .58 .72
3 .39 .47 .42 .56   
4 .49 .62 .9 1.9
}\javier

\pgfplotstableread{
0 .87
1 .67
2 .29
3 .47
4 .2
}\barea

\begin{tikzpicture}
\begin{axis}[ybar,
		 bar width=.55cm, 
		  enlarge x limits=.12,
        width=\textwidth,
        height=7cm,
        ymin=0,
        ymax=2,     
        legend style={font=\small, at={(0.02,1.3)},anchor=north west},   
        xtick=data,
        xticklabels = {
			$R(M_\nu)$,
            $|R(M_{\rm PS})|$,
            $R(M_{\rm T6})$,
            $R(M_{\rm sd, 1})$,
            $R(M_{\rm sd, 2})$
        },
        xticklabel style={yshift=-6ex},
        major x tick style = {opacity=0},
        minor x tick num = 1,
        minor tick length=2ex,
        every node near coord/.append style={yshift=0ex,
                anchor=east,
                rotate=90, color=black,opacity=1
        }
        ]
\addplot[draw=, opacity=0, fill=black!0, nodes near coords=$^{76}$Ge] table[x index=0,y index=1] \finnish;
\addplot[only marks, mark=triangle*,blue,opacity=.61,mark size = 3pt,every mark/.append style={xshift=-25pt}] table[x index=0,y index=1] \finnish;
\addplot[only marks, mark=square*,red,opacity=.61,mark size = 3pt, every mark/.append style={xshift=-25pt}] table[x index=0,y index=1] \horoiFINAL;
%\addplot[only marks, mark=o,red,opacity=1,mark size = 3pt,every mark/.append style={xshift=-25pt}] table[x index=0,y index=1] \horoiPrime;
\addplot[only marks, mark=*,black!60!green,opacity=.61,mark size = 3pt,every mark/.append style={xshift=-25pt}] table[x index=0,y index=1] \javier;
\addplot[only marks, mark=diamond*,orange,opacity=.61,mark size = 3pt,every mark/.append style={xshift=-25pt}] table[x index=0,y index=1] \barea;

\addplot[draw=black,opacity=0,fill=black!40, nodes near coords=$^{82}$Se] table[x index=0,y index=2] \finnish ; %Data2
\addplot[only marks, mark=square*,red,opacity=.61,mark size = 3pt,every mark/.append style={xshift=-8pt}] table[x index=0,y index=2] \horoiFINAL;
%\addplot[only marks, mark=o,red,opacity=1,mark size = 3pt,every mark/.append style={xshift=-8pt}] table[x index=0,y index=2] \horoiPrime;
\addplot[only marks, mark=triangle*,blue,opacity=.61,mark size = 3pt,every mark/.append style={xshift=-8pt}] table[x index=0,y index=2] \finnish;
\addplot[only marks, mark=*,black!60!green,opacity=.61,mark size = 3pt,every mark/.append style={xshift=-8pt}] table[x index=0,y index=2] \javier;

\addplot[draw=black,opacity=0,fill=black!60, nodes near coords=$^{130}$Te] table[x index=0,y index=3] \finnish; %Data3
\addplot[only marks, mark=square*,red,opacity=.61,mark size = 3pt,every mark/.append style={xshift=8pt}] table[x index=0,y index=3] \horoiFINAL;
%\addplot[only marks, mark=o,red,opacity=1,mark size = 3pt,every mark/.append style={xshift=8pt}] table[x index=0,y index=3] \horoiPrime;
\addplot[only marks, mark=triangle*,blue,opacity=.61,mark size = 3pt,every mark/.append style={xshift=8pt}] table[x index=0,y index=3] \finnish;
\addplot[only marks, mark=*,black!60!green,opacity=.61,mark size = 3pt,every mark/.append style={xshift=8pt}] table[x index=0,y index=3] \javier;

\addplot[draw=black,opacity=0,fill=black!60, nodes near coords=$^{136}$Xe] table[x index=0,y index=3] \finnish; %Data3
\addplot[only marks, mark=square*,red,opacity=.61,mark size = 3pt,every mark/.append style={xshift=25pt}] table[x index=0,y index=4] \horoiFINAL;
%\addplot[only marks, mark=o,red,opacity=1,mark size = 3pt,every mark/.append style={xshift=25pt}] table[x index=0,y index=4] \horoiPrime;
\addplot[only marks, mark=triangle*,blue,opacity=.61,mark size = 3pt,every mark/.append style={xshift=25pt}] table[x index=0,y index=4] \finnish;
\addplot[only marks, mark=*,black!60!green,opacity=.61,mark size = 3pt,every mark/.append style={xshift=25pt}] table[x index=0,y index=4] \javier;

\legend{  ,Hyv\"arinen et\ al.\ \cite{Hyvarinen:2015bda},Horoi et al.\ \cite{Horoi:2017gmj}, Men\'endez et al.\ \cite{Javier}, Barea et al.\ \cite{Barea:2015kwa,Barea}}
\end{axis}
 \end{tikzpicture}
\caption{\small Comparison of the NMEs obtained using the calculations of  Refs.~\cite{Hyvarinen:2015bda} (blue triangles), \cite{Horoi:2017gmj}  (red squares),  \cite{Javier} (green circles) and \cite{Barea:2015kwa,Barea} (orange diamonds). To show the different NMEs, $M_i$, on a similar scale we arbitrarily normalized the calculations to the results of Ref.\ \cite{Hyvarinen:2015bda}, i.e.\ $R(M_i) = M_i/M_i^\text{\cite{Hyvarinen:2015bda}}$. For $M_{\rm PS}$ we show the absolute value of the ratio. In this case, Ref.~\cite{Horoi:2017gmj} finds a negative ratio, while for Refs.\ \cite{Javier} and \cite{Barea:2015kwa,Barea} we find positive values. The same finding holds for  $M_{m_e,L}$ shown in Fig.\ \ref{FigNME2}.
}\label{FigNME}
\end{figure}
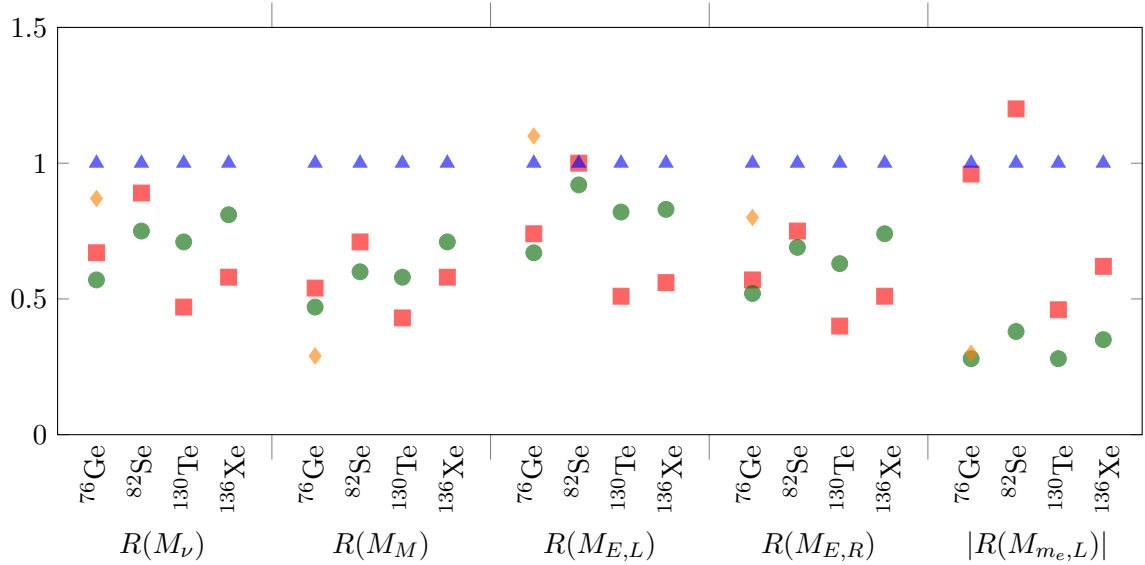
\begin{figure}
\center
\makeatletter
\pgfplotsset{
    calculate offset/.code={
        \pgfkeys{/pgf/fpu=true,/pgf/fpu/output format=fixed}
        \pgfmathsetmacro\testmacro{(\pgfplotspointmeta *10^\pgfplots@data@scale@trafo@EXPONENT@y)*\pgfplots@y@veclength)}
        \pgfkeys{/pgf/fpu=false}
    },
    every node near coord/.style={
        /pgfplots/calculate offset,
        yshift=-\testmacro
    },
}

\pgfplotstableread{
0 1 1 1 1
1 1 1 1 1 
2 1 1 1 1 
3 1 1 1 1 
4 1 1 1 1 
}\finnish

\pgfplotstableread{
0 .67 .89 .47 .58
1 .54 .71 .43 .58
2 .74 1 .51 .56
3 .57 .75 .4 .51
4 -2 -2.6 -1.05 -1.5
}\horoi

\pgfplotstableread{
0 .67 .89 .47 .58
1 .54 .71 .43 .58
2 .74 1 .51 .56
3 .57 .75 .4 .51
4 .35 .45 .2 .3
}\horoiPrime

\pgfplotstableread{
0 .67 .89 .47 .58
1 .54 .71 .43 .58
2 .74 1 .51 .56
3 .57 .75 .4 .51
4 .96 1.2 .46 .62
}\horoiFINAL

\pgfplotstableread{
0 .57 .75 .71 .81
1 .47 .6 .58 .71
2 .67 .92 .82 .83
3 .52 .69 .63 .74
4 .28 .38 .28 .35
}\javier

\pgfplotstableread{
0 .87
1 .29
2 1.1
3 .8
4 .3
}\barea
\begin{tikzpicture}
\begin{axis}[ybar,
		 bar width=.55cm, 
		  enlarge x limits=.12,
        width=\textwidth,
        height=7cm,
        ymin=0,
        ymax=1.5,     
        legend style={font=\small, at={(0.7,1.3)},anchor=north west},   
        xtick=data,
        xticklabels = {
			 $R(M_\nu)$,
            $R(M_{ M})$,
            $R(M_{ E,L})$,
            $R(M_{ E,R})$,
            $|R(M_{ m_e,L})|$
        },
        xticklabel style={yshift=-6ex},
        major x tick style = {opacity=0},
        minor x tick num = 1,
        minor tick length=2ex,
        every node near coord/.append style={yshift=0ex,
                anchor=east,
                rotate=90, color=black,opacity=1
        }
        ]

\addplot[draw=, opacity=0, fill=black!0, nodes near coords=$^{76}$Ge] table[x index=0,y index=1] \finnish; %Data1
\addplot[only marks, mark=square*,red,opacity=.61,mark size = 3pt, every mark/.append style={xshift=-25pt}] table[x index=0,y index=1] \horoiFINAL;
%\addplot[only marks, mark=o,red,opacity=1,mark size = 3pt, every mark/.append style={xshift=-25pt}] table[x index=0,y index=1] \horoiPrime;
\addplot[only marks, mark=triangle*,blue,opacity=.61,mark size = 3pt,every mark/.append style={xshift=-25pt}] table[x index=0,y index=1] \finnish;
\addplot[only marks, mark=*,black!60!green,opacity=.61,mark size = 3pt,every mark/.append style={xshift=-25pt}] table[x index=0,y index=1] \javier;
\addplot[only marks, mark=diamond*,orange,opacity=.61,mark size = 3pt,every mark/.append style={xshift=-25pt}] table[x index=0,y index=1]\barea;

\addplot[draw=black,opacity=0,fill=black!40, nodes near coords=$^{82}$Se] table[x index=0,y index=2] \finnish; %Data2
\addplot[only marks, mark=square*,red,opacity=.61,mark size = 3pt,every mark/.append style={xshift=-8pt}] table[x index=0,y index=2] \horoiFINAL;
%\addplot[only marks, mark=o,red,opacity=1,mark size = 3pt, every mark/.append style={xshift=-8pt}] table[x index=0,y index=2] \horoiPrime;
\addplot[only marks, mark=triangle*,blue,opacity=.61,mark size = 3pt,every mark/.append style={xshift=-8pt}] table[x index=0,y index=2] \finnish;
\addplot[only marks, mark=*,black!60!green,opacity=.61,mark size = 3pt,every mark/.append style={xshift=-8pt}] table[x index=0,y index=2] \javier;

\addplot[draw=black,opacity=0,fill=black!60, nodes near coords=$^{130}$Te] table[x index=0,y index=3] \finnish; %Data3
\addplot[only marks, mark=square*,red,opacity=.61,mark size = 3pt,every mark/.append style={xshift=8pt}] table[x index=0,y index=3] \horoiFINAL;
%\addplot[only marks, mark=o,red,opacity=1,mark size = 3pt, every mark/.append style={xshift=8pt}] table[x index=0,y index=3] \horoiPrime;
\addplot[only marks, mark=triangle*,blue,opacity=.61,mark size = 3pt,every mark/.append style={xshift=8pt}] table[x index=0,y index=3] \finnish;
\addplot[only marks, mark=*,black!60!green,opacity=.61,mark size = 3pt,every mark/.append style={xshift=8pt}] table[x index=0,y index=3] \javier;

\addplot[draw=black,opacity=0,fill=black!60, nodes near coords=$^{136}$Xe] table[x index=0,y index=3] \finnish; %Data3
\addplot[only marks, mark=square*,red,opacity=.61,mark size = 3pt,every mark/.append style={xshift=25pt}] table[x index=0,y index=4] \horoiFINAL;
%\addplot[only marks, mark=o,red,opacity=1,mark size = 3pt, every mark/.append style={xshift=25pt}] table[x index=0,y index=4] \horoiPrime;
\addplot[only marks, mark=triangle*,blue,opacity=.61,mark size = 3pt,every mark/.append style={xshift=25pt}] table[x index=0,y index=4] \finnish;
\addplot[only marks, mark=*,black!60!green,opacity=.61,mark size = 3pt,every mark/.append style={xshift=25pt}] table[x index=0,y index=4] \javier;

%\legend{  ,Horoi et al.\ \cite{Horoi:2017gmj},Horoi et al'.\ \cite{Horoi:2017gmj},Hyv\"arinen et\ al.\ \cite{Hyvarinen:2015bda}, Men\'endez et al.\ \cite{Javier}, Barea et al.\ \cite{Barea:2015kwa,Barea}}
\end{axis}
 \end{tikzpicture}
\caption{\small Continuation of the comparison between the NMEs of Refs.\ \cite{Hyvarinen:2015bda,Horoi:2017gmj,Javier,Barea:2015kwa,Barea}. Notation is the same as in Fig.\ \ref{FigNME}.}\label{FigNME2}
\end{figure}

\subsection{Matrix elements from different many-body methods}
\label{sec:MBmethods}

In Figs.~\ref{FigNME} and \ref{FigNME2}  we show results for 
the nine combinations of NMEs that determine the contribution of SM-EFT dimension-seven operators to \NLDBD, 
obtained by combining the results of Refs.\ \cite{Hyvarinen:2015bda} (blue triangles),
\cite{Horoi:2017gmj}  (red squares), \cite{Javier}  (green circles),  and \cite{Barea:2015kwa,Barea} (orange diamonds). 
The calculation of Ref.\ \cite{Hyvarinen:2015bda} is based on the quasiparticle random phase approximation (QRPA) method.
Refs.\ \cite{Horoi:2017gmj} and \cite{Javier} are shell model calculations. Refs. \cite{Barea:2015kwa,Barea} use the interacting boson model. 
Note that Refs. \cite{Hyvarinen:2015bda,Horoi:2017gmj,Javier} include short-range correlations
in various ways using CD-Bonn or AV-18 parameterizations. The choice of parameterization has a non-negligible effect for the $sd$ NMEs. In Table \ref{tab:NMEs}
we have used results using the CD-Bonn parameterization for \cite{Hyvarinen:2015bda,Horoi:2017gmj,Javier}.

In order to generate the results presented in Figs.~\ref{FigNME} and \ref{FigNME2} we made a few assumptions.
$M_{T6}$ and $M_{sd,\, 2}$ depend on the ratios of LEC $g^{\prime}_{T}/g_T$ and ${g_{27 \times 1}^{\pi\pi,\, \pi N}}/{g^{NN}_{27 \times 1}}$. In Fig.~\ref{FigNME}, we
assumed the unknown LECs to follow NDA,  $g^{\prime}_{T}= g_{27 \times 1}^{\pi N} = g_{27 \times 1}^{N N} = 1$, while $g_T$ and $g_{27 \times 1}^{\pi\pi}$ are given in Table \ref{Tab:LECs}.
Varying the size of $g^{\prime}_{T}$ has a limited effect on $M_{T6}$, while  $M_{sd, 2}$ is quite sensitive to the precise values of the LECs. We discuss this in more detail below.
In addition $M_{E,L}$, $M_{E,R}$, and $M_{m_e,L}$ depend on the matrix element $M^{AA}_{T}$, which is not evaluated in any of the references we use for the NMEs. Fortunately, this matrix element was computed 
in Ref. \cite{Muto:1989cd}, 
which found $M^{AA}_{T} = \{-0.92, -1.2, -0.86, -0.72\}$ for $^{76}$Ge, $^{82}$Se, $^{130}$Te  and $^{136}$Xe, respectively. 
For these  values of $M^{AA}_{T}$, the effect on the mentioned NMEs is mild. In addition, $M^{AA}_{T}$ mainly affects the limits on $C_{\rm VR}^{(6)}$, since the constraint on $C_{\rm VL}^{(6)}$ is dominated by  $M_M$. Nevertheless, it would be useful if $M^{AA}_{T}$ is included in future calculations. 

Figs. \ref{FigNME} and \ref{FigNME2} show that the nonstandard NMEs computed with different many-body methods differ by at most a factor of $2$-to-$3$. This level of agreement is similar to the one observed for the  light-neutrino-exchange mechanism \cite{Engel:2016xgb} -- see the spread in $M_{\nu}$ -- and leads to an uncertainty in the $0\nu\beta\beta$ rate of about one order of magnitude. 
The calculation of Ref.\ \cite{Horoi:2017gmj} yields values of  $M_{PS}$ which have very similar size, but opposite sign with respect to Refs.\ \cite{Hyvarinen:2015bda,Javier,Barea}.
The sign difference has no impact in the single-coupling scenario explored in Sec. \ref{Bounds}. 
It will affect scenarios in which several operators are turned on at the same time, but  in this case the effect is mitigated by the ignorance of the relative phase between the coefficients.
A similar argument applies to $\mathcal M_{m_e,L}$ and $\mathcal M_{m_e,R}$, which, using the results of Ref.\ \cite{Horoi:2017gmj} are found to have similar size, but different sign with respect to the other calculations.  
The uncertainty on the short-distance NME $M_{sd,2}$ is somewhat larger than for the other NMEs. This is not unexpected as such matrix elements depend on short-distance details of nuclear wave functions which are more model dependent then long-range aspects.
The relative sizes of the NME combination $M_{sd, 2}$ for various isotopes vary strongly between Refs. \cite{Hyvarinen:2015bda,Javier,Horoi:2017gmj}. Although we do not understand this behaviour in detail, it might be related to possible accidental cancellations between the various contributions to $M_{sd, 2}$.
In the next section we explore the consequences of these uncertainties on the constraints on the scale of BSM lepton-number-violating physics.

It is possible to further reduce the  set of relevant NMEs. $M_{T6}$ depends in principle on a linear combination of
$M_{F,sd}$ and $M_{GT}^{MM}+M_{T}^{MM}$, but the latter numerically dominates due to the large nucleon isovector magnetic moment. As such, the NME combinations $M_{T6}$ and $M_M$ are related by $M_{T6}/M_M \simeq -4/g_A$. This relation holds up to $\mathcal O(10\%)$ corrections for all sets of NMEs. Finally, the NME combination $M_{E,\{L,R\}}$ and $M_{m_e,\{L,R\}}$ only appear for the dimension-six vector operators $C_{\rm VL,VR}^{(6)}$. However, the contributions to the \NLDBD\ rate from $M_{m_e,\{L,R\}}$ are numerically suppressed with respect to those from $M_{E,\{L,R\}}$. This suppression can be partially understood from phase space factors as the electron mass is small with respect to the typical $Q$ value (compare $2 G_{02}$ to $G_{04}$ in Table~\ref{Tab:phasespace}). The above considerations imply that seven combinations of NMEs dominate \NLDBD\ in the SM-EFT.

\section{Single-coupling constraints}\label{Bounds}

\begin{table}
$
\renewcommand{\arraystretch}{1.5}\footnotesize
\begin{array}{c||cccc}
& \text{}^{76} \text{Ge} & \text{}^{82} \text{Se} & \text{}^{130} \text{Te} & \text{}^{136} \text{Xe} \\\hline
 m_{\beta\beta}(\mathrm{eV}) & 0.17 & 1.6 & 0.32 & 0.084
\\ 
 \hline
 C_{\text{SL}}^{(6)}& 270 & 130 & 220 & 350 \\
 C_{\text{SR}}^{(6)}& 270 & 130 & 220 & 350 \\
 C_{\text T}^{(6)}  & 240 & 110 & 200 & 300 \\
  C_{\text{VL}}^{(6)}& 180 & 83 & 150 & 220 \\
 C_{\text{VR}}^{(6)} & 33 & 17 & 29 & 44 \\
 C_{\text{VL}}^{(7)}& 8.1 & 3.8 & 6.8 & 11 \\
 C_{\text{VR}}^{(7)}& 8.1 & 3.8 & 6.8 & 11 \\
 C_1^{(9)}& 13 & 6.3 & 10 & 13 \\
 C_4^{(9)}& 43 & 21 & 38 & 55 \\
 C_5^{(9)} & 66 & 31 & 58 & 85 \\
\end{array}$
$\renewcommand{\arraystretch}{1.5}\footnotesize
\begin{array}{c||cccc}
 & \text{}^{76} \text{Ge} & \text{}^{82} \text{Se} & \text{}^{130} \text{Te} & \text{}^{136} \text{Xe}   \\\hline 
&0.19 & 1.4 & 0.49 & 0.1 \\
 \hline
 & 210 & 110 & 150 & 260 \\
 & 210 & 110 & 150 & 260 \\
 & 190 & 99 & 150 & 250 \\
 & 150 & 74 & 110 & 190 \\
 & 26 & 15 & 20 & 34 \\
 & 6.4 & 3.3 & 4.6 & 7.8 \\
 & 6.4 & 3.3 & 4.6 & 7.8 \\
 & 11 & 5.5 & 8.3 & 14 \\
 & 32 & 17 & 24 & 42 \\
& 50 & 26 & 37 & 64 \\
\end{array}$  $
\renewcommand{\arraystretch}{1.5}\footnotesize
\begin{array}{c||cccc}
& \text{}^{76} \text{Ge} & \text{}^{82} \text{Se} & \text{}^{130} \text{Te} & \text{}^{136} \text{Xe} \\\hline
& 0.3 & 2.2 & 0.45 & 0.1 \\
 \hline
& 200 & 100 & 180 & 290 \\
& 200 & 100 & 180 & 290 \\
 & 180 & 94 & 170 & 270 \\
& 140 & 70 & 120 & 200 \\
& 26 & 15 & 24 & 39 \\
& 6 & 3.2 & 5.4 & 8.9 \\
& 6 & 3.2 & 5.4 & 8.9 \\
& 10 & 5.4 & 9.7 & 16 \\
& 32 & 16 & 28 & 45 \\
& 49 & 24 & 44 & 70 \\
\end{array}$
\caption{
The table shows the upper limits on $|m_{\bt\bt}|$ 
and lower limits on the scales, $\Lambda_i$, related to the dimension-six, -seven, and -nine operators from the GERDA \cite{Agostini:2016iid}, NEMO \cite{Arnold:2015wpy,Arnold:2016ezh}, CUORE \cite{Alfonso:2015wka}, and KamLAND-Zen \cite{KamLAND-Zen:2016pfg} experiments, assuming $C_i (\mu =2 \hbox{ GeV})= v^3/\Lambda_i^3$. The left, middle, and right tables correspond to the matrix elements of Refs.~\cite{Hyvarinen:2015bda}, \cite{Horoi:2017gmj}, and \cite{Javier}, respectively. The lower limits on $\Lambda$ are shown in units of TeV. 
}\label{tab:limits}
\end{table}

In this section we discuss the constraints on the low-energy operators, as well as the fundamental dimension-seven operators that arise at the scale $\Lambda$. We start by considering the bounds from $0\nu\bt\bt$ experiments and discuss other relevant observables in Sect.~\ref{Others}. Throughout this section we will assume that only one operator is present at a time. We study scenarios involving multiple couplings in Sect.~\ref{two-coupling-analysis}. We apply the following experimental limits \cite{Agostini:2017iyd,KamLAND-Zen:2016pfg,Alfonso:2015wka,SeNEMO} (all at $90\%$ c.l.)
\begin{eqnarray}
&&\,T^{0\nu}_{1/2}({}^{76}\mathrm{Ge}) > 5.3\cdot10^{25}\,\mathrm{yr}\,,\qquad \,T^{0\nu}_{1/2}({}^{82}\mathrm{Se}) > 2.5\cdot10^{23}\,\mathrm{yr}\,, \nonumber\\
&&T^{0\nu}_{1/2}({}^{130}\mathrm{Te}) > 4.0\cdot10^{24}\,\mathrm{yr}\,,\qquad T^{0\nu}_{1/2}({}^{136}\mathrm{Xe}) > 1.1\cdot10^{26}\,\mathrm{yr}\, .
\end{eqnarray}

By inserting the phase-space factors of Table~\ref{Tab:phasespace} and the NMEs in Table~\ref{tab:comparison} into Eq.~\eqref{eq:T1/2}, we obtain limits on the coefficients of the $\Delta L=2$ operators.
In Table~\ref{tab:limits} we show bounds on  $m_{\bt\bt}$ and the low-energy dimension-six, -seven, and -nine operators of Eq.\ \eqref{lowenergy9}, which were 
derived using the NMEs of Refs.\ \cite{Hyvarinen:2015bda}, \cite{Horoi:2017gmj},  and \cite{Javier} in the left, middle, and right panels, respectively. 

Using NMEs from Ref.~\cite{Hyvarinen:2015bda} we find an upper bound  $m_{\beta\beta} < 0.084$ eV, and slightly weaker bounds for the other NMEs. The limits we obtain are in agreement with, for example, Ref.~\cite{Horoi:2017gmj}. All bounds are somewhat weaker than the most stringent bound reported in Ref.~\cite{KamLAND-Zen:2016pfg}, $m_{\beta\beta} < 0.061$ eV which is based on different NMEs than considered here.

For the non-standard operators, Table \ref{tab:limits} shows the constraints on the scale of new physics, $\Lambda$, assuming that $C_i (\mu =2 \hbox{ GeV})= v^3/\Lambda^3$ and only one coupling is turned on at a time. In addition, we assumed natural values for the unknown LECs, $g_T'=g_{27\times 1}^{\pi N}=g_{27\times 1}^{N N}=1$.
As expected from the discussion of the previous section, the most stringent constraints arise in the case of $C_{\rm SL,SR}^{(6)}$, reaching scales of $\Or(100\, {\rm TeV })$.
Although the power counting of Table \ref{TabPC} would predict the limit on $C_{\rm T,VL}^{(6)}$ to be weaker by $\epsilon_\chi^{2/3}$, the actual constraints are somewhat stronger than expected due to the large isovector magnetic moment. For most of the remaining couplings the limits closely follow what one would expect from the power counting.
For example, the limits on $C_{\rm VL,VR}^{(7)}$ and $C_{\rm VR}^{(6)}$ 
are weaker than the limits on $C_{\rm SL,SR}^{(6)}$ by  factors of $(\frac{\Lambda_\chi}{v}\epsilon_\chi\sq)^{1/3}\simeq 0.05$ and $\epsilon_\chi \simeq 0.15$, respectively, which agrees with Table~\ref{TabPC}. Finally, we would expect the limits on $C_{\rm VR}^{(6)}$ to be weaker than the limit on $C_{\rm VL}^{(6)}$ by roughly a factor $(\epsilon_\chi/(1+\kappa_1)^2)^{1/3}\simeq 0.2$ which agrees fairly well with the actual results. Here we took into account by hand the large nucleon magnetic moment.

\begin{figure}
\begin{center}
\includegraphics[width=10cm]{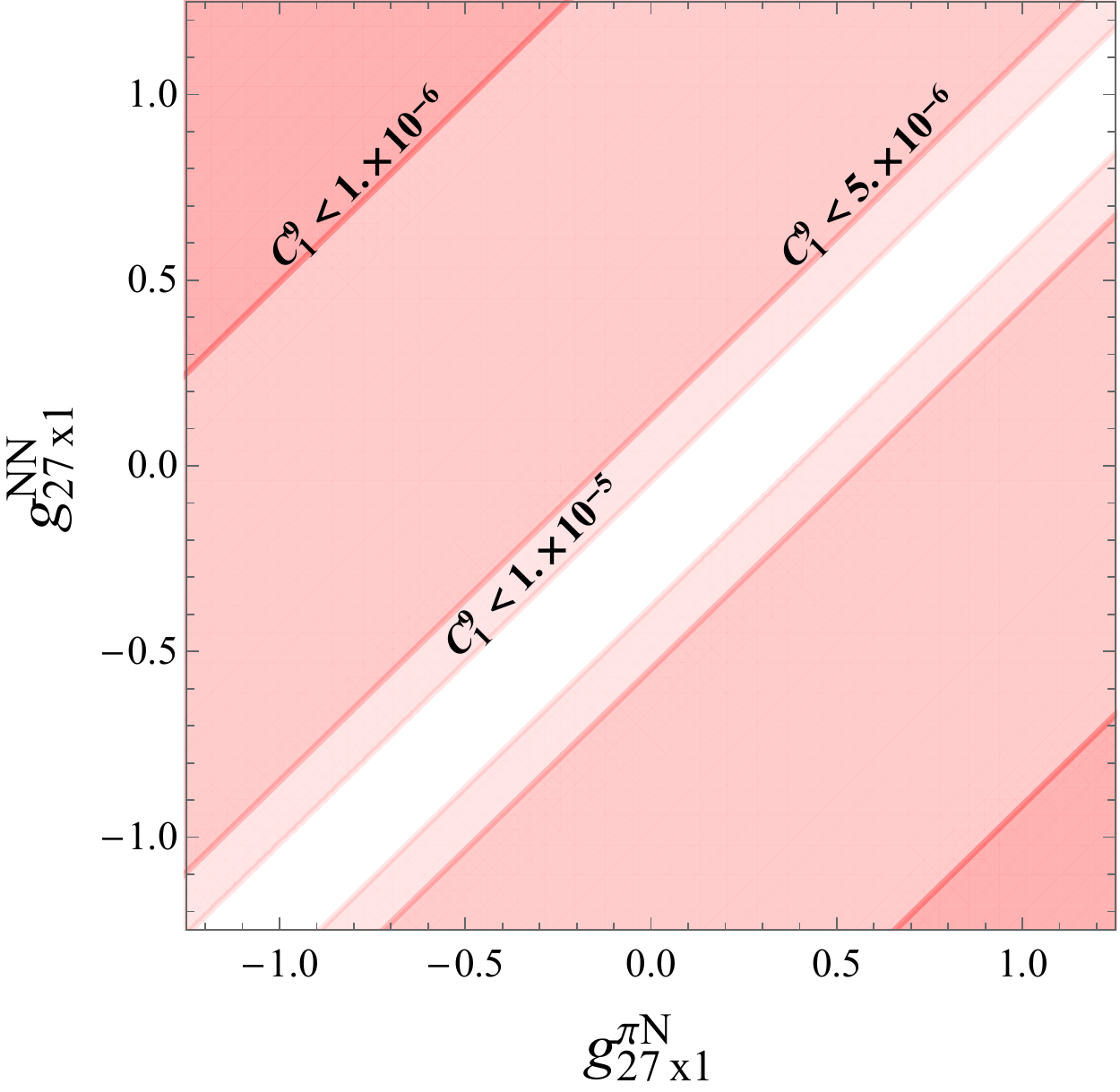}
\end{center}
\caption{Constraints on the coupling $C_1^{(9)}(\mu=2\,{\rm GeV})$ as a function of the unknown LECs $g_{27\times 1}^{\pi N}$ and $g_{27\times 1}^{N N}$. Here we show the constraints derived using the NMEs of Ref.\ \cite{Hyvarinen:2015bda} and the experimental limit on the half-life of $^{136}$Xe \cite{KamLAND-Zen:2016pfg}.
}\label{fig:lecPlot}
\end{figure}

The case of $C_{1}^{(9)}$ requires additional explanation. From the power counting we would expect this coupling to contribute at the same order as $C_{\rm VL,VR}^{(7)}$. 
 However, the  matrix element $M_{sd,\,2}$ receives several contributions proportional to unknown LECs, $g_{27\times 1}^{\pi N}$ and $g_{27\times 1}^{NN}$.
As a result, the contribution of $C_{1}^{(9)}$ can vary substantially depending on the values and signs of these LECs. 
This is illustrated in Fig.\ \ref{fig:lecPlot} where we show the constraint on $C_{1}^{(9)}$ as a function of $g_{27\times 1}^{\pi N}$ and $g_{27\times 1}^{NN}$. By varying the LECs in a natural range, the bound on $C_{1}^{(9)}$ can decrease or increase by a factor of $\Or(10)$. In fact, there exists a small, fine-tuned, region where the limit on $C_{1}^{(9)}$ disappears. Although such a near-exact cancellation is not expected, and is sensitive to higher-order corrections,
the limits on the scale $\Lambda$ for $C_1^{(9)}$ appearing in Table~\ref{tab:limits} should be taken as an order-of-magnitude estimate, at least until the values of $g_{27\times 1}^{\pi N, NN}$ are further constrained.
 In contrast, varying the sign of the only other unknown LEC, $g_T'$, only leads to $\Or(10\%)$ effects in the limits on $\Lambda$ for $C_{\rm T}^{(6)}$.

\begin{figure}[h!]
\pgfplotstableread[row sep=\\,col sep=&]{
interval & horoi & finnish &Javier\\
    c1    & 16& 13 &17 \\
    c2   & 160 & 200 &180 \\
    c3    &26 & 20 &28\\
    c4    &72 & 95& 78\\
    c5    &250 & 320 &270\\
    c6    &130 & 150 & 140\\
    c7    &300 & 410 & 340\\
    c8    &30 & 39 &35\\
    }\mydata
\begin{tikzpicture}
    \begin{axis}[
            ybar,ymode=log,
           log origin=infty    ,         
            bar width=.45cm, 
            width=\textwidth,
            height=.25\textwidth,
            legend style={at={(0.5,1.40)},
                anchor=north,legend columns=-1},
           symbolic  x coords={c1,c2,c3,c4,c5,c6,c7,c8},
                                 enlarge x limits=.1,
            xtick=data,
            xticklabel style={align=center,text width=1cm, font=\small},
            xticklabels={$\mathcal C_{LHD}^{(1)}$,$\mathcal C_{LHDe}$,$\mathcal C_{LHW}$,$\mathcal C_{LL\bar d uD}^{(1)}$,$\mathcal C_{LLQ\bar d H}^{(1)}$,$\mathcal C_{LLQ\bar d H}^{(2)}$,$\mathcal C_{LL\bar Q u H}$,$\mathcal C_{Leu\bar dH}$},
            tick label style={font=\small},
            ytick={10,100,1000},
            ylabel={$\Lambda\,$ (TeV)} , 
       nodes near coords/.append style={font=\small},                    
              ymin=10,ymax=1000,
                          ytick pos=left,
            nodes near coords align={vertical},            
            point meta=rawy  ]			
        \addplot +[black,fill=blue, fill opacity=.51, nodes near coords={\pgfmathprintnumber[fixed,fixed zerofill,fixed relative, precision=2]{\pgfplotspointmeta}}]  table[x=interval,y=finnish]{\mydata};
         \addplot   +[black,fill=red, fill opacity=.51,nodes near coords={\pgfmathprintnumber[fixed,fixed zerofill, fixed relative, precision=2]{\pgfplotspointmeta}}]  table[x=interval,y=horoi]{\mydata};
          \addplot   +[black,fill=black!60!green, fill opacity=.51,nodes near coords={\pgfmathprintnumber[fixed,fixed zerofill, fixed relative, precision=2]{\pgfplotspointmeta}}]  table[x=interval,y=Javier]{\mydata};
       \legend{Hyv\"arinen et\ al.\ \cite{Hyvarinen:2015bda}\,\,\,\,\,\,\,\,,Horoi et al.\ \cite{Horoi:2017gmj}\,\,\,\,\,\,\,\,, Men\'endez et al.\ \cite{Javier}};
    \end{axis}
\end{tikzpicture}
\
\caption{ Constraints from the KamLAND-Zen experiment \cite{KamLAND-Zen:2016pfg} on the scale of the dimension-seven operators. We assume $\mathcal C_i(\mu=\Lambda) = 1/\Lambda^3$ and only turn on one operator at a time.}
\label{fig:limitsChart}
\end{figure}
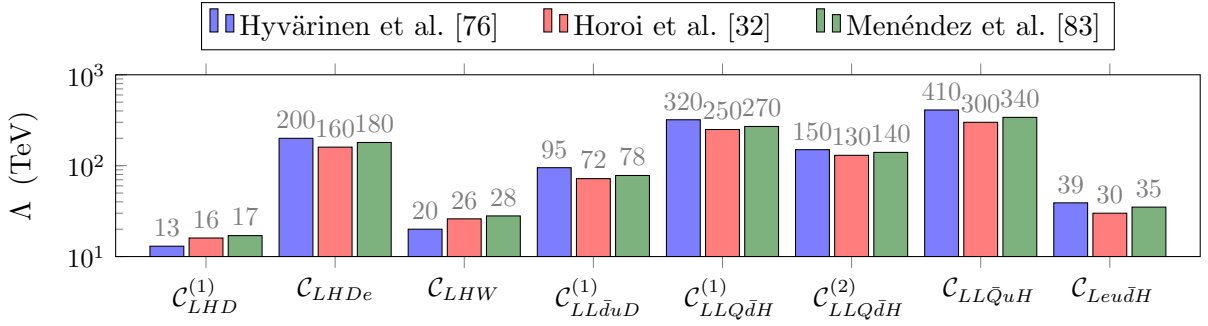

Although the above constraints are useful to test the power counting, the fundamental $\Delta L=2$ operators of interest are the dimension-seven operators of Table \ref{dim7}. We present the limits on these couplings in Table \ref{tab:limits2}, where the left, middle, and right panels again employ the NMEs of \cite{Hyvarinen:2015bda}, \cite{Horoi:2017gmj},  and \cite{Javier}, respectively. The bounds on the scale of new physics are obtained by assuming a single coupling is present at the high scale, and $\mathcal{C}_i(\mu = \Lambda)=1/\Lambda^3$. 
The strongest limits are derived in the case of $\mathcal C_{LLQ\bar d H}^{(1)}$ and $\mathcal C_{LL\bar Q u H}$ because  these operators mainly induce the stringently constrained $C_{\rm SL,SR}^{(6)}$.
Instead, the weakest limits are obtained in cases where only the low-energy dimension-seven and -nine operators are induced. This is the case, for example, for $\mathcal C_{LHD}^{(1)}$ and $\mathcal C_{LHW}$, which both mainly contribute to $C_{\rm VL}^{(7)}$ and $C_{1}^{(9)}$. 
Since these operators induce $C_1^{(9)}$, the corresponding limits are sensitive to the  values of the unknown LECs, $g_{27\times 1}^{\pi N, NN}$. In Fig.~\ref{fig:limitsChart} we present the same information in a different format, focusing on the bounds on the dimension--7 operators arising from the KamLAND-Zen experiment \cite{KamLAND-Zen:2016pfg}.

It should be noted that the Wilson coefficients will in general depend on a dimensionless coupling, $c_i$, in addition the scale $\Lambda$, i.e.\ $\mathcal C_i=c_i/\Lambda^3$.
The presence of these $c_i$ implies that the limits on $\Lambda$ in Table \ref{tab:limits2} (where we assumed $c_i=1$) do not necessarily correspond to constraints on particle masses in any given BSM theory. In particular, in weakly coupled BSM theories, $c_i< 1$, the limits on the masses of particles could be significantly weaker than those on $\Lambda$ given in Table \ref{tab:limits2}. Thus,  the stringent bounds on $\Lambda$ derived above do not necessarily imply that the responsible BSM physics is out of reach of collider searches.
Apart from a simple rescaling of the limits in Fig.~\ref{fig:limitsChart}, dimensionless couplings, $c_i\neq 1$, would change the starting point of the RGEs. However, the numerical impact of such a change in $\Lambda$ is rather minimal. For example, changing the starting point of the RG from $\Lambda= 50$ TeV to $\Lambda= 100$ TeV, changes the running of the $\mathcal C_i$ by no more than $10\%$.

An alternative way to present the limits is shown in Table \ref{tab:limits3}, where we show the bounds on the dimensionless couplings, $c_i = \Lambda^3 \mathcal C_i(\Lambda)$. Here we picked the scale $\Lambda$ to be $10$ TeV, and derived constraints using  several calculations for the NMEs \cite{Hyvarinen:2015bda,Horoi:2017gmj,Javier,Barea:2015kwa,Barea}. The bounds in Table \ref{tab:limits3} are inversely proportional to these NMEs, $c_i\propto M_i^{-1}$, while the limits on the scales have a much weaker dependence, $\Lambda\propto M_i^{1/3}$. As a result, the variation between different nuclear calculations is more pronounced in Table \ref{tab:limits3} than in Table \ref{tab:limits2}.
\begin{table}
$
\renewcommand{\arraystretch}{1.5}\footnotesize
\begin{array}{c||cccc}
& \text{}^{76} \text{Ge} & \text{}^{82} \text{Se} & \text{}^{130} \text{Te} & \text{}^{136} \text{Xe} \\\hline
 \mathcal{C}_{{LHD}}^{\text{(1)}}& 15 & 6.9 & 11 & 13 \\
 \mathcal{C}_{{LHDe}}  & 160 & 73 & 130 & 200 \\
 \mathcal{C}_{{LHW}}  & 23 & 11 & 17 & 20 \\
 \mathcal{C}_{{LLduD}}^{\text{(1)}} & 74 & 35 & 65 & 95 \\
 \mathcal{C}_{{LLQdH}}^{\text{(1)}}  & 240 & 110 & 200 & 320 \\
 \mathcal{C}_{{LLQdH}}^{\text{(2)}} & 120 & 58 & 100 & 150 \\
 \mathcal{C}_{{LLQuH}}  & 310 & 150 & 260 & 410 \\
 \mathcal{C}_{{Leu\bar dH}} & 29 & 15 & 26 & 39 \\
\end{array}$ 
$\renewcommand{\arraystretch}{1.5}\footnotesize
\begin{array}{c||cccc}
 & \text{}^{76} \text{Ge} & \text{}^{82} \text{Se} & \text{}^{130} \text{Te} & \text{}^{136} \text{Xe}   \\\hline 
 & 13 & 6.6 & 9.9& 16\\
& 130 & 65 & 98 & 160 \\
& 20 & 11 & 16 & 26 \\
 & 56 & 29 & 42 & 72 \\
& 200 & 100 & 140& 250\\
 & 99 & 51 & 77 & 130 \\
& 250 & 130 & 180 & 300 \\
& 24 & 14 & 18 & 30 \\
\end{array}$
$
\renewcommand{\arraystretch}{1.5}\footnotesize
\begin{array}{c||cccc}
& \text{}^{76} \text{Ge} & \text{}^{82} \text{Se} & \text{}^{130} \text{Te} & \text{}^{136} \text{Xe} \\\hline
 & 12 & 5.9 & 11 & 17 \\
 & 120 & 61 & 110 & 180 \\
 & 18 & 9.4 & 17 & 28 \\
& 54 & 27 & 49 & 78 \\
 & 180 & 93 & 160 & 270 \\
 & 94 & 48 & 85 & 140 \\
& 230 & 120 & 210 & 340 \\
 & 23 & 13 & 22 & 35 \\
\end{array}$
\caption{The table shows the lower limits on the scale  of the dimension-seven couplings, from the GERDA \cite{Agostini:2016iid}, NEMO \cite{Arnold:2015wpy,Arnold:2016ezh}, CUORE \cite{Alfonso:2015wka}, and KamLAND-Zen \cite{KamLAND-Zen:2016pfg} experiments, assuming $\mathcal C_i(\mu=\Lambda) = 1/\Lambda^3$. The left, middle, and right tables correspond to the matrix elements of Refs.~\cite{Hyvarinen:2015bda}, \cite{Horoi:2017gmj}, and \cite{Javier}, respectively. The limits on $\Lambda$ are shown in units of TeV. }\label{tab:limits2}
\end{table}

\begin{table}
$
\renewcommand{\arraystretch}{1.5}\footnotesize
\begin{array}{c||cccc}
 \text{}^{76} \text{Ge} &$ \cite{Hyvarinen:2015bda}$&$ \cite{Horoi:2017gmj}$&$ \cite{Javier}$ &$\cite{Barea:2015kwa,Barea}$\\\hline
 \mathcal{C}_{{LHD}}^{\text{(1)}} & 3.3\times 10^{-1} & 4.7\times 10^{-1} & 6.5\times 10^{-1} & 2.1\times 10^{-1} \\
 \mathcal{C}_{{LHDe}} & 2.6\times 10^{-4} & 4.7\times 10^{-4} & 5.5\times 10^{-4} & 9.0\times 10^{-4} \\
 \mathcal{C}_{{LHW}} & 8.2\times 10^{-2} & 1.2\times 10^{-1} & 1.6\times 10^{-1} & 5.3\times 10^{-2} \\
 \mathcal{C}_{{LLduD}}^{\text{(1)}} & 2.4\times 10^{-3} & 5.8\times 10^{-3} & 6.2\times 10^{-3} & 5.2\times 10^{-3} \\
 \mathcal{C}_{{LLQdH}}^{\text{(1)}} & 8.1\times 10^{-5} & 1.4\times 10^{-4} & 2.0\times 10^{-4} & 1.2\times 10^{-4} \\
 \mathcal{C}_{{LLQdH}}^{\text{(2)}} & 5.4\times 10^{-4} & 1.0\times 10^{-3} & 1.2\times 10^{-3} & 2.4\times 10^{-3} \\
 \mathcal{C}_{{LLQuH}} & 3.8\times 10^{-5} & 7.6\times 10^{-5} & 9.0\times 10^{-5} & 5.6\times 10^{-5} \\
 \mathcal{C}_{{LeudH}} & 4.0\times 10^{-2} & 7.9\times 10^{-2} & 7.7\times 10^{-2} & 6.8\times 10^{-2} \\
\end{array}$
$
\renewcommand{\arraystretch}{1.5}\footnotesize
\begin{array}{c||cccc}
 \text{}^{136} \text{Xe} &$ \cite{Hyvarinen:2015bda}$&$ \cite{Horoi:2017gmj}$&$ \cite{Javier}$\\\hline
&  4.9\times 10^{-1} & 2.3\times 10^{-1} & 1.9\times 10^{-1} \\
& 1.3\times 10^{-4} & 2.3\times 10^{-4} & 1.8\times 10^{-4} \\
& 1.2\times 10^{-1} & 5.8\times 10^{-2} & 4.7\times 10^{-2} \\
& 1.2\times 10^{-3} & 2.7\times 10^{-3} & 2.1\times 10^{-3} \\
& 3.7\times 10^{-5} & 7.5\times 10^{-5} & 6.2\times 10^{-5} \\
& 2.6\times 10^{-4} & 4.5\times 10^{-4} & 3.6\times 10^{-4} \\
& 1.7\times 10^{-5} & 4.1\times 10^{-5} & 2.8\times 10^{-5} \\
& 1.7\times 10^{-2} & 3.5\times 10^{-2} & 2.3\times 10^{-2} \\
\end{array}$
\caption{The table shows the limits on the dimensionless couplings, $ c_i$, of the dimension-seven operators, from the GERDA \cite{Agostini:2016iid} and KamLAND-Zen \cite{KamLAND-Zen:2016pfg} experiments. Here we assume $ c_i(\mu=\Lambda) = \mathcal C_i(\Lambda) \,\Lambda^3$ and choose the scale of BSM physics to be  $\Lambda = 10$ TeV. The columns from left to right, correspond to the matrix elements of Refs.~\cite{Hyvarinen:2015bda}, \cite{Horoi:2017gmj}, \cite{Javier}, and, in the case of $^{76}$Ge, \cite{Barea:2015kwa,Barea}, respectively.  }\label{tab:limits3}
\end{table}

\subsection{Other constraints}\label{Others}

Although  $0\nu\bt\bt$ leads to stringent constraints on the $\mathcal C_i$ couplings, reaching scales of $\Or(100\, {\rm TeV})$, it is interesting to see how these compare to constraints from other probes.
In particular, all operators in Table~\ref{dim7} induce radiative corrections to the neutrino masses. In Sec.~\ref{ref:neutrino-masses} we therefore discuss the naturalness 
bounds that can extracted from the neutrino masses. We find that for several operators they are stronger than the bounds from \NLDBD. 

Considering additional probes is particularly important for the operators $\mathcal C_{LHB}$ and $\mathcal C_{LL\bar e H}$, which do not induce \NLDBD\ at tree level, 
and $\mathcal C^{(2)}_{LHD}$, whose contribution to \NLDBD\ is suppressed by the electron energy, and was not considered in Secs.~\ref{sec:2btOperators} and \ref{Master}. 
We address the contributions of these operators to the neutrino masses in Sec. \ref{ref:neutrino-masses}, and take into account bounds from the neutrino transition magnetic moments in Sec. \ref{ref:magnetic-moments},
and from non-standard muon decays in Sec. \ref{ref:muon-decay}.

\subsubsection{Neutrino mass}
\label{ref:neutrino-masses}
The operators in Table~\ref{dim7} can generate neutrino masses. The tree-level contribution is
\begin{equation}\label{eq:nuMass}
(\dt m_{\nu})_{ij} = -  \frac{v}{2} \, (v^3 \mathcal C_{LH, ij})\,.
\end{equation}
The other $\mathcal C_i$ do no contribute at tree level, but can  contribute to $\mathcal C_{LH}$ through RG effects between $\mu=\Lambda$ and $\mu = m_W$.
The complete neutrino mass is  a combination of the contributions of the dimension-seven operators and the Weinberg operator. In total we have $m_\nu = m_\nu^{(0)} +\dt m_{\nu} $, where $m_\nu^{(0)}$ is the contribution from the Weinberg operator. Since $m_\nu^{(0)}$ is unknown we can only set constraints if we assume that the dimension-five and -seven contributions are not unnaturally large compared to the total neutrino mass. That is, we assume there is no large cancellation between $m_\nu^{(0)}$ and $\dt m_\nu$. To get an idea of these naturalness limits we will, somewhat arbitrarily, impose $|\dt m_{\nu}| \lesssim 1$ eV.

From Eq.~\eqref{eq:nuMass}, we can already estimate the constraint on $\mathcal C_{LH}$. Assuming $\mathcal C_{LH}(\mu=\Lambda)=1/\Lambda^3$, we get $\Lambda> 1200$ TeV. For the other dimension-seven operators that contribute at loop level, we require the evolution between $\mu=\Lambda$ and $\mu = m_W$. The relevant one-loop RGE is given by
\bea
\frac{d \mathcal C_{LH}}{d\ln\mu} &=& \frac{1}{(4\pi)\sq}\bigg[6 g^{ 4}\,\mathcal C_{LHW}-\frac{3}{2}g^4 \mathcal C_{LHD}^{(1)}-\frac{3}{4}(3
g^4+2g\sq g^{\prime\, 2}+g^{\prime\, 4})\mathcal C_{LHD}^{(2)}+3\sqrt{2}\frac{m_e}{v}g\sq\, i\, \mathcal C_{LHDe}\nn\\
&&+4\sqrt{2}N_C\left(\frac{m_d}{v}\right)^3 \mathcal C_{LLQ\bar d H}^{(1)}
-8\sqrt{2}N_C\left(\frac{m_u}{v}\right)^3 \mathcal C_{LL\bar Qu H}
+8\sqrt{2}\left(\frac{m_e}{v}\right)^3 \mathcal C_{LL\bar eH}\bigg]\,\, .
\label{eq:nuMassRG}\eea
The above expression provides us with $\mathcal C_{LH}(\mu=m_W)$, which together with Eq.\ \eqref{eq:nuMass} and  $|\dt m_{\nu}| \lesssim 1$ eV,  leads to the  constraints
\bea\label{eq:nuMassLimits}
\mathcal C_{LHD}^{(1)}:&\quad \Lambda > 280 \, {\rm TeV}\,,\qquad\mathcal C_{LHD}^{(2)}&:\quad \Lambda > 350 \, {\rm TeV}\,,\nn\\\mathcal C_{LHDe}:&\quad \Lambda > 6 \, {\rm TeV}\,,\qquad \mathcal \quad C_{LHW}&:\quad \Lambda > 460 \, {\rm TeV}\,,
\eea
where we again assumed $\mathcal C_i = 1/\Lambda^3$. Contributions of the operators appearing in the second line of Eq.\ \eqref{eq:nuMassRG} are  severely suppressed by three powers of small Yukawa couplings. The corresponding limits are well below the electroweak scale such that we do not obtain sensible constraints.

Here we only considered contributions to the neutrino masses through corrections to the dimension-seven coupling $\mathcal C_{LH}$. In principle, one could consider corrections directly to the dimension-five coupling, $\mathcal C^{(5)}$, in Eq.\ \eqref{eq:1stLag} as well. Below the scale $\Lambda$, the $SU(2)$-invariant dimension-seven operators do not mix with this dimension-five operator. However, assuming the dimension-five term is not protected by symmetry considerations, one might expect  the BSM interactions that induce the $\mathcal C_i$ appearing in Eq.\ \eqref{eq:nuMassRG}, to contribute to  $\mathcal C^{(5)}$ as well. These contributions would result from matching the BSM theory to the EFT and, if they arise from loop diagrams, could in principle scale as $\mathcal C^{(5)}\sim  \frac{1}{(4\pi)^2} \frac{1}{\Lambda}$, in which case they would dominate over those in Eq.\ \eqref{eq:nuMassRG} by a factor of $\Lambda^2/v^2$. 
Such contributions would lead to more stringent limits than those in Eq.\ \eqref{eq:nuMassLimits}.   On the other hand, it is possible to realize smaller contributions to the neutrino masses than those induced by Eq.\ \eqref{eq:nuMassRG} if there is a fine-tuned cancellation  at work.
Which of these scenarios is realized, as well as the mentioned matching contributions, depend strongly on the specific BSM theory above the scale $\Lambda$.
Here we refrain from estimating such model-dependent effects and only consider the terms that are calculable within the EFT framework. Nevertheless, one should keep in mind that specific BSM theories could give larger contributions to the neutrino masses than those captured by Eq.\ \eqref{eq:nuMassRG}.

It is certainly possible to avoid the above naturalness limits by allowing for some amount of fine-tuning between, for example, dimension-five and -seven contributions to the neutrino mass. Nevertheless, taken at face value, the contributions to $\dt m_\nu$ can lead to very stringent constraints. This is certainly true for $\mathcal C_{LH}$ and $\mathcal C_{LHD}^{(2)}$, for which the limits reach $\Or(100 \,{\rm TeV})$ or more, while these couplings would be left unconstrained by $0\nu\bt\bt$. Note that these naturalness limits even exceed the $0\nu\bt\bt$ constraints for $\mathcal C_{LHW}$ and $\mathcal C_{LHD}^{(1)}$, while $0\nu\bt\bt$ is  more constraining for $\mathcal C_{LHDe}$ (as well as for $\mathcal C_{LL\bar QuH}$ and $\mathcal C_{LL Q\bar d H}^{(1)}$).

Of the remaining operators,  $\mathcal C_{LHB}$ does not contribute at one loop as it is anti-symmetric in flavor space, while $\mathcal C^{(2)}_{LLQ\bar d H}$, $\mathcal C_{LL\bar d u D}$, and $\mathcal C_{Leu \bar d H}$, mix with $\mathcal C_{LH}$ at two loops and require, respectively, one, two, and three Yukawa insertions. The $0\nu\bt\bt$ limits are more stringent in these cases, and we do not consider the contributions to $\dt m_\nu$.

\subsubsection{Magnetic moments}
\label{ref:magnetic-moments}

Apart from neutrino masses, the operators in  Table \ref{dim7} also induce contributions to the magnetic moment of the neutrinos. These magnetic moments can be constrained by neutrino-electron scattering in solar and reactor experiments \cite{Bell:2006wi,Giunti:2014ixa,Canas:2015yoa}, or through astrophysical limits from globular clusters \cite{Raffelt:1998xu}. As we are mainly interested in an order-of-magnitude estimate, here we will employ the limits of Ref.\ \cite{Canas:2015yoa} from the scattering of solar neutrinos.

Tree-level contributions of the dimension-seven operators to the magnetic moments are
\bea
\mu_{i j} =  \frac{1}{2 v} \, \left(v^3 \mathcal C_{LHB,ij} -v^3 \frac{\mathcal C_{LHW,ij}-\mathcal C_{LHW,ji}}{2}\right)\,,
\eea
where $\mu$ and $\mathcal C_{LHB}$ are anti-symmetric in flavor space. Following the notation of Ref.~\cite{Canas:2015yoa}, the transition magnetic moments can be parametrized by three complex parameters, $\Lambda_i$, as follows,
\bea
\left(U^T_{}\mu U_{}\right)_{ij} = -\frac{1}{4e}\epsilon_{ijk}\Lambda_k\,\,,
\eea
where the PMNS matrix, $U$, appears due to the rotation to the mass basis. The constraints  derived in Ref.\ \cite{Canas:2015yoa} are
\bea
|\Lambda_1|\leq 5.6\cdot 10^{-11} \,\mu_B\,,\qquad |\Lambda_2|\leq 4.0\cdot10^{-11} \,\mu_B\,,\qquad |\Lambda_3|\leq 3.1\cdot10^{-11} \,\mu_B\,\,.
\eea
In principle, a detailed analysis should take into account the flavor structure of $\mathcal C_{LHB,LHW}$ as well as the unknown phases in $U$. As we are mainly interested the order-of-magnitude of the limits, we take the following estimate
\bea
|\mathcal C_{LHB}-\mathcal C_{LHW}|\lesssim \frac{1}{4m_e v\sq} 10^{-10}\rightarrow \Lambda > 11 \,{\rm TeV}\,.
\eea
For $\mathcal C_{LHW}$ this limit is weaker than both the limit from $0\nu\bt\bt$ as well as the naturalness constraint from the neutrino mass. However, the neutrino magnetic moments do provide the most stringent limit on $\mathcal C_{LHB}$, whose contributions to $0\nu\bt\bt$ and the neutrino mass are suppressed.

\subsubsection{Muon decay}
\label{ref:muon-decay}

The operator $\mathcal O_{LL\bar e H}$ does not contribute to \NLDBD\ at tree level,
and its contribution to the neutrino mass in Eq.~\eqref{eq:nuMassRG}  is suppressed by three powers of the electron Yukawa coupling, leaving the coefficient $\mathcal C_{LL\bar e H}$
poorly constrained. In this section we discuss the constraints on $\mathcal C_{LL\bar e H}$ from non-standard muon decays.
After electroweak symmetry breaking, the $\Delta L = 2$ Lagrangian relevant for muon decay is  
\begin{eqnarray}
\mathcal L &=& - \frac{4 G_F}{\sqrt{2}} \Bigg\{  C^{\mu e}_{\textrm{S}}\,  \bar{\mu}_R e_L \, \nu^T_{L,\, e} C \nu_{L,\, \mu}  + C^{e \mu }_{\textrm{S}}\,  \bar{e}_R \mu_L \, \nu^T_{L,\, e} C \nu_{L,\, \mu}  \nn \\
& &  + \frac{1}{4}  C^{\mu e}_{\textrm{T}}\,  \bar{\mu}_R \sigma^{\mu \nu} e_L \, \nu^T_{L,\, e} C \sigma_{\mu \nu} \nu_{L,\, \mu}  
     + \frac{1}{4}  C^{e \mu}_{\textrm{T}}\,  \bar{e}_R \sigma^{\mu \nu}  \mu_L \, \nu^T_{L,\, e} C \sigma_{\mu \nu}  \nu_{L,\, \mu}
\Bigg\} + \textrm{h.c.}\,\,,
\end{eqnarray}
where the coefficients $C_{\rm S}$ and $C_{\rm T}$ are
\begin{eqnarray}
C^{\mu e}_{\textrm{S}}  &=&\frac{v^3}{4\sqrt{2}} \left(  \mathcal C^{\mu\mu\, e e}_{LL\bar{e} H} + 2\mathcal C^{\mu e\, \mu e}_{LL\bar{e} H} + 3 \mathcal C^{\mu e\, e \mu}_{LL\bar{e} H}  \right), \quad  
C^{\mu e}_{\textrm{T}}  ={\color{white}-}\frac{v^3}{4\sqrt{2}} \left(  \mathcal C^{\mu\mu\, e e}_{LL\bar{e} H} -  \mathcal C^{\mu e\, e \mu}_{LL\bar{e} H}  \right)\,, \nn \\
C^{e \mu}_{\textrm{S}}  &=&\frac{v^3}{4\sqrt{2}} \left(  \mathcal C^{e e \,\mu\mu}_{LL\bar{e} H} + 2\mathcal C^{e\mu \, e\mu }_{LL\bar{e} H} + 3 \mathcal C^{e\mu \, \mu e}_{LL\bar{e} H}  \right), \quad  
C^{e \mu}_{\textrm{T}}  = -\frac{v^3}{4\sqrt{2}} \left(   \mathcal C^{e e \,\mu\mu}_{LL\bar{e} H} -   \mathcal C^{e\mu \, \mu e}_{LL\bar{e} H}  \right).  
\end{eqnarray}
$C^{\mu e}_{\rm S, T}$, and its hermitian  $C^{\mu e *}_{\rm S, T}$, mediate, respectively, the $\Delta L = 2$ decays $\mu^+ \rightarrow e^+ \bar{\nu}_e \bar{\nu}_\mu$  and $\mu^- \rightarrow e^-  {\nu}_e {\nu}_\mu$,
while $C^{e \mu}_{\rm S, T}$ and $C^{e \mu *}_{\rm S, T}$ induce  $\mu^- \rightarrow e^- \bar{\nu}_e \bar{\nu}_\mu$  and $\mu^+ \rightarrow e^+  {\nu}_e {\nu}_\mu$.

The experimental analysis of Ref. \cite{Armbruster:2003pq} searched for $\bar{\nu}_e$ in the decay products of a $\mu^+$ at rest, by looking for the charged current processes 
$p \, \bar{\nu}_e \rightarrow e^+ n$ and $^{12}{\rm C} \, \bar{\nu}_e \rightarrow e^+\, n\, ^{11}$B following the decay of the muon. The  muonic neutrino is not identified,  
and thus the experiment constrains $\mu^+ \rightarrow e^+ \bar{\nu}_e ( \bar{\nu}  + \nu)$. The experimental setup is such that the contribution of neutrino oscillations, $\bar \nu_\mu \rightarrow \bar \nu_e$, 
is negligible \cite{Armbruster:2003pq}. If, in addition, we assume  that there are no $\Delta L =0$ lepton-flavor violating operators,
which would for example induce $\mu^+ \rightarrow e^+ \bar{\nu}_e  {\nu}_\mu$, the limits on the branching ratio can be used to put bounds on $C^{\mu\, e}_{\rm S,\, T}$. 

In terms of $C^{\mu e}_{\rm S,\, T}$, the branching ratio is 
\begin{eqnarray}
\textrm{BR} \left(\mu^+ \rightarrow e^+ \bar{\nu}_e \bar{\nu}_\mu\right) = \frac{\Gamma\left(\mu^+ \rightarrow e^+ \bar{\nu}_e \bar{\nu}_\mu\right)}{\Gamma\left(\mu^+ \rightarrow e^+ {\nu}_e \bar{\nu}_\mu\right)}
= \frac{1}{4} \left|C^{\mu e}_{\textrm{S}}\right|^2 + \frac{3}{4} \left|C^{\mu e}_{\textrm{T}}\right|^2.
\end{eqnarray}
The dependence of the decay rate on the $\bar{\nu}_e$ energy is determined by the Michel parameter $\tilde{\rho}$, which, at tree level, is $\tilde{\rho} = 3/4$ for the scalar, and $\tilde{\rho}= 1/4$ for the tensor operator. 

With this information, we can use the 90\% C.L. limits on the branching ratio \cite{Armbruster:2003pq}
\begin{equation}
\textrm{BR} \left(\mu^+ \rightarrow e^+ \bar{\nu}_e \bar{\nu}_\mu, \tilde{\rho} = 0.75 \right) < 0.9 \cdot 10^{-3}, \qquad \
\textrm{BR} \left(\mu^+ \rightarrow e^+ \bar{\nu}_e \bar{\nu}_\mu, \tilde{\rho} = 0.25 \right) < 1.3 \cdot 10^{-3},
\end{equation}
to obtain $|C^{\mu e}_{\rm S}| < 0.06$ and $|C^{\mu e}_{\rm T}| < 0.04$, corresponding to a scale of around $350$ GeV for the operator $\mathcal O_{LL\bar e H}$.

\section{Two-coupling analysis}
\label{two-coupling-analysis}
\begin{figure}
\begin{center}
\includegraphics[width=7cm]{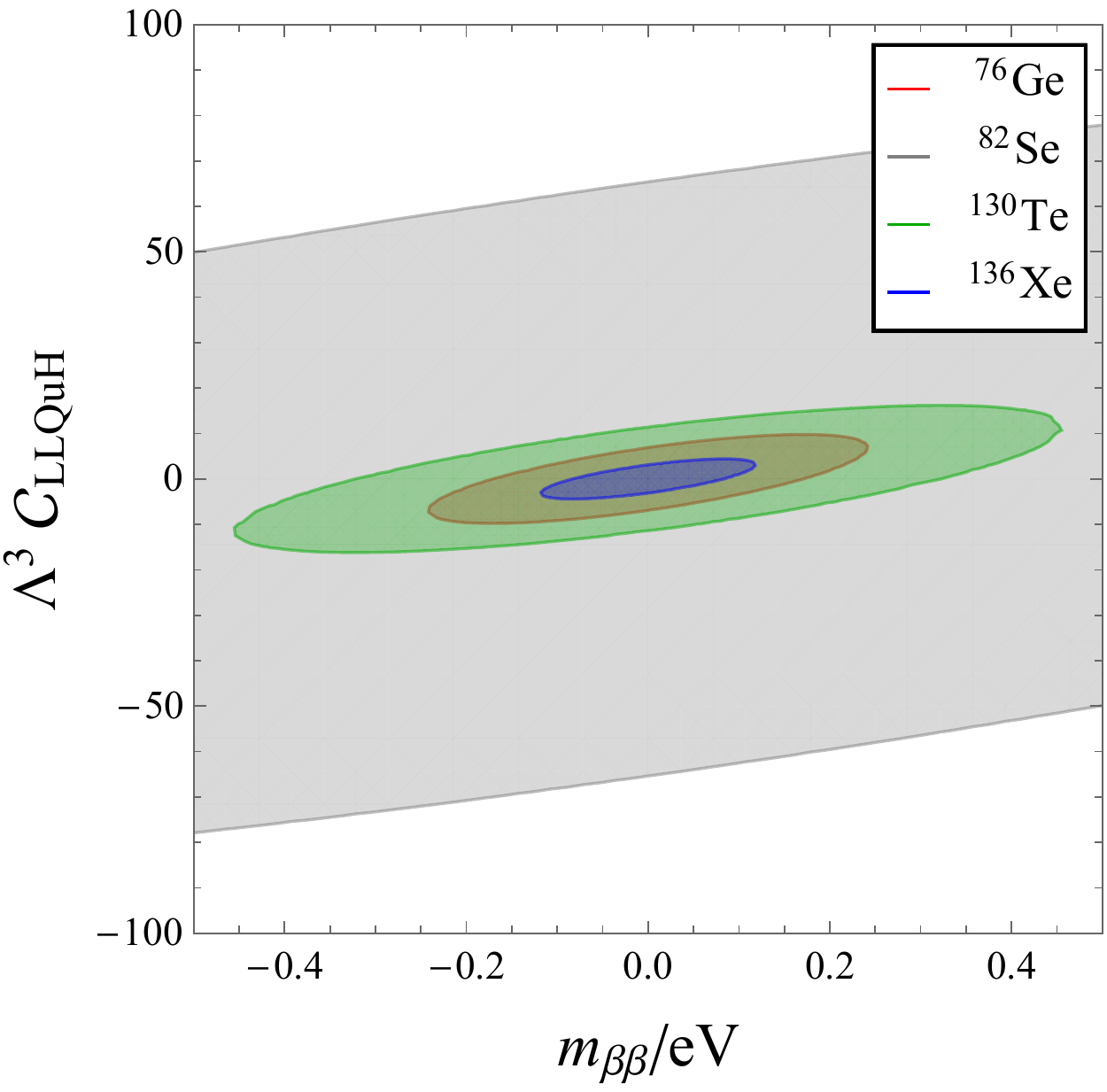}
\includegraphics[width=7cm]{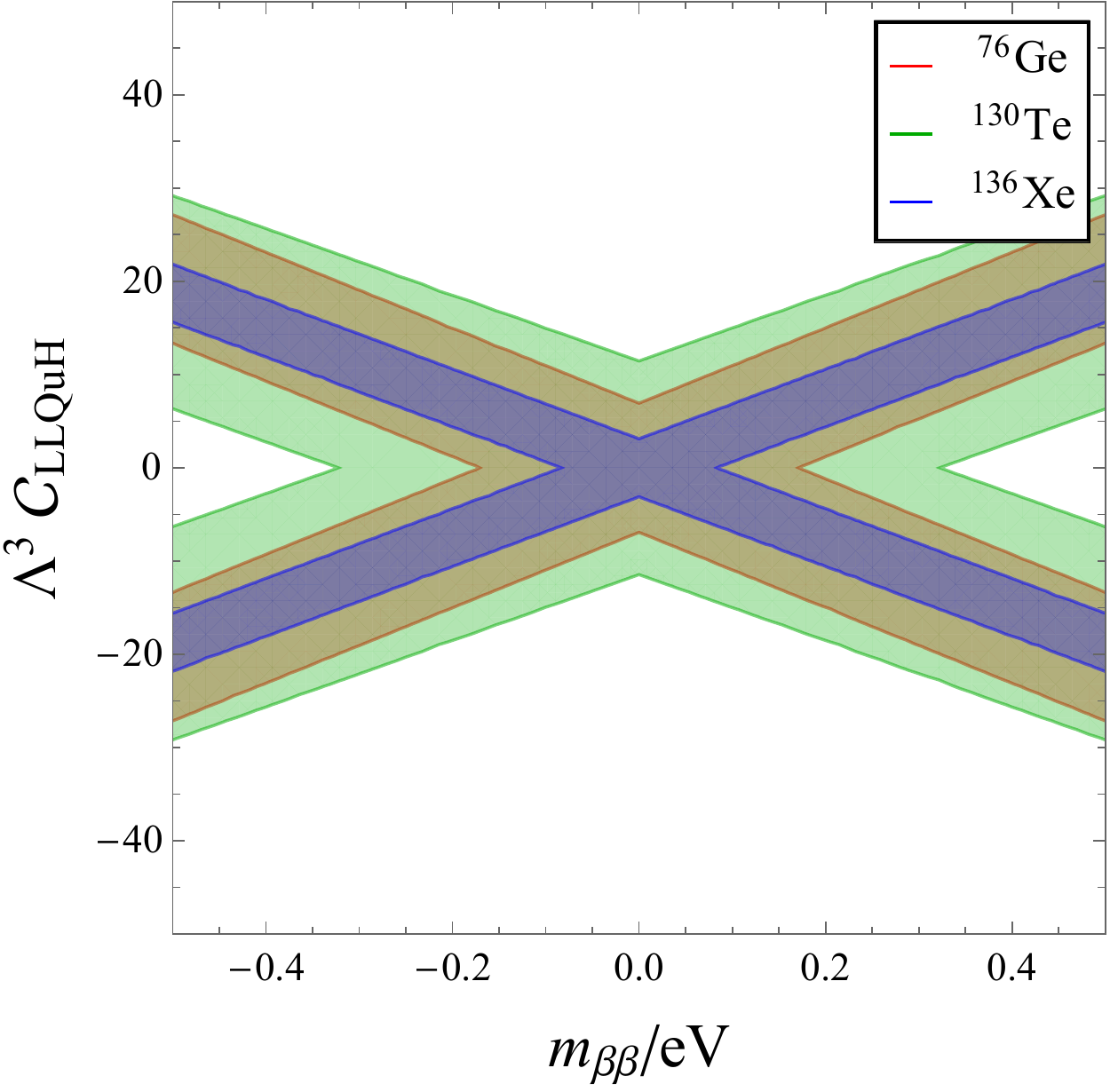}
\end{center}
\caption{Constraints
in the $m_{\bt\bt}$-$ \Lambda^3 \mathcal C_{LL\bar QuH}$ plane using the NMEs of \cite{Hyvarinen:2015bda} and assuming $\Lambda = 600$ TeV. The left panel assumes ${\rm Arg}\,\mathcal C_{LL\bar QuH}=3/4\pi$, while in the right panel we marginalize over the phase of $\mathcal C_{LL\bar QuH}$.
}\label{contourPlot}
\end{figure}

The single-coupling limits of section \ref{Bounds} clearly show the constraining power of the $0\nu\bt\bt$ experiments, as they reach scales of $\Or(100\, {\rm TeV})$. However, in realistic lepton-number-violating scenarios one would generally expect to generate multiple $\Delta L=2$ couplings at the scale of new physics. In this section,  we discuss  scenarios in which both $m_{\bt\bt}$ and a dimension-seven operator are turned on simultaneously. We study how such scenarios differ from the well-known light-Majorana neutrino case. Finally, in section \ref{sec:differential}, we briefly consider the possibility of distinguishing different $\Delta L=2$ operators using the energy and/or angular distributions of the electrons emitted in $0\nu\bt\bt$. 

We begin with showing the limits in the $|m_{\bt\bt}|-\Lambda^3 \, \mathcal C_{LL\bar QuH}$ plane in Fig.\ \ref{contourPlot}. Here we assumed $\Lambda= 600$ TeV and used the NMEs of Ref.\ \cite{Hyvarinen:2015bda}. In the left panel we take a specific value for the relative phase between the dimension-seven coupling and $m_{\bt\bt}$, namely, Arg$\,(\mathcal C_{LL\bar QuH} m_{\bt\bt}^*)=3/4\pi$. As one can see, in this case the experimental limits form ellipses in the $m_{\bt\bt}-\Lambda^3 \, \mathcal C_{LL\bar QuH}$ plane. For a generic relative phase the picture is qualitatively the same. However, specific values of the relative phase, namely, $0$ and $\pi$, allow for cancellations between the dimension-seven and $m_{\bt\bt}$ contributions. As a result, free directions appear once we marginalize over the relative phase. This is clearly shown in the right panel of Fig.\ \ref{contourPlot}.

\begin{figure}
\begin{center}
\includegraphics[width=7cm]{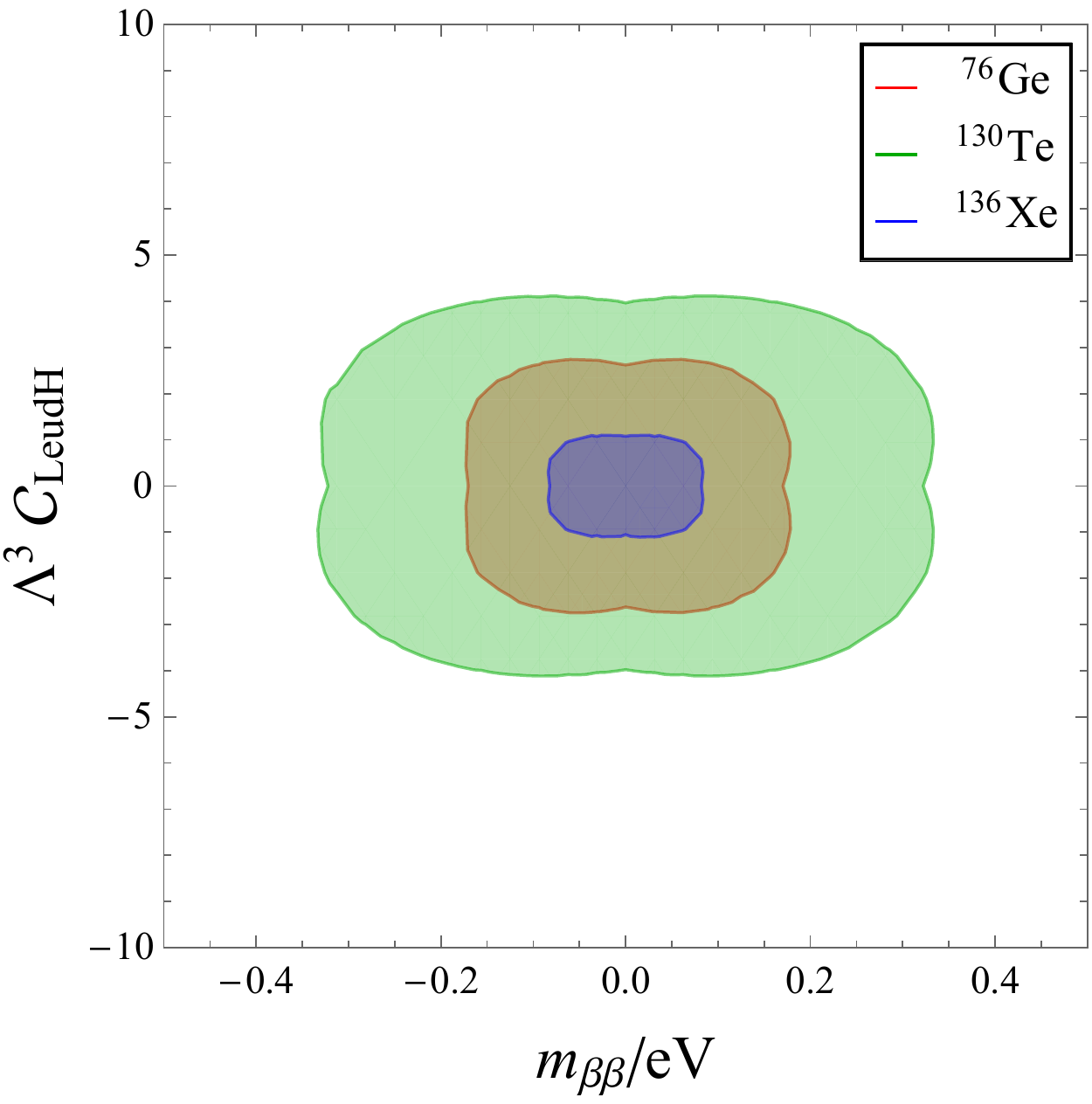}
\end{center}
\caption{Constraints
in the $m_{\bt\bt}$-$ \Lambda^3 \mathcal C_{Le u \bar dH}$ plane using the NMEs of \cite{Hyvarinen:2015bda} and assuming $\Lambda = 40$ TeV. Here the phase of $\mathcal C_{Le u \bar d H}$ is marginalized over.
}\label{contourPlot2}
\end{figure}

These free directions appear in part because $\mathcal C_{LL\bar QuH}$ contributes to the same leptonic structure as $m_{\bt\bt}$ (see e.g.\ Eq.~\eqref{MSM1}). As such, we also consider operators that generate different leptonic structures. We  show the $m_{\bt\bt}$-$ \Lambda^3 \mathcal C_{Le u \bar dH}$ plane in Fig.~\ref{contourPlot2}, now assuming $\Lambda = 40$ TeV. Although we marginalized over the relative phase, no free directions appear because the different leptonic structure prohibit a (complete) cancellation between $m_{\bt\bt}$ and the dimension-seven contribution.
Finally, both Fig.~\ref{contourPlot} and \ref{contourPlot2} illustrate that the different nuclei considered here do not have very different sensitivities, i.e.\ the ellipses and bands all have roughly the same slope. This is a generic feature that does not depend on the dimension-seven coupling under consideration. Unfortunately this implies that it will be difficult to unravel the underlying $\Delta L=2$ mechanism from just nonzero \NLDBD\ total decay rates.

\begin{figure}
\begin{center}
\includegraphics[trim={1.7cm 0 0 0},clip,width=7.5cm]{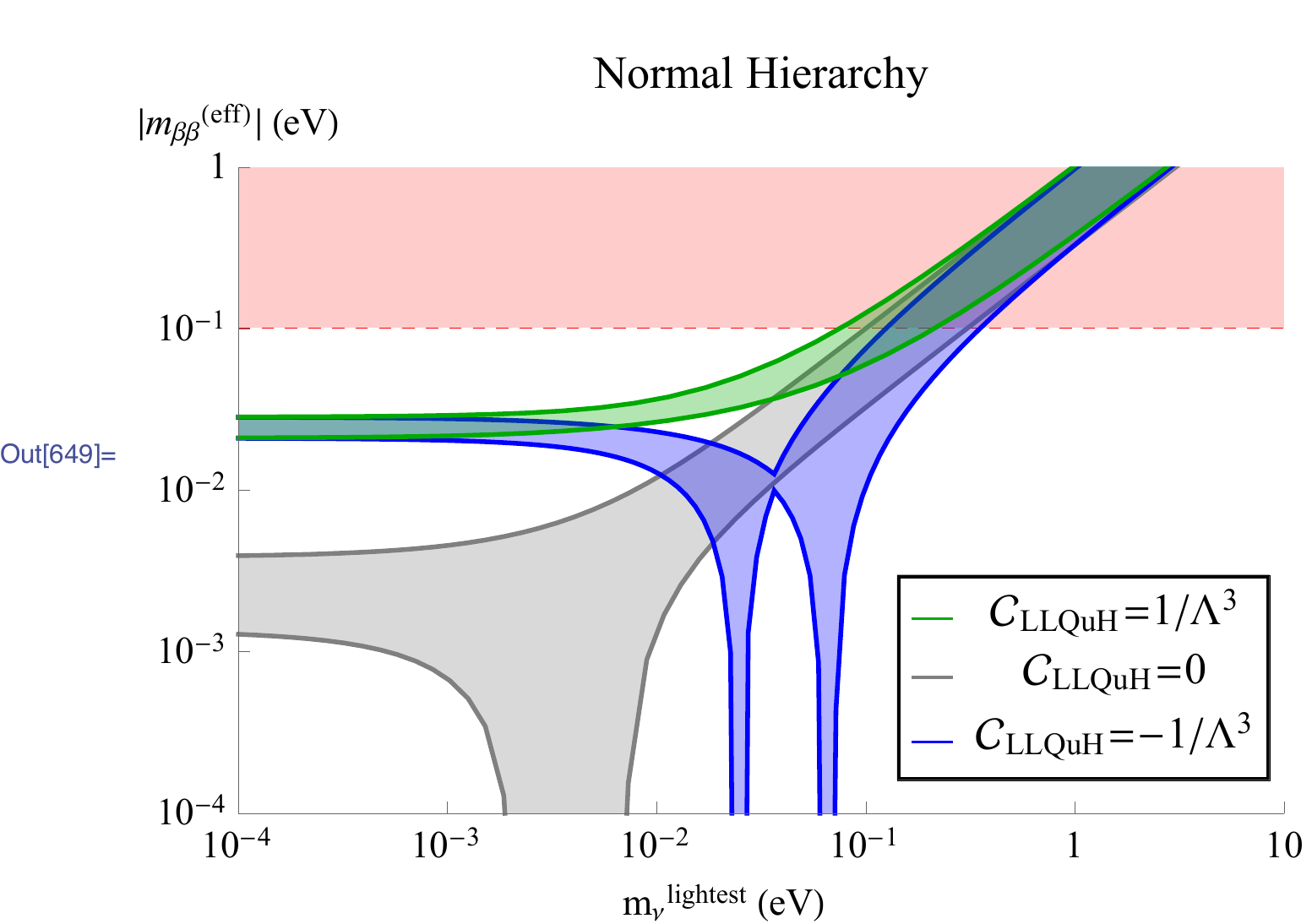}\quad
\includegraphics[trim={1.7cm 0 0 0},clip,width=7.5cm]{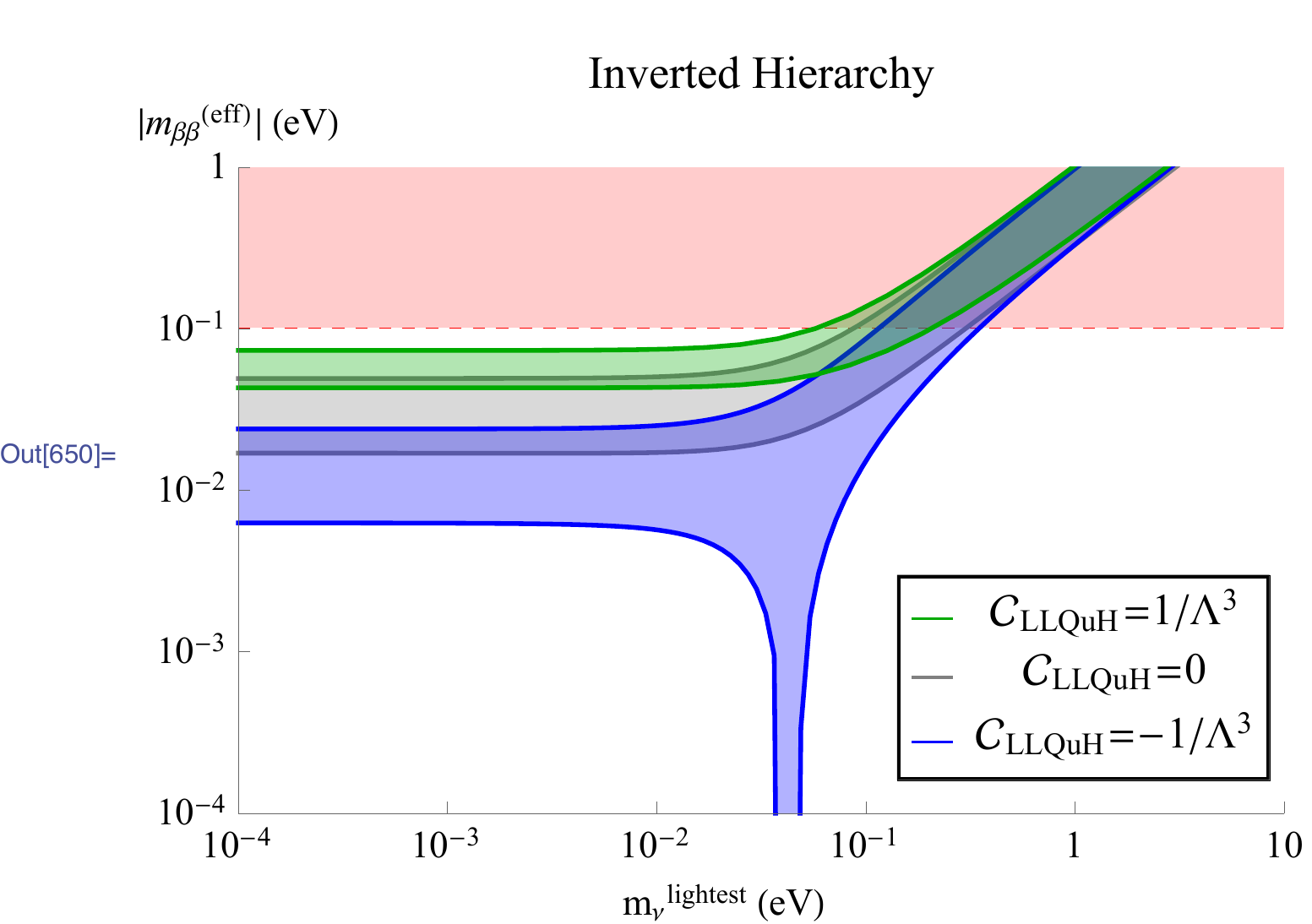}\\
\includegraphics[trim={1.7cm 0 0 0},clip,width=7.5cm]{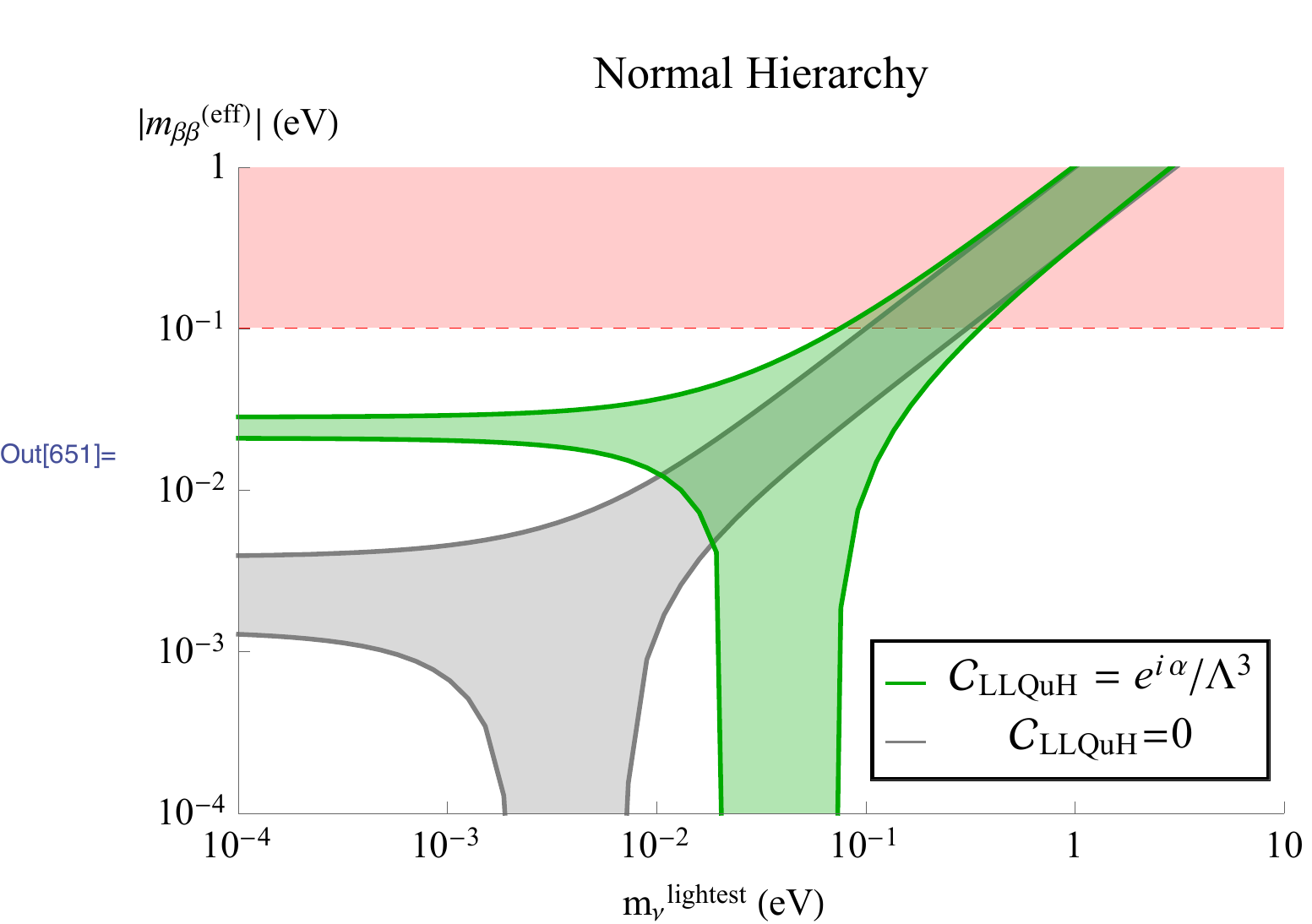}\quad
\includegraphics[trim={1.7cm 0 0 0},clip,width=7.5cm]{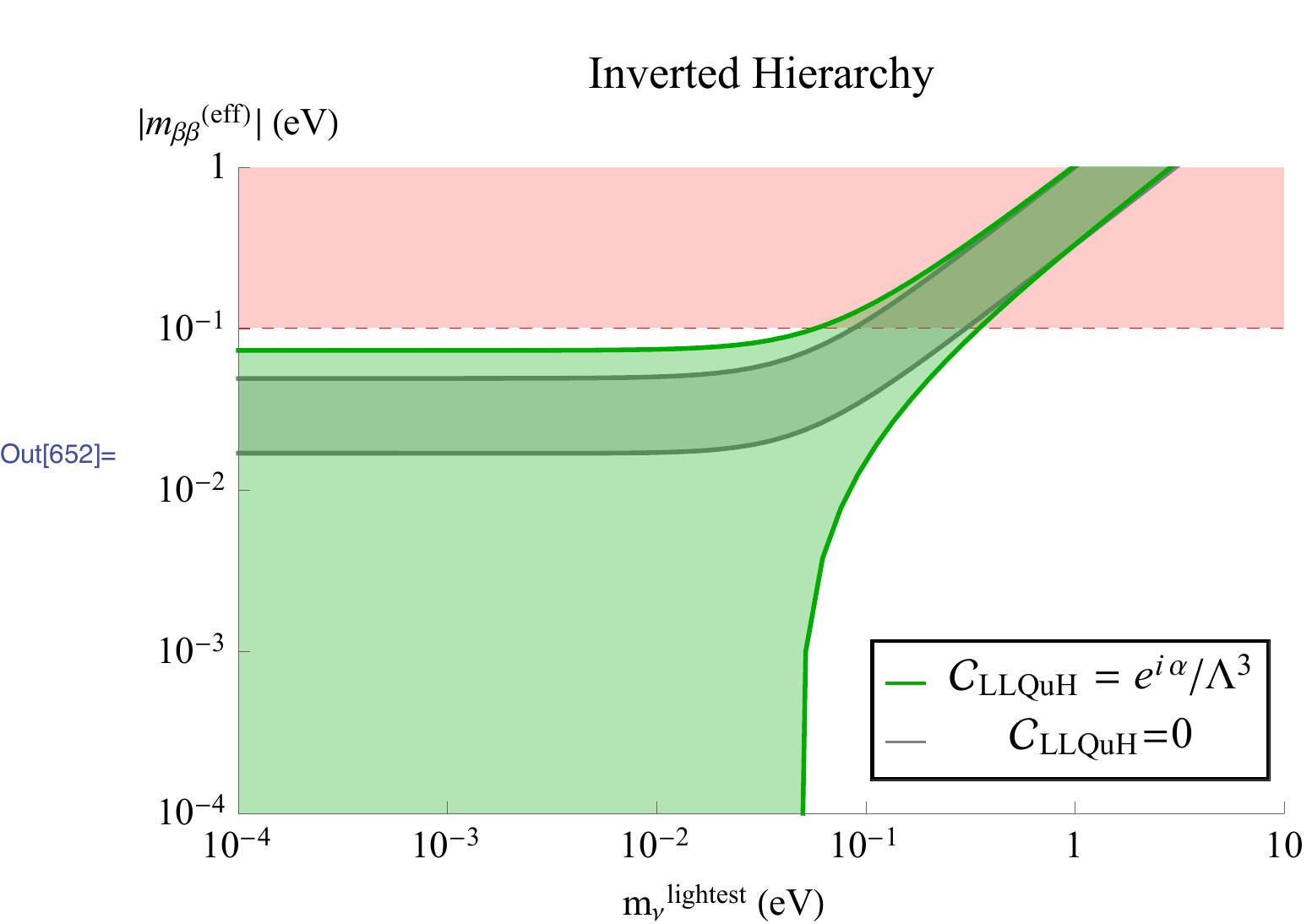}\\
\includegraphics[trim={1.7cm 0 0 0},clip,width=7.5cm]{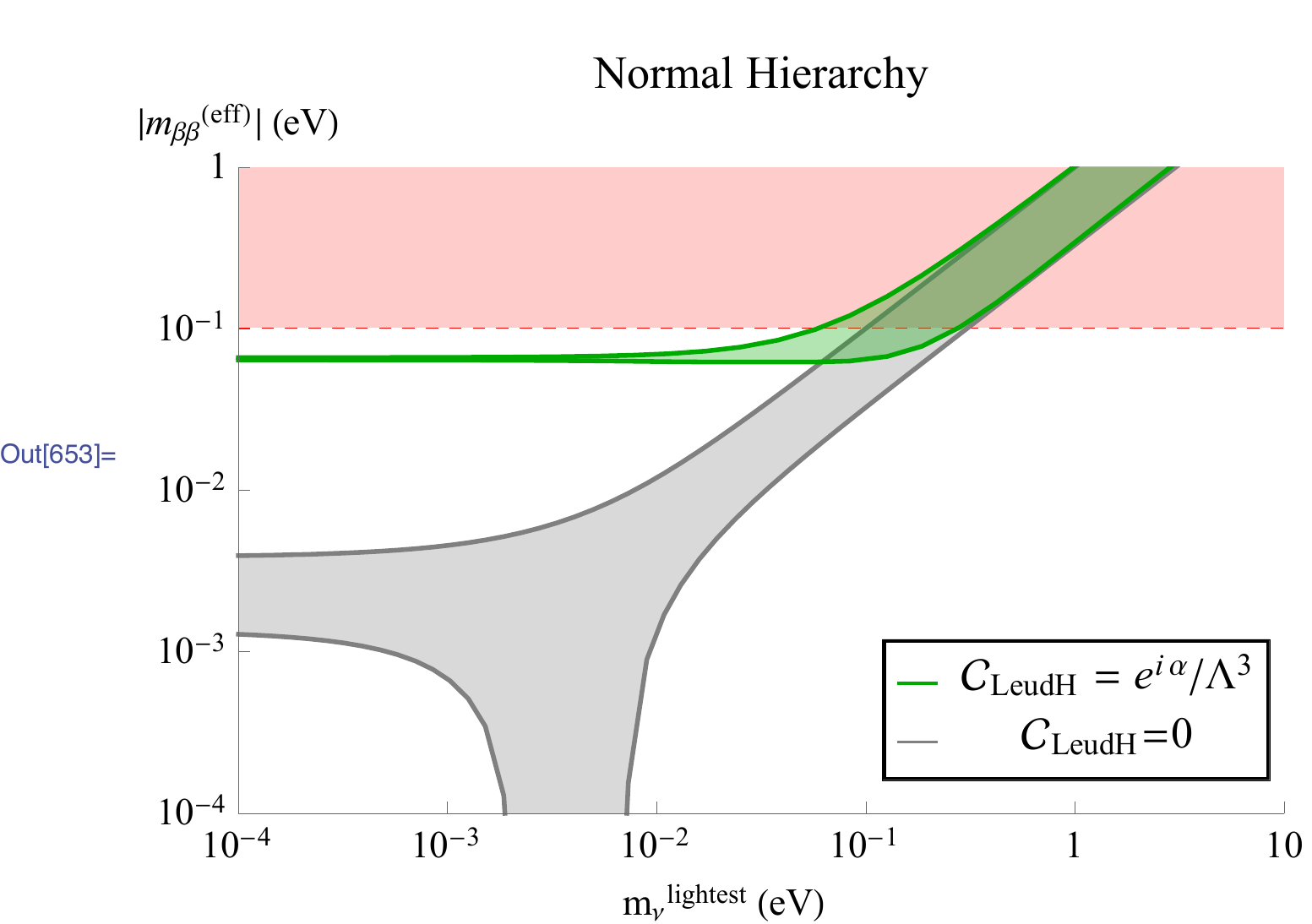}\quad
\includegraphics[trim={1.7cm 0 0 0},clip,width=7.5cm]{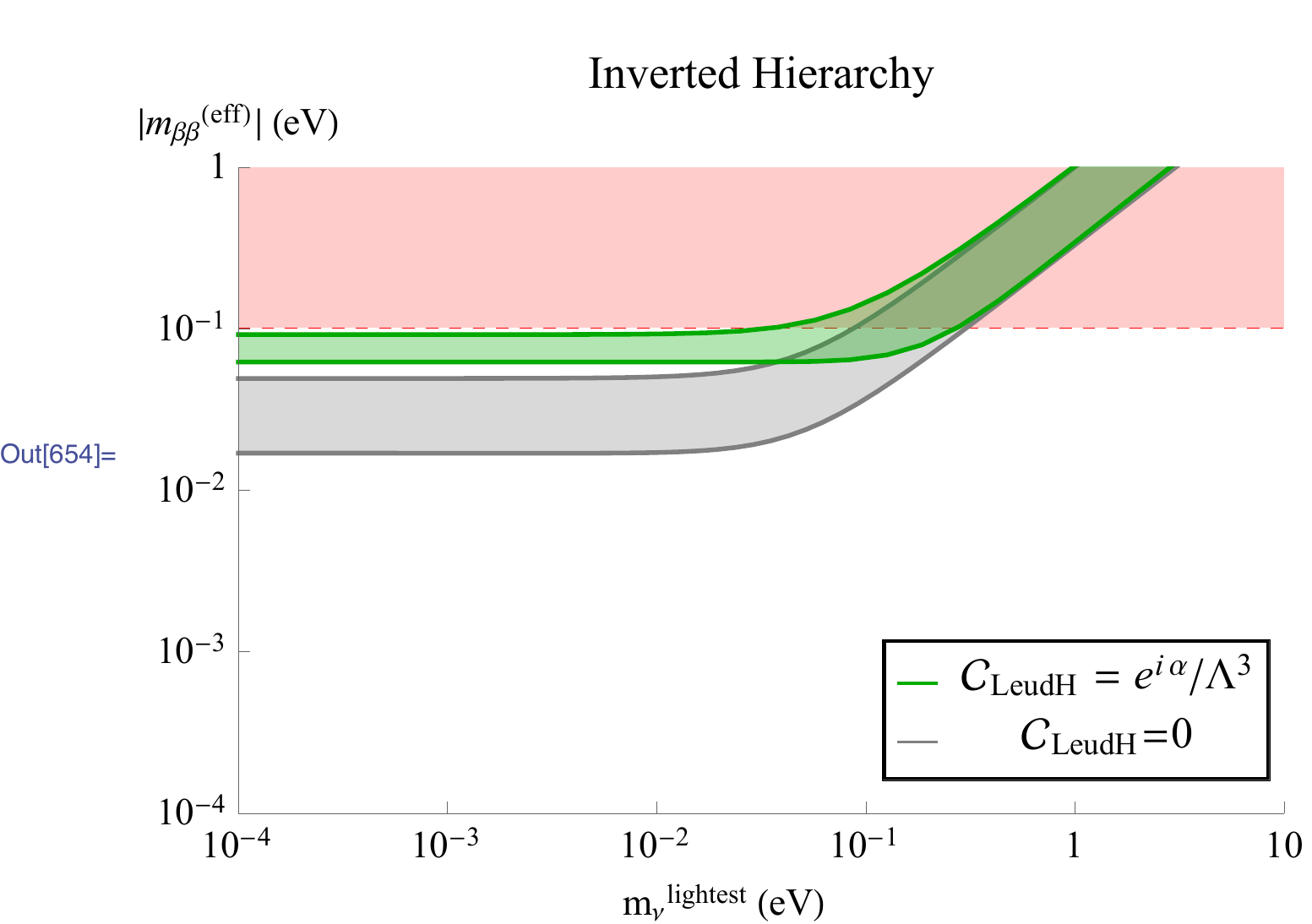}
\end{center}
\caption{The  left (right) column shows the allowed values for the effective parameter $|m_{\bt\bt}^{\rm eff} | $ (defined in Eq.\ \eqref{eq:mbbEff}) as a function of $m_\nu^{\rm lightest}$ for the normal (inverted) hierarchy. 
The gray bands depict the  case with all dimension-seven operators set to zero, while the red horizontal line shows the $0\nu\bt\bt$ limit from $^{136}$Xe. In the top panels, the green and blue bands show the allowed values for the case that $\mathcal C_{LL\bar QuH}=1/\Lambda^3$  and $\mathcal C_{LL\bar QuH}=-1/\Lambda^3$, respectively, assuming $\Lambda = 600$ TeV. 
The middle panels show the same scenarios after marginalizing over the possible phase of $\mathcal C_{LL\bar QuH}$. I.e.\ we take $\mathcal C_{LL\bar QuH}=e^{i\alpha}/\Lambda^3$ and marginalize over $\al$. Finally, the bottom panels show $\mathcal C_{Leu\bar dH}=e^{i\alpha}/\Lambda^3$ marginalized over $\al$, and assuming $\Lambda = 40$ TeV.
}\label{mbbPlot}
\end{figure}

It is interesting to consider the impact of the dimension-seven operators on the interpretation of $0\nu\bt\bt$ measurements.  $0\nu\bt\bt$ experiments are often interpreted as constraints on $m_{\bt\bt}$, however, in the presence of $\Delta L=2$ operators, they are actually sensitive to a combination of dimension-seven couplings and $m_{\bt\bt}$. This combination can be defined as, 
\bea\label{eq:mbbEff}
m_{\bt\bt}^{(\rm eff)} =\frac{m_e}{g_A\sq  V_{ud}\sq M_{\nu}} \left(\frac{T^{0\nu}_{1/2}}{G_{01}}\right)^{-1/2}\,,
\eea
which reduces to  $m_{\bt\bt}$ in case of vanishing dimension-seven operators.

To see how the dimension-seven operators affect $m_{\bt\bt}^{\rm (eff)}$ we turn on $m_{\bt\bt}$ and a  dimension-seven coupling, and show the resulting allowed values of $m_{\bt\bt}^{\rm (eff)}$ as a function of the lightest neutrino mass in Fig.~\ref{mbbPlot}.
The allowed areas are obtained by using the standard parametrization in terms of the neutrino masses, $m_{\nu_i}$, the sines (cosines) of the neutrino mixing angles, $s_{ij}$ ($c_{ij}$), and the Dirac phase $\dt_{13}$,
\bea
m_{\bt\bt} = m_{\nu_1} c_{12}\sq c_{13}\sq +m_{\nu_2} e^{2i \lambda_1}s_{12}\sq c_{13}\sq +m_{\nu_3} e^{2i (\lambda_2-\dt_{13})}s_{13}\sq \,.
\eea
We then marginalize over the Majorana phases, $\lambda_{1,2}$, and the experimentally allowed values of the Dirac phase, while setting the mixing angles to their central values \cite{Olive:2016xmw}.
The top-left (-right) panel of Fig.\ \ref{mbbPlot} depicts the normal (inverted) hierarchy  for several values of $\mathcal C_{LL\bar QuH}$. Blue, gray, and green bands assume $\mathcal C_{LL\bar QuH}=\{-1,0,1\}\cdot \Lambda^{-3}$, respectively, with $\Lambda= 600$ TeV. The current limit on  $m_{\bt\bt}^{\rm (eff)}$ from $^{136}$Xe is depicted by the red shaded area.

The usual light-Majorana-neutrino scenario with $\mathcal C_i=0$ (shown in gray)  allows for a vanishing $m_{\bt\bt}$ in the  normal hierarchy, while this is not possible in the inverted case. However, the blue bands show that a nonzero dimension-seven operator ($\mathcal C_{LL\bar QuH}=-1/\Lambda^3$ in this case) could alter this picture, as  $m_{\bt\bt}^{\rm (eff)}$ can go to zero for both hierarchies. Thus, a vanishing $0\nu\bt\bt$ signal is possible even in the case where the neutrinos are Majorana particles that follow an inverted hierarchy. 
In contrast, if $\mathcal C_{LL\bar QuH}=+1/\Lambda^3$ is chosen (green bands), both the normal and inverted hierarchies require $m_{\bt\bt}^{\rm (eff)}$ to be nonzero and a finite $0\nu\bt\bt$ must exist at some level. 
We show similar plots in the middle row of Fig.\ \ref{mbbPlot}, where the green band is obtained from marginalizing over the phase of $\mathcal C_{LL\bar QuH}$. For a wide range of $m_{\nu}^{\rm lightest}$, the effective parameter $m_{\bt\bt}^{\rm (eff)}$ and thus the \NLDBD\ rate, can go to zero even for an inverted hierarchy. 

$\mathcal C_{LL\bar QuH}$ generates the same leptonic structures as $m_{\bt\bt}$ and it is interesting to look at a coupling that induces a different phase-space factor. In the bottom row of Fig.\ \ref{mbbPlot}, we depict the allowed region for $m_{\bt\bt}^{\rm (eff)}$ assuming that $m_{\bt\bt}$ and $\mathcal C_{Leu\bar dH}$ are both turned on. In this case  the effective parameter $m_{\bt\bt}^{\rm (eff)}$ is always nonzero and the allowed $m_{\bt\bt}^{\rm (eff)}$ region simply shifts upwards for the normal  and inverted hierarchies (left and right panels, respectively). 

\subsection{Pinpointing the $\Delta L=2$ mechanism}\label{sec:differential}
\begin{figure}[t]
\begin{center}
\includegraphics[trim={1.7cm 0 0 0},clip,width=7.5cm]{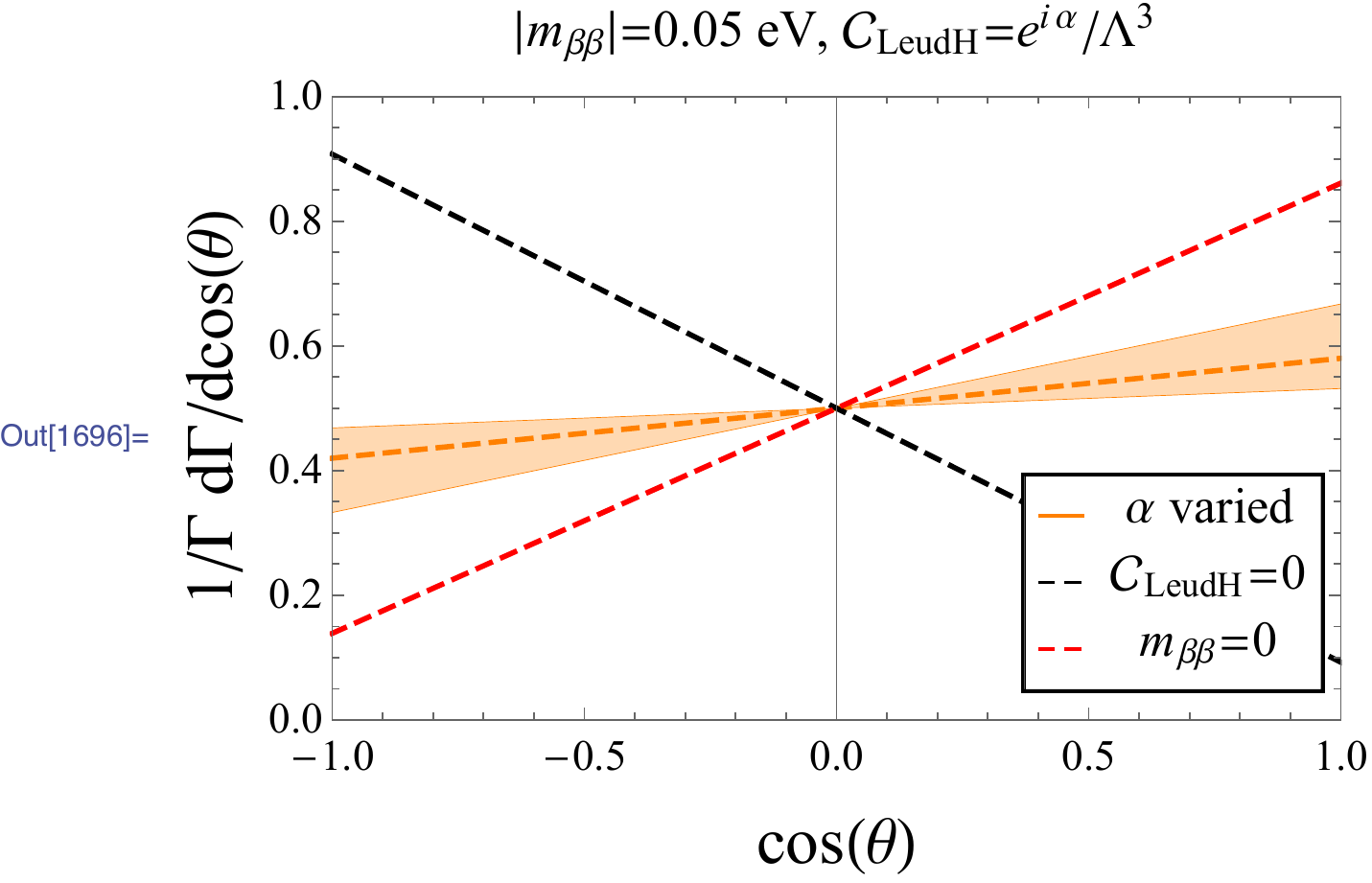}
\includegraphics[trim={1.7cm 0 0 0},clip,width=7.5cm]{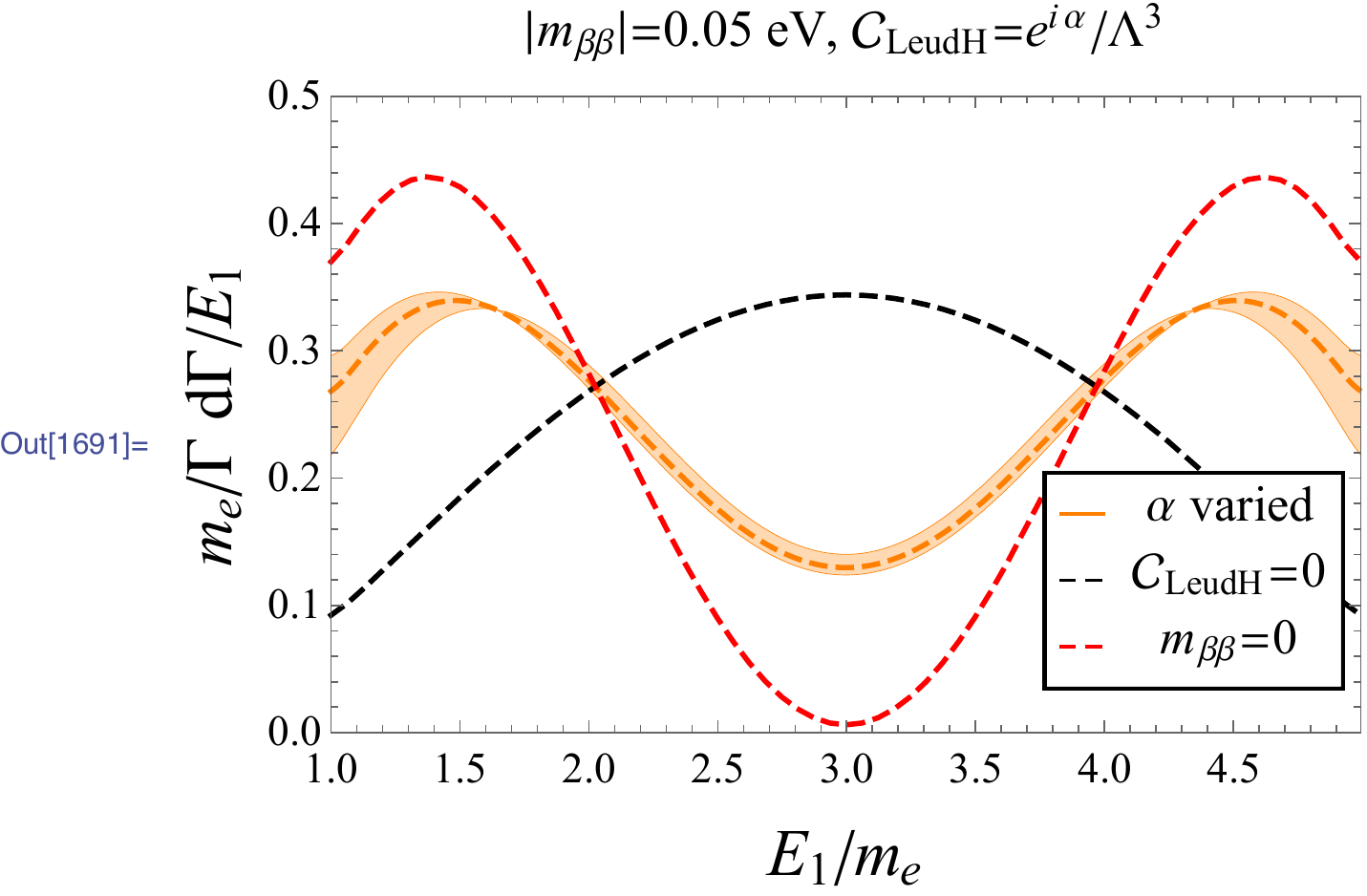}
\end{center}
\caption{The left and right panels show, respectively, the angular and energy dependence of the inverse half-life for $^{76}$Ge. Here the dashed black  and red lines show the case where only $m_{\bt\bt}$ or $\mathcal C_{Leu\bar dH}$ are nonzero, respectively. Instead the orange bands show the scenario in which $|m_{\bt\bt}| = 0.05$ eV and $\mathcal C_{Leu\bar d H}=e^{i\al}/\Lambda^3$ with $\Lambda = 40$ TeV, while we varied over $\al$.
}\label{Fig:PhaseSpace}
\end{figure}

In the best-case scenario in which a $0\nu\bt\bt$ signal is measured, it would be crucial to identify the underlying $\Delta L=2$ mechanism.  
Of course, a nonzero value of $T_{1/2}^{0\nu}$ could be generated by any of the dimension-five or -seven couplings and additional information is required to disentangle them. In principle, one could think of using measurements of $T_{1/2}^{0\nu}$ in different nuclei. Although the NMEs generally show similar patterns for different nuclei, leading to degenerate sensitivities, this is not always the case for the phase space factors. In particular, $G_{02}$ has an increased sensitivivity to the $Q$ value compared to the other phase space factors (see Eq.\ \eqref{eq:b0k}). This means that, $^{128}$Te, which has a rather small $Q$ value, will have a significantly smaller value of $G_{02}$ than $^{76}$Ge. As $C_{\rm VR}^{(6)}$ contributes proportional to $G_{02}$, this in turn implies that $^{128}$Te is less sensitive to $C_{\rm VR}^{(6)}$ compared to $^{76}$Ge \cite{Deppisch:2006hb, Gehman:2007qg}. 
This in principle provides a way to disentangle $C_{\rm VR}^{(6)}$ from the other operators, by measuring the decay rates in several isotopes.
 However, as discussed above, the nuclei considered here ($^{76}$Ge, $^{82}$Se, $^{130}$Te, and $^{136}$Xe) have very similar sensitivities to the dimension-seven couplings, something which is even worsened once nuclear and hadronic uncertainties are taken into account. It would therefore be difficult to pinpoint the underlying $\Delta L=2$ mechanism from just \NLDBD\ total rates of the nuclei under consideration here.
Similar conclusions were reached in Refs. \cite{Fogli:2009py,Lisi:2015yma}.
 
Additional information could come from $\Delta L=2$ signals at colliders such as the LHC. There are certainly scenarios in which colliders can compete with the $0\nu\bt\bt$ measurements \cite{Peng:2015haa}. While the limits derived in Sect.~\ref{Bounds} already put some of the operators at very high scales of $\Or(100\, {\rm TeV})$, two effects, in combination, may mitigate these bounds and make collider searches competitive with $0\nu\bt\bt$ experiments. First, in specific models the Wilson coefficients $\mathcal{C}_i$ may naturally be suppressed by small Yukawa couplings, allowing for a smaller scale $\Lambda$ consistent with the 0$\nu \beta \beta$ bounds obtained here.\footnote{The authors thank F. Deppisch for this observation.} In addition, for a fixed mass scale, we saw in Sect.~\ref{sec:MBmethods} that the uncertainty in the values of the nuclear matrix elements can lead to an order-of-magnitude variation in the predicted $0\nu \beta \beta$ rate. This is the appropriate measure for comparison, since in the contact limit , the production rate at a collider experiment has the same scaling with $\Lambda$ as the $0\nu \beta \beta$ rate, yet is unaffected by uncertainties in the nuclear matrix elements. The rate at a collider may be even higher if intermediate particles can be produced on-shell.
It therefore remains an open question whether direct searches at the LHC or a future collider would be able to see a signal from the fundamental $\Delta L=2$ operators.

As such, here we focus on additional observables that can be measured by the $0\nu\bt\bt$ experiments \cite{Arnold:2010tu}, namely, the angular and energy distributions of the electrons produced in $0\nu\bt\bt$. These distributions are determined by the leptonic structures in Eq.\ \eqref{eq:TotAmp}. Several dimension-seven operators generate different leptonic structures such that the angular and energy distributions carry information about the $\mathcal C_i$.
Unfortunately, only the low-energy couplings $C_{\rm VL}^{(6)}$ and $C_{\rm VR}^{(6)}$ induce  leptonic structures different from the one generated by $m_{\bt\bt}$. These vector couplings are induced by the high-energy dimension-seven couplings $\mathcal C_{LHDe}$ and $\mathcal C_{Leu\bar dH}$. Consequently, all other dimension-seven couplings induce the same lepton structure as $m_{\bt\bt}$ and will be  degenerate with $m_{\bt\bt}$ and each other.

Thus, the angular and energy distributions can in principle be used to disentangle $\mathcal C_{LHDe}$ and $\mathcal C_{Leu\bar d H}$ from the remaining couplings. These two couplings induce a dependence on $\cos \theta$ whose slope has the opposite sign of the one induced by $m_{\bt\bt}$. In addition, although $\mathcal C_{LHDe}$ gives rise to an energy dependence that is very similar to $m_{\bt\bt}$, the energy distribution of $\mathcal C_{Leu\bar d H}$ is significantly different. This is illustrated in Fig.~\ref{Fig:PhaseSpace} which shows the angular and energy dependence in the left and right panels, respectively. The different lines correspond to the case of nonzero $m_{\bt\bt}$ (dashed black), nonzero $\mathcal C_{Leu\bar  dH}$ (dashed red), and a scenario where both couplings are turned on (orange band). In the latter scenario we set $|m_{\bt\bt}| = 0.05$ eV and $\mathcal C_{Leu\bar d H}=e^{i\al}/\Lambda^3$ with $\Lambda = 40$ TeV, while we varied over the relative phase $\al$. As can be seen from the left panel, the slope of the $\cos\theta$ dependence does indeed differ by a sign between $m_{\bt\bt}$ and $\mathcal C_{Leu\bar dH}$. Once both couplings are turned on the resulting slope lies somewhere in between the two extremes. Although many couplings could induce the same $\cos \theta$ dependence as $m_{\bt\bt}$, the opposite slope can only point to either $\mathcal C_{Leu\bar dH}$ or  $\mathcal C_{LHDe}$.

The energy dependence is shown in the right panel of Fig.\ \ref{Fig:PhaseSpace}. Again there is a clear difference between the case in which only $m_{\bt\bt}$ is turned on (dashed black) or only $\mathcal C_{Leu\bar d H}$ is nonzero (dashed red). As one would expect, including both couplings (orange band) gives a combination of the two dashed lines. It should be noted that only $\mathcal C_{Leu\bar dH}$  is able to induce an energy dependence that significantly differs from the $m_{\bt\bt}$ case, while the $\mathcal C_{LHDe}$ case looks very similar to that of $m_{\bt\bt}$.

\section{Summary, conclusions, and outlook}\label{conclusions}
In this work we have investigated neutrinoless double beta decay in the framework of the Standard Model effective field theory. In principle, the dominant contribution to \NLDBD\ arises from the dimension-five Weinberg operator which is only suppressed by one power of the scale of beyond-the-SM physics. However, in several models competing contributions arise from higher-dimensional operators and we therefore extended the analysis to include all $\Delta L=2$ operators of dimension seven. 

In the first part of this work we classified the different dimension-seven operators and studied how they manifest at a relatively low-energy scale of a few GeV. We studied the evolution of the operators to lower energies by considering renormalization-group running and threshold effects from integrating out relatively heavy SM fields such as the Higgs and electroweak gauge bosons. This analysis gives rise to a set of effective dimension-six, -seven, and -nine $\Delta L=2$ operators  that we evolve to slightly above the QCD scale using their renormalization group equations. All operators scale as $1/\Lambda^3$, where $\Lambda$ is the scale of BSM physics, and their effective dimension is determined by powers of the electroweak scale. 

In the second part we applied the framework of chiral effective field theory to construct the effective $\Delta L=2$ hadronic Lagrangian. For each effective operator at the quark-gluon level we build the chiral Lagrangian up to the order where we find the first non-vanishing contribution to the \NLDBD\ decay rate. Depending on the effective operator under consideration, the chiral Lagrangian consists of pionic, pion-nucleon, and/or nucleon-nucleon interactions.  Armed with the chiral Lagrangian we calculated effective two-nucleon \NLDBD\ operators in a consistent power-counting scheme, and derived, within the same scheme, a Master formula for the \NLDBD\ decay rate.
Our results contain several new aspects
\begin{itemize}
\item We used up-to-date hadronic input for several low-energy constants that connect $\Delta L=2$ quark-gluon operators to $\Delta L=2$ chiral operators. While remarkable progress has been made in recent years on several of the LECs, others, in particular those associated to $\Delta L=2$ pion-nucleon and nucleon-nucleon interactions, are still unknown. In the future it will be important to further constrain or compute these LECs. For illustrative purposes, we show in Fig.\ \ref{fig:lecPlot} how the current bound on the Wilson coefficient $C^{(9)}_1$ is affected by the uncertainty on the unknown LECs.
\item We introduced a power-counting-scheme for \NLDBD\ operators which includes, apart from the standard $\chi$EFT counting rules, the additional scales associated with \NLDBD\,: the so-called ``closure energy" and the $Q$ value of the reaction. We showed that up to leading order in the power counting, the rate does not depend on the closure energy. In addition, we find that the leading-order rate only depends on several nuclear moments (scalar, vector, axial, and tensor) and not on the associated radii which are often included. These considerations greatly reduce the number of nuclear matrix elements that needs to be calculated. We confirmed these power-counting predictions by explicit comparison with several sets of nuclear matrix elements calculated in the literature.
\item Based on the extended $\chi$EFT power counting we identified nine combinations of nuclear matrix elements, which determine the leading-order \NLDBD\ rate up-to-and-including dimension-seven operators in the SM-EFT. Two combinations of nuclear matrix elements turned out to be numerically suppressed due to factors beyond the power-counting scheme (the large size of the nucleon isovector magnetic moment and the smallness of the electron mass with respect to the reaction $Q$ values.)  As such, the \NLDBD\ rate is dominated by a relatively small set of nuclear matrix elements.

\item We find that the nuclear matrix elements that are needed to constrain the contributions of dimension-seven operators  
can be lifted from existing calculations of  \NLDBD .  With the exception of $M^{AA}_{T}$, the required matrix elements can be deduced from calculations of light- and heavy-Majorana-neutrino exchange, provided that the various components, $M_{GT,T \,(sd)}^{ij}$ in Eq.\ \eqref{MSMstandard}, are listed separately and the  calculations include the contributions from weak magnetism and induced pseudoscalar form factor.

\item The matrix element $M^{AA}_T$ is important in constraining $C^{(6)}_{VR}$, but is not evaluated in any of the recent nuclear matrix element literature. Here we used the value computed in Ref.~\cite{Muto:1989cd}. It would be preferable if in the future this matrix element is reported along with the other $M^{ij}_{F,GT,M}$ nuclear matrix elements such that all nuclear physics input to the $0\nu \beta \beta$ rate is internally consistent.

\item We have compared different sets of nuclear matrix elements obtained with various many-body methods. We find that uncertainties on the non-standard matrix elements, based on the spread of the results, are of similar size as the uncertainty on the light-Majorana-neutrino-exchange matrix elements. Typically the matrix elements vary at most by factors of two-to-three (and several are in much better agreement) depending on the chosen nuclear method. However, the sign and relative sizes of the matrix elements are in good agreement with each other and the chiral power counting.   

\end{itemize} 

In the 
final 
phenomenological part of this work, we studied the constraints on the fundamental $\Delta L=2$ operators. The above-described framework provides essentially a dictionary between high-scale $\Delta L=2$ physics and low-energy \NLDBD\ measurements such that constraints on the scale of BSM physics can be immediately obtained. We obtain several interesting conclusions:
\begin{itemize}
\item Depending on the $\Delta L=2$ operator under consideration, the limits on the BSM scale varies from $\Lambda > 10$ TeV to $\Lambda > 400$ TeV. For most operators these limits on the scale $\Lambda$ are not too much affected by hadronic and nuclear uncertainties, except for operators which mainly induce so-called short-distance contributions to \NLDBD\ which depend on unknown LECs associated to $\Delta L=2$ pion-nucleon and nucleon-nucleon interactions. LQCD calculations of these LECs, along the lines of Refs.~\cite{Nicholson:2016byl,Shanahan:2017bgi}, could improve this situation. Several dimension-seven SM-EFT operators do not contribute to \NLDBD\ at a significant level. We studied  complementary observables, such as the neutrino mass and magnetic moment, and muon decay, that can be used to probe such couplings.
\item We find that \NLDBD\ experiments with different isotopes (we studied ${}^{76}$Ge, ${}^{82}$Se, ${}^{130}$Te, and ${}^{136}$Xe) are rather degenerate with respect to the different $\Delta L=2$ mechanism they are sensitive to. We have illustrated this in Figs.~\ref{contourPlot} and \ref{contourPlot2} where it can be seen that different isotopes probe roughly the same combination of $\Delta L=2$ operators. 
\item The inclusion of non-zero dimension-seven $\Delta L=2$ couplings can affect the standard interpretation of (the absence of) \NLDBD\ signals in terms of 
light Majorana-neutrino exchange.   In this framework,
 it is possible to rule out the inverted ordering of the neutrino mass spectrum  with sufficiently sensitive \NLDBD\ experiments. The upper panels of Fig.~\ref{mbbPlot} illustrates that this is no longer necessarily true once dimension-seven operators are included in the analysis, although some fine-tuning is required to suppress the  \NLDBD\ rate. At the same time, the inclusion of dimension-seven operators can lead to a non-zero  \NLDBD\ rate for all values of the lightest neutrino mass even for a normal hierarchy.
\item While total  \NLDBD\ rates of different isotopes have little discriminating power with respect to the underlying source, additional information could be obtained by angular and energy differential rates. As shown in Fig.~\ref{Fig:PhaseSpace}, the differential rates can potentially separate several $\Delta L=2$ dimension-seven operators from the dimension-five and other dimension-seven operators. This is particularly relevant for BSM models, such as left-right symmetric models, that induce low-energy vector-like $\Delta L=2$ operators.
\end{itemize}

Our work can be extended in several ways. First of all, in several models also $\Delta L=2$ dimension-nine operators provide relevant \NLDBD\ contributions. We aim to extend the framework to include these operators in future work.  This will enable one
to match specific UV-complete models to the effective field theory framework. 
In particular, this would allow for a global analysis of Standard Model extensions involving lepton-number violation, including \NLDBD\  
and high-energy probes at the LHC  or future high-energy colliders.

\section*{Acknowledgements}
We thank the  Institute for Nuclear Theory at the University of Washington for its hospitality and the Department of Energy for partial support during the completion of this work. We thank Frank Deppisch for a discussion on the LHC and our bounds on dimension-seven operators.  We are very grateful to Javier Men\'endez for providing us with updated shell-model nuclear matrix elements before publication,
and for comments on the manuscript. 
We are indebted to Jose Barea for providing us unpublished results for the nuclear matrix elements in the interacting boson model. 
We thank Mihai Horoi and Andrei Neacsu for several interesting discussions, and for clarifications on the nuclear matrix elements of Ref. \cite{Horoi:2017gmj}. 
VC and EM  acknowledge support by the US DOE Office of Nuclear Physics and by the LDRD program at Los Alamos National Laboratory.
MG acknowledges support by the US DOE Office of High Energy Physics and by the LDRD program at Los Alamos National Laboratory.
WD and JdV  acknowledge  support by the Dutch Organization for Scientific Research (NWO) 
through a RUBICON  and VENI grant, respectively.

\newpage
\appendix

\section{Comparison with other operator bases}\label{AppEps}
 
In this Appendix we compare our operator basis and Wilson coefficients 
to the one previously used in the literature. 
The basis introduced in Refs.~\cite{Pas:1999fc,Pas:2000vn,Deppisch:2012nb} 
contains at the hadronic scale operators of dimension six (long range part), 
related to the ones in \eqref{lowenergy6}   and dimension nine (short range part), related to the ones in  \eqref{lowenergy9}.
They do not consider operators of dimension seven (see \eqref{lowenergy7})  which naturally arise in our analysis based on $SU(2)\times U(1)$ gauge invariance. 

The effective couplings $\epsilon^{\alpha}_\beta$  parameterizing long-range contributions to \NLDBD \ are related to our dimension-six Wilson coefficients as follows:
\begin{equation}
\epsilon^{V+A}_{V \mp A} = \frac{1}{2} C_{\rm VL, VR}^{(6)} ~,  \qquad 
\epsilon^{S+P}_{S \mp P} = \frac{1}{2} C_{\rm SL, SR}^{(6)} ~,  \qquad 
\epsilon^{T_R}_{T_R} = \frac{1}{2} C_{\rm T}^{(6)} ~. 
\end{equation}
The operator corresponding $\epsilon^{T_R}_{T_L}$ in  Ref.~\cite{Pas:1999fc, Bonnet:2012kh}  vanishes identically, due to the identity 
$\sigma_{\mu \nu} (1\pm \gamma_5) \otimes \sigma^{\mu \nu} (1 \mp \gamma_5) \equiv 0$, 
so we have five dimension-six coefficients rather than six.  

For the short-range effective couplings associated to dimension-nine six-fermion operators,  \cite{Pas:2000vn,Deppisch:2012nb} 
the effective couplings $\epsilon_i^{xyz}$ (with $x,y,z$  labeling the chirality of the two hadronic densities and the leptonic current, in that order)  
the  mapping goes as follows: 

\begin{equation}
\epsilon_3^{LLR} = \frac{1}{2} \frac{m_N}{v} \, C_1^{(9)} ~, \qquad 
\epsilon_3^{LRR} = \frac{1}{2} \frac{m_N}{v} \, C_4^{(9)} ~, \qquad 
\epsilon_1^{RLR} = -  \frac{m_N}{v} \, C_5^{(9)} ~.
\end{equation}

\section{RG evolution}\label{AppRG}
In this appendix we briefly discuss the scale dependence of the couplings mentioned in sections \ref{sec:operators} and \ref{sec:lowE}. The running of the dimension-seven operators between the high scale, $\Lambda$, and the electroweak scale is given by
\bea
 \vec C(\mu) &=&U(\mu,\, \Lambda) \cdot \vec C(\Lambda),\qquad \vec C= (\mathcal  C_{LL\bar QuH},\,\mathcal C_{LLQ\bar d H}^{(1)},\,\mathcal C_{LLQ\bar d H}^{(2)})^T,\nn\\
   U(\mu,\, \Lambda) &=& \bma\left(  \frac{\al_s(\Lambda)}{\al_s(\mu)}\right)^{-3 C_F/\bt_0} &0&0\\
 0& \left(  \frac{\al_s(\Lambda)}{\al_s(\mu)}\right)^{-3 C_F/\bt_0} &0\\
 0 & \frac{1}{2}\left[\left(  \frac{\al_s(\Lambda)}{\al_s(\mu)}\right)^{C_F/\bt_0}-\left(  \frac{\al_s(\Lambda)}{\al_s(\mu)}\right)^{-3 C_F/\bt_0}\right] &\left(  \frac{\al_s(\Lambda)}{\al_s(\mu)}\right)^{C_F/\bt_0}
 \ema
\eea
while the remaining couplings are scale independent at one loop in QCD. Here $\bt_0 = \frac{1}{3}(11 N_c-2n_f)$, with $n_f$ the number of active flavors, 
and recall $C_F=(N_c^2-1)/(2 N_c)$. The couplings $\mathcal  C_{LL\bar QuH}$ and $\mathcal C_{LLQ\bar d H}^{(1)}$ decrease in the ultra-violet (UV), 
whereas the behavior of $\mathcal C_{LLQ\bar d H}^{(2)}$ depends on the initial values.
The couplings at the electroweak scale are then given by
$\vec C(m_W) = U^{(n_f=5)}(m_W,\, m_t) U^{(n_f=6)}(m_t,\, \Lambda)\cdot \vec C(\Lambda)\,\,.$
Numerically, using the one-loop running of $\al_s$, this results in
\bea\vec C(m_W) =\bma 1.3 &0&0\\
 0& 1.3 &0\\
 0 & -0.21 &0.91
 \ema \cdot\vec C(10 \,{\rm TeV}) = 
 \bma 1.5 &0&0\\
 0& 1.5 &0\\
 0 & -0.29&0.88
 \ema \cdot \vec C(100 \,{\rm TeV})\,\,.
\eea

Below the electroweak scale we match onto the dimension-six, -seven, and -nine operators in Eqs.\ \eqref{lowenergy6}, \eqref{lowenergy7}, and \eqref{lowenergy9}. The RGEs for the dimension-six operators are solved by
\bea
C_{\rm SL(SR)}^{(6)}(\mu) &=& \left(  \frac{\al_s(m_W)}{\al_s(\mu)}\right)^{-3 C_F/\bt_0}C_{\rm SL(SR)}^{(6)}(m_W),\qquad C_{\rm T}^{(6)}(\mu)=\left(  \frac{\al_s(m_W)}{\al_s(\mu)}\right)^{C_F/\bt_0}C_{\rm T}^{(6)}(m_W)\,\,.\nn
\eea
The couplings $C_{\rm SL(SR)}^{(6)}$ decrease in the UV while the tensor coupling $C_{\rm T}^{(6)}$ increases.
The dimension-seven operators do not run, while for the dimension-nine operators we have,
\bea
 \vec C'(\mu) &=&U(\mu,\, m_W) \cdot \vec C'(m_W),\qquad \vec C'= (C_1^{(9)},\,C_4^{(9)},\,C_5^{(9)})^T,\nn\\
   U(\mu,\, m_W) &=& 
\bma 
\left(  \frac{\al_s(m_W)}{\al_s(\mu)}\right)^{3(1-1/N_c)/\bt_0} &0 &0\\
0& \left(\frac{\al_s(m_W)}{\al_s(\mu)}\right)^{3/(N_c\bt_0)}&0\\
 0& \lambda \left[\left(  \frac{\al_s(m_W)}{\al_s(\mu)}\right)^{-6 C_F/\bt_0}-\left(  \frac{\al_s(m_W)}{\al_s(\mu)}\right)^{3/(N_c\bt_0)}\right] &\left(  \frac{\al_s(m_W)}{\al_s(\mu)}\right)^{-6C_F/\bt_0}
 \ema\,\,.\nn
\eea
where $\lambda=1/(2 C_F + 1/N_c)=1/N_c$. Here the couplings $C_1^{(9)}$ and  $C_4^{(9)}$ increase in the UV, and the behavior of $C_5^{(9)}$ depends on the boundary values.
Taking into account the bottom mass threshold, we obtain for the evolution between $\mu=m_W$ and $\mu=2$ GeV,
\bea
C_{\rm SL(SR)}^{(6)}(2 \,{\rm GeV}) &=& 1.5\,C_{\rm SL(SR)}^{(6)}(m_W),\qquad C_{\rm T}^{(6)}(2 \,{\rm GeV})=0.87\,C_{\rm T}^{(6)}(m_W),\\
\vec C'(2\, {\rm GeV}) &=& \bma 0.82 &0&0\\
0 & 0.90 &0 \\
0&0.45 &2.3\ema \cdot \vec C'(m_W)\,\,. 
\eea
The remaining operators in Eqs.\ \eqref{lowenergy6}, \eqref{lowenergy7}, and \eqref{lowenergy9} are scale independent at one loop in QCD.

\section{Recoil matrix elements}\label{AppRecoil}

The  tensor $C^{(6)}_{\rm{T}}$ and vector operators $C^{(6)}_{\rm VL, VR}$ induce, at lowest order in $\chi$PT, 
two-nucleon operators whose matrix elements vanish in $0^+ \rightarrow 0^+$ transitions.
For example, $C^{(6)}_{\rm{T}}$ induces contributions proportional to 
\begin{equation}\label{AppC1}
C^{(6)}_{T}   \frac{\vec q \cdot \left( \boldsigma^{(1)} - \boldsigma^{(2)} \right)}{\vec q\sq}.
\end{equation}
The operator in Eq.~\eqref{AppC1} is pseudoscalar, and, consequently, its  matrix element vanishes in $0^+ \rightarrow 0^+$  transitions.
Similar considerations apply to the LO operators induced by    $C^{(6)}_{\rm VL, VR}$.

The most important transition operators induced by $C^{(6)}_{\rm{T}}$ and  $C^{(6)}_{\rm VL, VR}$ were discussed in Secs. \ref{MEtensor} 
and \ref{MEvector}. At the order we are working, corrections  proportional to the nucleon recoil momentum can become important. 
In addition to the neutrino potential defined in Sec. \ref{MEtensor}, 
we find that the tensor operator gives 
\begin{eqnarray}\label{eq:Trec}
& &V(\vec q^2) =  2 \tau^{(1) +} \tau^{(2) + } \, 2 G_F^2 \,   m_N C^{(6)}_{\textrm{T}} \frac{1}{\vec q^2}  \bar u(k_1) \, P_RC \bar u^T(k_2)\, \\
& & \Bigg\{  \frac{g_A g_T}{m_N^2} \left(\boldsigma^{(1)} \cdot \vec q\, \boldsigma^{(2)}\cdot (\vec P_1 - \vec P_2) +  \boldsigma^{(1)} \cdot \vec (\vec P_1 - \vec P_2)\, \boldsigma^{(2)}\cdot \vec q 
-\boldsigma^{(1)} \cdot \boldsigma^{(2)}\,  \vec q \cdot (\vec P_1 - \vec P_2) \right) \nonumber \\
& & + i \frac{g_V g_T}{m_N^2} \left(\vec q \times (\vec P_1 - \vec P_2)\right) \cdot (\boldsigma^{(1)} + \boldsigma^{(2)})
 \Bigg\}, \nonumber  
\end{eqnarray}
where $\vec P_1 = \vec p_1 + \vec p_1^\prime$ and $\vec P_2 = \vec p_2 + \vec p_2^\prime$.
Similarly, $C^{(6)}_{\textrm{VL}}$ gives 
\begin{eqnarray}\label{eq:AVrec}
& & V(\vec q^2) =     \tau^{(1) +} \tau^{(2) +} \,  \,  G_F^2 \, m_N C_{6,\textrm{VL}} \,     \frac{1}{\vec q\sq} \, \bar u(k_1) C \gamma_\alpha \gamma_5 \bar u^T(k_2)  \, i \frac{g_A g_V}{2 m^2_N}  \nonumber \\
& &  \left( 
\left( \vec q \times \left(\vec P_1 - \vec P_2\right)\right) \cdot   ( \boldsigma^{(1)} + \boldsigma^{(2)} ) 
- \left( \vec q \times \left(\vec P_1 + \vec P_2\right)\right) \cdot ( \boldsigma^{(1)} - \boldsigma^{(2)} ) 
   \right). 
\end{eqnarray}
The neutrino potentials in Eqs. \eqref{eq:Trec} and \eqref{eq:AVrec} enter the amplitude at $\mathcal O(\Lambda_\chi \epsilon^2_\chi)$.
The NMEs in Eq.~\eqref{eq:Trec} have not been calculated in the literature. 
Compared with the second term in Eq.~\eqref{tensor}, they are not enhanced by the large nucleon isovector magnetic moment. Therefore we expect their contribution 
to be numerically somewhat smaller. 
In the case of Eq.~\eqref{eq:AVrec}, the second term was included in the analysis of Refs. \cite{Muto:1989cd,Horoi:2017gmj}, 
where it was found to be much smaller than the magnetic term in Eq.~\eqref{eq:vlvr}. For this reason, we neglected it in our formulae for the decay rate in Section \ref{Master}.

\section{Conversion of nuclear matrix elements }\label{MEconversion}

\begin{table}
\center
\renewcommand{\arraystretch}{1.5}
\begin{tabular}{||c||c|c|c ||  }
\hline
NMEs & Ref.\ \cite{Hyvarinen:2015bda,Barea:2015kwa,Barea} & Ref.\ \cite{Javier} & Ref.\ \cite{Horoi:2017gmj}\\\hline
$M_F$		&$M_F$		&   $M_F$ 		&$M_{F,F\omega,Fq}$
\\
$M_{GT}^{AA}$ 	&$M_{GT}^{AA}$	&  $M_{GT}^{AA}$ 	&$M_{GT\omega,GTq}$
\\
$M_{GT}^{AP}$	&$M_{GT}^{AP}$	&$M_{GT}^{AP}$		&$4\frac{m_e}{B}M_{GT\pi\nu}+\frac{1}{3}M_{GT2\pi}$
\\
$M_{GT}^{PP}$	&$M_{GT}^{PP}$	&$M_{GT}^{PP}$		&$-\frac{1}{6}M_{GT2\pi}$
\\
$M_{GT}^{MM}$	&$ r_M^2 M_{GT}^{MM}$&
$  M_{GT}^{MM}$
& $r_M\frac{g_M}{2 g_Ag_V R_A m_N}M_R= \frac{g_M\sq}{6 g_A\sq R_A m_N} M_{GT'}$
\\
$M_{T}^{AA}$&\ding{55}&\ding{55} &\ding{55}
\\
$M_{T}^{AP}$&$M_{T}^{AP}$&$M_{T}^{AP}$&$4\frac{m_e}{B}M_{T\pi\nu}+\frac{1}{3}M_{T2\pi}$
\\
$M_{T}^{PP}$&$M_{T}^{PP}$&$M_{T}^{PP}$&$-\frac{1}{6}M_{T2\pi}$
\\
$M_{T}^{MM}$&$r_M^2 M_{T}^{MM}$&$M_{T}^{MM}$&$-\frac{g_M\sq}{12 g_A\sq R_A m_N} M_T'$
\\\hline
$M_{F,sd}$&  $\frac{m_e m_N}{m_\pi^2} M_{F,sd}$&$\frac{m_e m_N}{m_\pi^2} M_{F,sd}$&$ \frac{m_e m_N}{m_\pi^2} M_{FN}=\frac{m_N}{R_A m_\pi^2}M_F'$
\\
$M^{AA}_{GT,sd}$&  $\frac{m_e m_N}{m_\pi^2} M^{AA}_{GT,sd}$&$\frac{m_e m_N}{m_\pi^2} M^{AA}_{GT,sd}$&$ \frac{m_e m_N}{m_\pi^2} M_{GTN}=\frac{m_N}{R_A m_\pi^2}M_{GT}'$\\
$M_{GT,sd}^{AP}$&$\frac{m_e m_N}{m_\pi^2} M_{GT,sd}^{AP}$&$\frac{m_e m_N}{m_\pi^2} M_{GT,sd}^{AP}$&$\frac{2}{3} M_{GT1\pi}$
\\
$M_{GT,sd}^{PP}$&$ \frac{m_e m_N}{m_\pi^2} M_{GT,sd}^{PP}$&$ \frac{m_e m_N}{m_\pi^2} M_{GT,sd}^{PP}$&$\frac{1}{6}(M_{GT2\pi}-2M_{GT1\pi})$
\\
$M_{T,sd}^{AP}$&$\frac{m_e m_N}{m_\pi^2} M_{T,sd}^{AP}$&$\frac{m_e m_N}{m_\pi^2} M_{T,sd}^{AP}$&$\frac{2}{3}M_{T1\pi}$
\\
$M_{T,sd}^{PP}$&$\frac{m_e m_N}{m_\pi^2} M_{T,sd}^{PP}$&$\frac{m_e m_N}{m_\pi^2} M_{T,sd}^{PP}$&$\frac{1}{6}(M_{T2\pi}-2M_{T1\pi})$
\\
\hline \hline
\end{tabular}
\caption{Comparison of the different notations used in Refs.\ \cite{Hyvarinen:2015bda,Muto:1989cd,Horoi:2017gmj}. For each row the expressions in the different columns equal one another in the limit that $\bar E\to 0$. Furthermore, $B=\frac{m_\pi\sq}{m_u+m_d}$, where Ref.~\cite{Horoi:2017gmj} uses $m_u+m_d = 11.6$ MeV.
$g_{M}^{}$ has different definitions in various papers. Here we use $g_M =1 + \kappa_1$ and introduce the ratio $r_M = (1 + \kappa_1)/\kappa_1$.}
\label{tab:NMEs}
\end{table}

In this appendix, we provide the conversion between the NMEs defined in Sec. \ref{NME} and 
those of the original papers \cite{Hyvarinen:2015bda,Muto:1989cd,Horoi:2017gmj,Barea:2015kwa,Barea,Javier}.

For the matrix elements involving the exchange of a light neutrino, our definitions match those in Refs.  \cite{Hyvarinen:2015bda,Barea:2015kwa,Barea,Javier}.
The only exceptions are $M^{MM}_{GT, T}$, for which  Refs.  \cite{Hyvarinen:2015bda,Barea:2015kwa,Barea}
used $g_M(0) = \kappa_1 = 3.7$ rather than $g_M(0) = 1 + \kappa_1$. In Section \ref{NME}, we thus rescaled these matrix elements by powers of $r_M = (1 + \kappa_1)/\kappa_1$.
For the Gamow-Teller and tensor matrix elements, Ref.~\cite{Horoi:2017gmj} does not separately provide the $AA$, $AP$, $PP$ and $MM$ components.   
However, we can reconstruct the needed NMEs from linear combinations of other matrix elements computed in Ref. \cite{Horoi:2017gmj}, as detailed in Table \ref{tab:NMEs}.
The definitions of the NMEs in the third column of Table \ref{tab:NMEs} are given in  Ref.\ \cite{Horoi:2017gmj}~\footnote{The relation between $M_{GT}^{MM}$ and $M_R$ given in Table \ref{tab:NMEs} takes into account a  factor of $1/3$ that is missing from the definition of $H_R$ in Eq.\ (21v) of (the first arXiv version of) Ref.\ \cite{Horoi:2017gmj}. We thank M.\ Horoi for clarification on this issue.}.

The relations we use are  valid at LO in the chiral expansion, when one can take $\bar E \rightarrow 0$ and neglect subleading effects as the difference between the axial and vector form factors.
We discussed some  checks of these assumptions in Sec. \ref{NME}. Additional consistency checks can be performed with the NMEs of  Ref.~\cite{Horoi:2017gmj}.
In the limit $\bar{E} \rightarrow 0$, one would expect $M_{F}=M_{F\omega} =M_{Fq}$ and $M_{GT\omega} =M_{GTq}$.
These relations are respected  to a few percent for $M_{F}$ and $M_{F\omega}$, while $M_{Fq}$ appears to be  $\sim 50\%$ smaller than $M_{F\omega}$. 
The relation between the GT elements holds to $20\%$.
Furthermore, we can use the complete $GT$ and $T$ matrix elements computed in  Ref.~\cite{Horoi:2017gmj} to verify whether $M_{GT}=M_{GT}^{AA}+M_{GT}^{AP}+M_{GT}^{PP}+M_{GT}^{MM}$ and $M_{T}=M_{T}^{AP}+M_{T}^{PP}+M_{T}^{MM}$. 
The agreement is within $20\%$ for the $GT$ elements and for most of the $T$ matrix elements.
In the main body of the paper, to obtain $M_{F}$, $M_{GT}^{AA}$, $M_{GT}^{MM}$, $M_{F,\, sd}$, and $M^{AA}_{GT,\, sd}$ from the results of Ref. \cite{Horoi:2017gmj}
we used, respectively, $M_{F}$, $M_{GT\omega}$, $M_{GT'}$, $M_{FN}$, and $M_{GTN}$.

The long-distance matrix element $M_{T}^{AA}$ is not defined in Refs.   \cite{Hyvarinen:2015bda,Barea:2015kwa,Barea,Javier},
since it does not appear in the standard scenario of light Majorana neutrino exchange.  Ref.\  \cite{Horoi:2017gmj}
computes similar tensor matrix elements, which are needed in neutrino exchange diagram when the neutrino is emitted from a $\Delta L=2$ vector or axial current, as in the second diagram of Fig. \ref{twobody}.
We were however not able to relate $M_{Tq}$ of Ref.  \cite{Horoi:2017gmj} to $M^{AA}_{T}$, even in the $\bar E \rightarrow 0$ limit.
$M^{AA}_T$ is related to $M_T$ of Ref. \cite{Muto:1989cd} by $M^{AA}_{T} = 3/2 M_T$. With the values of Ref. \cite{Muto:1989cd}, $M^{AA}_{T}$ has only a small effects on the bounds on $C^{(6)}_{\rm VR}$,
and can be safely neglected.

For the short-distance matrix elements, which do not involve  neutrino exchange, our definitions differ from Refs. \cite{Hyvarinen:2015bda,Barea:2015kwa,Barea,Javier}
only in the overall normalization. To keep the power counting of the NMEs manifest, we normalized them to $m_\pi^2$ rather than $m_e m_N$.    
Ref. \cite{Horoi:2017gmj} computed the pion-exchange matrix elements $M_{GT1\pi}$, $M_{GT2\pi}$, $M_{T1\pi}$, $M_{T2\pi}$, which are related to $M^{AP, PP}_{GT, sd}$
and $M^{AP, PP}_{T, sd}$  by the equations in Tab. \ref{tab:NMEs}

\bibliographystyle{h-physrev3} 
\bibliography{bibliography}

\begin{thebibliography}{10}

\bibitem{Weinberg:1979sa}
S.~Weinberg,
\newblock Phys. Rev. Lett. {\bf 43}, 1566 (1979).
%%CITATION = PRLTA,43,1566;%%

\bibitem{Schechter:1981bd}
J.~Schechter and J.~W.~F. Valle,
\newblock Phys. Rev. {\bf D25}, 2951 (1982).
%%CITATION = PHRVA,D25,2951;%%

\bibitem{Davidson:2008bu}
S.~Davidson, E.~Nardi, and Y.~Nir,
\newblock Phys. Rept. {\bf 466}, 105 (2008), 0802.2962.
%%CITATION = ARXIV:0802.2962;%%

\bibitem{Gando:2012zm}
KamLAND-Zen, A.~Gando {\em et~al.},
\newblock Phys. Rev. Lett. {\bf 110}, 062502 (2013), 1211.3863.
%%CITATION = ARXIV:1211.3863;%%

\bibitem{Agostini:2013mzu}
GERDA, M.~Agostini {\em et~al.},
\newblock Phys. Rev. Lett. {\bf 111}, 122503 (2013), 1307.4720.
%%CITATION = ARXIV:1307.4720;%%

\bibitem{Albert:2014awa}
EXO-200, J.~B. Albert {\em et~al.},
\newblock Nature {\bf 510}, 229 (2014), 1402.6956.
%%CITATION = ARXIV:1402.6956;%%

\bibitem{Alfonso:2015wka}
CUORE, K.~Alfonso {\em et~al.},
\newblock Phys. Rev. Lett. {\bf 115}, 102502 (2015), 1504.02454.
%%CITATION = ARXIV:1504.02454;%%

\bibitem{Andringa:2015tza}
SNO+, S.~Andringa {\em et~al.},
\newblock Adv. High Energy Phys. {\bf 2016}, 6194250 (2016), 1508.05759.
%%CITATION = ARXIV:1508.05759;%%

\bibitem{Arnold:2015wpy}
NEMO-3, R.~Arnold {\em et~al.},
\newblock Phys. Rev. {\bf D92}, 072011 (2015), 1506.05825.
%%CITATION = ARXIV:1506.05825;%%

\bibitem{Elliott:2016ble}
S.~R. Elliott {\em et~al.},
\newblock {Initial Results from the MAJORANA DEMONSTRATOR},
\newblock 2016, 1610.01210.
%%CITATION = ARXIV:1610.01210;%%

\bibitem{Arnold:2016ezh}
NEMO-3, R.~Arnold {\em et~al.},
\newblock Phys. Rev. {\bf D93}, 112008 (2016), 1604.01710.
%%CITATION = ARXIV:1604.01710;%%

\bibitem{Agostini:2017iyd}
M.~Agostini {\em et~al.},
\newblock Nature {\bf 544}, 47 (2017), 1703.00570.
%%CITATION = ARXIV:1703.00570;%%

\bibitem{KamLAND-Zen:2016pfg}
KamLAND-Zen, A.~Gando {\em et~al.},
\newblock Phys. Rev. Lett. {\bf 117}, 082503 (2016), 1605.02889,
\newblock [Addendum: Phys. Rev. Lett.117,no.10,109903(2016)].
%%CITATION = ARXIV:1605.02889;%%

\bibitem{Rodejohann:2011mu}
W.~Rodejohann,
\newblock Int. J. Mod. Phys. {\bf E20}, 1833 (2011), 1106.1334.
%%CITATION = ARXIV:1106.1334;%%

\bibitem{Mohapatra:1974hk}
R.~N. Mohapatra and J.~C. Pati,
\newblock Phys. Rev. D {\bf 11}, 566 (1975).
%%CITATION = PHRVA,D11,566;%%

\bibitem{Senjanovic:1975rk}
G.~Senjanovic and R.~N. Mohapatra,
\newblock Phys. Rev. D {\bf 12}, 1502 (1975).
%%CITATION = PHRVA,D12,1502;%%

\bibitem{Mohapatra:1983aa}
R.~N. Mohapatra,
\newblock NATO Sci. Ser. B {\bf 122}, 219 (1985).
%%CITATION = PRINT-84-0012 (MARYLAND);%%

\bibitem{Doi:1985dx}
M.~Doi, T.~Kotani, and E.~Takasugi,
\newblock Prog. Theor. Phys. Suppl. {\bf 83}, 1 (1985).
%%CITATION = PTPSA,83,1;%%

\bibitem{Tello:2010am}
V.~Tello, M.~Nemevsek, F.~Nesti, G.~Senjanovic, and F.~Vissani,
\newblock Phys. Rev. Lett. {\bf 106}, 151801 (2011), 1011.3522.
%%CITATION = ARXIV:1011.3522;%%

\bibitem{Ge:2015yqa}
S.-F. Ge, M.~Lindner, and S.~Patra,
\newblock JHEP {\bf 10}, 077 (2015), 1508.07286.
%%CITATION = ARXIV:1508.07286;%%

\bibitem{Buchmuller:1985jz}
W.~Buchm{\"u}ller and D.~Wyler,
\newblock Nucl. Phys. B {\bf 268}, 621 (1986).
%%CITATION = NUPHA,B268,621;%%

\bibitem{Grzadkowski:2010es}
B.~Grzadkowski, M.~Iskrzynski, M.~Misiak, and J.~Rosiek,
\newblock JHEP {\bf 1010}, 085 (2010), 1008.4884.
%%CITATION = ARXIV:1008.4884;%%

\bibitem{Lehman:2014jma}
L.~Lehman,
\newblock Phys. Rev. {\bf D90}, 125023 (2014), 1410.4193.
%%CITATION = ARXIV:1410.4193;%%

\bibitem{Prezeau:2003xn}
G.~Prezeau, M.~Ramsey-Musolf, and P.~Vogel,
\newblock Phys. Rev. {\bf D68}, 034016 (2003), hep-ph/0303205.
%%CITATION = HEP-PH/0303205;%%

\bibitem{Graesser:2016bpz}
M.~L. Graesser,
\newblock JHEP {\bf 08}, 099 (2017), 1606.04549.
%%CITATION = ARXIV:1606.04549;%%

\bibitem{deGouvea:2007qla}
A.~de~Gouvea and J.~Jenkins,
\newblock Phys. Rev. {\bf D77}, 013008 (2008), 0708.1344.
%%CITATION = ARXIV:0708.1344;%%

\bibitem{Kobach:2016ami}
A.~Kobach,
\newblock Phys. Lett. {\bf B758}, 455 (2016), 1604.05726.
%%CITATION = ARXIV:1604.05726;%%

\bibitem{Liao:2016qyd}
Y.~Liao and X.-D. Ma,
\newblock Phys. Rev. {\bf D96}, 015012 (2017), 1612.04527.
%%CITATION = ARXIV:1612.04527;%%

\bibitem{Pas:1999fc}
H.~Pas, M.~Hirsch, H.~V. Klapdor-Kleingrothaus, and S.~G. Kovalenko,
\newblock Phys. Lett. {\bf B453}, 194 (1999).
%%CITATION = PHLTA,B453,194;%%

\bibitem{Deppisch:2012nb}
F.~F. Deppisch, M.~Hirsch, and H.~Pas,
\newblock J. Phys. {\bf G39}, 124007 (2012), 1208.0727.
%%CITATION = ARXIV:1208.0727;%%

\bibitem{Helo:2016vsi}
J.~C. Helo, M.~Hirsch, and T.~Ota,
\newblock JHEP {\bf 06}, 006 (2016), 1602.03362.
%%CITATION = ARXIV:1602.03362;%%

\bibitem{Horoi:2017gmj}
M.~Horoi and A.~Neacsu,
\newblock (2017), 1706.05391.
%%CITATION = ARXIV:1706.05391;%%

\bibitem{Vergados:1986td}
J.~D. Vergados,
\newblock Phys. Lett. {\bf B184}, 55 (1987).
%%CITATION = PHLTA,B184,55;%%

\bibitem{Faessler:1996ph}
A.~Faessler, S.~Kovalenko, F.~Simkovic, and J.~Schwieger,
\newblock Phys. Rev. Lett. {\bf 78}, 183 (1997), hep-ph/9612357.
%%CITATION = HEP-PH/9612357;%%

\bibitem{Pas:2000vn}
H.~Pas, M.~Hirsch, H.~V. Klapdor-Kleingrothaus, and S.~G. Kovalenko,
\newblock Phys. Lett. {\bf B498}, 35 (2001), hep-ph/0008182.
%%CITATION = HEP-PH/0008182;%%

\bibitem{Bonnet:2012kh}
F.~Bonnet, M.~Hirsch, T.~Ota, and W.~Winter,
\newblock JHEP {\bf 03}, 055 (2013), 1212.3045,
\newblock [Erratum: JHEP04,090(2014)].
%%CITATION = ARXIV:1212.3045;%%

\bibitem{Babu:2001ex}
K.~S. Babu and C.~N. Leung,
\newblock Nucl. Phys. {\bf B619}, 667 (2001), hep-ph/0106054.
%%CITATION = HEP-PH/0106054;%%

\bibitem{Bell:2006wi}
N.~F. Bell, M.~Gorchtein, M.~J. Ramsey-Musolf, P.~Vogel, and P.~Wang,
\newblock Phys. Lett. {\bf B642}, 377 (2006), hep-ph/0606248.
%%CITATION = HEP-PH/0606248;%%

\bibitem{Liao:2016hru}
Y.~Liao and X.-D. Ma,
\newblock JHEP {\bf 11}, 043 (2016), 1607.07309.
%%CITATION = ARXIV:1607.07309;%%

\bibitem{Armbruster:2003pq}
B.~Armbruster {\em et~al.},
\newblock Phys. Rev. Lett. {\bf 90}, 181804 (2003), hep-ex/0302017.
%%CITATION = HEP-EX/0302017;%%

\bibitem{Larin:1991tj}
S.~A. Larin and J.~A.~M. Vermaseren,
\newblock Phys. Lett. {\bf B259}, 345 (1991).
%%CITATION = PHLTA,B259,345;%%

\bibitem{Collins:1984xc}
J.~C. Collins,
\newblock {\em {Renormalization }}volume~26 of {\em Cambridge Monographs on
  Mathematical Physics} (Cambridge University Press, Cambridge, 1986).
%%CITATION = INSPIRE-209810;%%

\bibitem{Arbelaez:2016zlt}
C.~Arbel\'aez, M.~Gonz\'alez, M.~Hirsch, and S.~Kovalenko,
\newblock Phys. Rev. {\bf D94}, 096014 (2016), 1610.04096.
%%CITATION = ARXIV:1610.04096;%%

\bibitem{Buras:2000if}
A.~J. Buras, M.~Misiak, and J.~Urban,
\newblock Nucl. Phys. {\bf B586}, 397 (2000), hep-ph/0005183.
%%CITATION = HEP-PH/0005183;%%

\bibitem{Buras:2001ra}
A.~J. Buras, S.~Jager, and J.~Urban,
\newblock Nucl. Phys. {\bf B605}, 600 (2001), hep-ph/0102316.
%%CITATION = HEP-PH/0102316;%%

\bibitem{Weinberg:1978kz}
S.~Weinberg,
\newblock Physica {\bf A96}, 327 (1979).
%%CITATION = PHYSA,A96,327;%%

\bibitem{Gasser:1983yg}
J.~Gasser and H.~Leutwyler,
\newblock Annals Phys. {\bf 158}, 142 (1984).
%%CITATION = APNYA,158,142;%%

\bibitem{Bernard:1995dp}
V.~Bernard, N.~Kaiser, and U.-G. Mei{\ss}ner,
\newblock Int. J. Mod. Phys. {\bf E4}, 193 (1995), hep-ph/9501384.
%%CITATION = HEP-PH/9501384;%%

\bibitem{Weinberg:1990rz}
S.~Weinberg,
\newblock Phys. Lett. {\bf B251}, 288 (1990).
%%CITATION = PHLTA,B251,288;%%

\bibitem{Weinberg:1991um}
S.~Weinberg,
\newblock Nucl. Phys. {\bf B363}, 3 (1991).
%%CITATION = NUPHA,B363,3;%%

\bibitem{Ordonez:1992xp}
C.~Ordonez and U.~van Kolck,
\newblock Phys. Lett. {\bf B291}, 459 (1992).
%%CITATION = PHLTA,B291,459;%%

\bibitem{Epelbaum:2008ga}
E.~Epelbaum, H.-W. Hammer, and U.-G. Mei{\ss}ner,
\newblock Rev. Mod. Phys. {\bf 81}, 1773 (2009), 0811.1338.
%%CITATION = ARXIV:0811.1338;%%

\bibitem{Manohar:1983md}
A.~Manohar and H.~Georgi,
\newblock Nucl. Phys. {\bf B234}, 189 (1984).
%%CITATION = NUPHA,B234,189;%%

\bibitem{Kaplan:1998tg}
D.~B. Kaplan, M.~J. Savage, and M.~B. Wise,
\newblock Phys. Lett. {\bf B424}, 390 (1998), nucl-th/9801034.
%%CITATION = NUCL-TH/9801034;%%

\bibitem{Bedaque:2002mn}
P.~F. Bedaque and U.~van Kolck,
\newblock Ann. Rev. Nucl. Part. Sci. {\bf 52}, 339 (2002), nucl-th/0203055.
%%CITATION = NUCL-TH/0203055;%%

\bibitem{Cirigliano:2017ymo}
V.~Cirigliano, W.~Dekens, M.~Graesser, and E.~Mereghetti,
\newblock Phys. Lett. {\bf B769}, 460 (2017), 1701.01443.
%%CITATION = ARXIV:1701.01443;%%

\bibitem{Nicholson:2016byl}
A.~Nicholson {\em et~al.},
\newblock {Neutrinoless double beta decay from lattice QCD},
\newblock in {\em {Proceedings, 34th International Symposium on Lattice Field
  Theory (Lattice 2016): Southampton, UK, July 24-30, 2016}}, 2016, 1608.04793.
%%CITATION = ARXIV:1608.04793;%%

\bibitem{Olive:2016xmw}
Particle Data Group, C.~Patrignani {\em et~al.},
\newblock Chin. Phys. {\bf C40}, 100001 (2016).
%%CITATION = CHPHD,C40,100001;%%

\bibitem{Savage:1998yh}
M.~J. Savage,
\newblock Phys. Rev. {\bf C59}, 2293 (1999), nucl-th/9811087.
%%CITATION = NUCL-TH/9811087;%%

\bibitem{Bhattacharya:2016zcn}
T.~Bhattacharya {\em et~al.},
\newblock Phys. Rev. {\bf D94}, 054508 (2016), 1606.07049.
%%CITATION = ARXIV:1606.07049;%%

\bibitem{Jenkins:1990jv}
E.~E. Jenkins and A.~V. Manohar,
\newblock Phys.Lett. {\bf B255}, 558 (1991).
%%CITATION = PHLTA,B255,558;%%

\bibitem{Luke:1992cs}
M.~E. Luke and A.~V. Manohar,
\newblock Phys. Lett. {\bf B286}, 348 (1992), hep-ph/9205228.
%%CITATION = HEP-PH/9205228;%%

\bibitem{Adler:1975he}
S.~L. Adler {\em et~al.},
\newblock Phys. Rev. {\bf D11}, 3309 (1975).
%%CITATION = PHRVA,D11,3309;%%

\bibitem{Hirsch:1995cg}
M.~Hirsch, H.~V. Klapdor-Kleingrothaus, and S.~G. Kovalenko,
\newblock Phys. Lett. {\bf B372}, 181 (1996), hep-ph/9512237,
\newblock [Erratum: Phys. Lett.B381,488(1996)].
%%CITATION = HEP-PH/9512237;%%

\bibitem{AWL}
V.~Cirigliano, W.~Dekens, E.~Mereghetti, and A.~Walker-Loud,
\newblock in preparation .

\bibitem{Engel:2016xgb}
J.~Engel and J.~Menendez,
\newblock Rept. Prog. Phys. {\bf 80}, 046301 (2017), 1610.06548.
%%CITATION = ARXIV:1610.06548;%%

\bibitem{Simkovic:1999re}
F.~Simkovic, G.~Pantis, J.~D. Vergados, and A.~Faessler,
\newblock Phys. Rev. {\bf C60}, 055502 (1999), hep-ph/9905509.
%%CITATION = HEP-PH/9905509;%%

\bibitem{Tomoda:1990rs}
T.~Tomoda,
\newblock Rept. Prog. Phys. {\bf 54}, 53 (1991).
%%CITATION = RPPHA,54,53;%%

\bibitem{Gonzalez-Alonso:2013ura}
M.~Gonz\'alez-Alonso and J.~Martin~Camalich,
\newblock Phys. Rev. Lett. {\bf 112}, 042501 (2014), 1309.4434.
%%CITATION = ARXIV:1309.4434;%%

\bibitem{Brantley:2016our}
D.~A. Brantley {\em et~al.},
\newblock (2016), 1612.07733.
%%CITATION = ARXIV:1612.07733;%%

\bibitem{Bhattacharya:2015wna}
PNDME, T.~Bhattacharya {\em et~al.},
\newblock Phys. Rev. {\bf D92}, 094511 (2015), 1506.06411.
%%CITATION = ARXIV:1506.06411;%%

\bibitem{Rajan:2017lxk}
G.~Rajan, J.~Yong-Chull, L.~Huey-Wen, Y.~Boram, and B.~Tanmoy,
\newblock (2017), 1705.06834.
%%CITATION = ARXIV:1705.06834;%%

\bibitem{Menendez:2008jp}
J.~Menendez, A.~Poves, E.~Caurier, and F.~Nowacki,
\newblock Nucl. Phys. {\bf A818}, 139 (2009), 0801.3760.
%%CITATION = ARXIV:0801.3760;%%

\bibitem{Barea:2009zza}
J.~Barea and F.~Iachello,
\newblock Phys. Rev. {\bf C79}, 044301 (2009).
%%CITATION = PHRVA,C79,044301;%%

\bibitem{Menendez:2011qq}
J.~Menendez, D.~Gazit, and A.~Schwenk,
\newblock Phys. Rev. Lett. {\bf 107}, 062501 (2011), 1103.3622.
%%CITATION = ARXIV:1103.3622;%%

\bibitem{Hyvarinen:2015bda}
J.~Hyv\"arinen and J.~Suhonen,
\newblock Phys. Rev. {\bf C91}, 024613 (2015).
%%CITATION = PHRVA,C91,024613;%%

\bibitem{Muto:1989cd}
K.~Muto, E.~Bender, and H.~V. Klapdor,
\newblock Z. Phys. {\bf A334}, 187 (1989).
%%CITATION = ZEPYA,A334,187;%%

\bibitem{Vergados:2012xy}
J.~D. Vergados, H.~Ejiri, and F.~Simkovic,
\newblock Rept. Prog. Phys. {\bf 75}, 106301 (2012), 1205.0649.
%%CITATION = ARXIV:1205.0649;%%

\bibitem{Bilenky:2014uka}
S.~M. Bilenky and C.~Giunti,
\newblock Int. J. Mod. Phys. {\bf A30}, 1530001 (2015), 1411.4791.
%%CITATION = ARXIV:1411.4791;%%

\bibitem{Stefanik:2015twa}
D.~Stefanik, R.~Dvornicky, F.~Simkovic, and P.~Vogel,
\newblock Phys. Rev. {\bf C92}, 055502 (2015), 1506.07145.
%%CITATION = ARXIV:1506.07145;%%

\bibitem{Kotila:2012zza}
J.~Kotila and F.~Iachello,
\newblock Phys. Rev. {\bf C85}, 034316 (2012), 1209.5722.
%%CITATION = ARXIV:1209.5722;%%

\bibitem{Stoica:2013lka}
S.~Stoica and M.~Mirea,
\newblock Phys. Rev. {\bf C88}, 037303 (2013), 1307.0290.
%%CITATION = ARXIV:1307.0290;%%

\bibitem{Javier}
J.~Menendez,
\newblock private communication .

\bibitem{Barea:2015kwa}
J.~Barea, J.~Kotila, and F.~Iachello,
\newblock Phys. Rev. {\bf C91}, 034304 (2015), 1506.08530.
%%CITATION = ARXIV:1506.08530;%%

\bibitem{Barea}
J.~Barea,
\newblock private communication .

\bibitem{SeNEMO}
D.~Waters,
\newblock {Latest Results from NEMO-3 \& Status of the SuperNEMO experiment},
\newblock in {\em XXVII International Conference on Neutrino Physics and
  Astrophysics, London, July 2016}.

\bibitem{Agostini:2016iid}
M.~Agostini {\em et~al.},
\newblock Nucl. Part. Phys. Proc. {\bf 273-275}, 1876 (2016).
%%CITATION = INSPIRE-1467802;%%

\bibitem{Giunti:2014ixa}
C.~Giunti and A.~Studenikin,
\newblock Rev. Mod. Phys. {\bf 87}, 531 (2015), 1403.6344.
%%CITATION = ARXIV:1403.6344;%%

\bibitem{Canas:2015yoa}
B.~C. Canas, O.~G. Miranda, A.~Parada, M.~Tortola, and J.~W.~F. Valle,
\newblock Phys. Lett. {\bf B753}, 191 (2016), 1510.01684,
\newblock [Addendum: Phys. Lett.B757,568(2016)].
%%CITATION = ARXIV:1510.01684;%%

\bibitem{Raffelt:1998xu}
G.~G. Raffelt,
\newblock Phys. Rev. Lett. {\bf 81}, 4020 (1998), astro-ph/9808299.
%%CITATION = ASTRO-PH/9808299;%%

\bibitem{Deppisch:2006hb}
F.~Deppisch and H.~Pas,
\newblock Phys. Rev. Lett. {\bf 98}, 232501 (2007), hep-ph/0612165.
%%CITATION = HEP-PH/0612165;%%

\bibitem{Gehman:2007qg}
V.~M. Gehman and S.~R. Elliott,
\newblock J. Phys. {\bf G34}, 667 (2007), hep-ph/0701099,
\newblock [Erratum: J. Phys.G35,029701(2008)].
%%CITATION = HEP-PH/0701099;%%

\bibitem{Fogli:2009py}
G.~L. Fogli, E.~Lisi, and A.~M. Rotunno,
\newblock Phys. Rev. {\bf D80}, 015024 (2009), 0905.1832.
%%CITATION = ARXIV:0905.1832;%%

\bibitem{Lisi:2015yma}
E.~Lisi, A.~Rotunno, and F.~Simkovic,
\newblock Phys. Rev. {\bf D92}, 093004 (2015), 1506.04058.
%%CITATION = ARXIV:1506.04058;%%

\bibitem{Peng:2015haa}
T.~Peng, M.~J. Ramsey-Musolf, and P.~Winslow,
\newblock Phys. Rev. {\bf D93}, 093002 (2016), 1508.04444.
%%CITATION = ARXIV:1508.04444;%%

\bibitem{Arnold:2010tu}
SuperNEMO, R.~Arnold {\em et~al.},
\newblock Eur. Phys. J. {\bf C70}, 927 (2010), 1005.1241.
%%CITATION = ARXIV:1005.1241;%%

\bibitem{Shanahan:2017bgi}
P.~E. Shanahan {\em et~al.},
\newblock Phys. Rev. Lett. {\bf 119}, 062003 (2017), 1701.03456.
%%CITATION = ARXIV:1701.03456;%%

\end{thebibliography}

\end{document}